\documentclass[12pt]{article}
\usepackage[english]{babel}
\usepackage[margin=1in]{geometry}
\usepackage{color-edits}
\usepackage{setspace} 

\usepackage{amsmath}
\usepackage{amssymb, amsthm, mathrsfs}
\usepackage{mathtools}
\usepackage{newtxtext,newtxmath}
\usepackage{bbm}

\usepackage{pdfpages}
\usepackage{pgfplots}
\pgfplotsset{compat=1.15}
\definecolor{contestcol}{RGB}{31,60,155}   
\definecolor{hybridcol}{RGB}{224,130,20}   
\definecolor{innovcol}{RGB}{20,140,70}    
\definecolor{bdcol}{RGB}{120,40,150}      
\definecolor{plaincol}{RGB}{30,140,70}   

\usepackage{graphicx,multicol}
\usepackage{caption,subcaption}
\usepackage{verbatim}
\usepackage{multirow}
\usepackage{rotating}
\usepackage{tabularx}
\usepackage{xcolor}
\usepackage{array}
\usepackage{makecell}
\usepackage[thinc]{esdiff}

\usepackage{nicefrac}
\usepackage{ifthen}
\usepackage{booktabs}
\usepackage{bm}
\usepackage{natbib}
\usepackage{enumerate}
\usepackage{thm-restate}
\usepackage{csquotes}
\usepackage[capposition=top]{floatrow}
\usepackage{soul}

\usepackage{tikz}
\usetikzlibrary{patterns}
\usetikzlibrary{calc}
\usetikzlibrary{decorations.pathreplacing}
\usetikzlibrary{arrows}
\usetikzlibrary{arrows.meta, positioning}
\usetikzlibrary{shapes.geometric} 

\newcommand{\given}{\,|\,}

\newcommand{\prob}[2][]{\mathbb{P}\ifthenelse{\not\equal{}{#1}}{_{#1}}{}\!\left({\def\givenn{\middle|}#2}\right)}
\newcommand{\expect}[2][]{\mathbb{E}\ifthenelse{\not\equal{}{#1}}{_{#1}}{}\!\left[{\def\givenn{\middle|}#2}\right]}

\newcommand{\tparen}{\big}
\newcommand{\tprob}[2][]{\text{\bf P}\ifthenelse{\not\equal{}{#1}}{_{#1}}{}\tparen[{\def\given{\tparen|}#2}\tparen]}
\newcommand{\texpect}[2][]{\text{\bf E}\ifthenelse{\not\equal{}{#1}}{_{#1}}{}\tparen[{\def\given{\tparen|}#2}\tparen]}

\newcommand{\sprob}[2][]{\text{\bf P}\ifthenelse{\not\equal{}{#1}}{_{#1}}{}[#2]}
\newcommand{\sexpect}[2][]{\text{\bf E}\ifthenelse{\not\equal{}{#1}}{_{#1}}{}[#2]}

\theoremstyle{definition}

\newtheorem{example}{Example}
\newtheorem{assumption}{Assumption}

\theoremstyle{plain}
\newtheorem{theorem}{Theorem}
\newtheorem{corollary}{Corollary}
\newtheorem{lemma}{Lemma}
\newtheorem{proposition}{Proposition}
\newtheorem{claim}{Claim}

\theoremstyle{remark}

\usepackage[nocomma]{optidef}

\graphicspath{ {./images/} }

\newcommand{\calD}{\mathcal{D}}

\newcommand{\calH}{\mathcal{H}}

\newcommand{\calM}{\mathcal{M}}

\newcommand{\calP}{\mathcal{P}}

\newcommand{\calT}{\mathcal{T}}

\newcommand{\indicate}[1]{{\bf 1}\left[#1\right]}

\newcommand{\rate}{\lambda}
\newcommand{\cost}{c}
\newcommand{\totalcost}{\bar{\cost}}

\newcommand{\projset}{Y}

\newcommand{\strategy}{\alpha}
\newcommand{\Strategy}{A}

\newcommand{\taus}{\boldsymbol{\tau}}

\newcommand{\private}{p}

\newcommand{\mech}{\calM}
\newcommand{\Hybrid}{\calH}
\newcommand{\contract}{C}
\newcommand{\contracts}{\boldsymbol{\contract}}
\newcommand{\jointcontest}{J}
\newcommand{\rankcontest}{R}
\newcommand{\hybrid}{S}
\newcommand{\hybrids}{\boldsymbol{\hybrid}}

\newcommand{\Payment}{B}  

\newcommand{\payment}{b}  

\newcommand{\compensation}{Z} 

\newcommand{\share}{\pi}
\newcommand{\shares}{\boldsymbol{\share}}
\newcommand{\successtime}{s}
\newcommand{\successtimes}{\boldsymbol{\successtime}}
\newcommand{\rank}{r}

\newcommand{\outcome}{o}

\newcommand{\disclosure}{d}

\newcommand{\drule}{D}

\newcommand{\dset}{\calD}

\newcommand{\nosuccess}{\varnothing}

\newcommand{\stopset}{\calT}

\newcommand{\val}{V}

\newcommand{\obj}{\text{Obj}}

\newcommand{\continuation}{\Gamma}

\usepackage{titlesec}

\titleformat{\section}
  {\normalfont\Large\bfseries}
  {\thesection.}{1em}{}

\titlespacing*{\section}
  {-1em}{-8pt}{-3pt}

\titleformat{\subsection}
  {\normalfont\large\bfseries}
  {\thesubsection.}{1em}{}

\titlespacing*{\subsection}
  {-1em}{-3pt}{-3pt}

\usepackage[colorlinks=true, urlcolor=blue, linkcolor=blue, citecolor=magenta]{hyperref}
\usepackage[capitalize]{cleveref}

\usepackage{color-edits}
\addauthor{xq}{blue}    
\usepackage{xcolor}
\newcommand{\proofedit}[1]{{\color{blue}#1}}
\renewcommand{\proofedit}[1]{#1}

\begin{document}

\begin{spacing}{1.5}


\begin{titlepage}  
\singlespace 
\pagestyle{empty}

\begin{center}
\large \bf Contests Versus Contracts in Innovation Procurement
\end{center}

\vfill

\begin{flushleft}

Xiaoyun Qiu\\
{\em  Lead author. Department of Economics, Dartmouth College, Hanover NH, USA\\
Email: xiaoyun.qiu@dartmouth.edu}\\[2ex]

Christopher M. Snyder\\
{\em Department of Economics, Dartmouth College, Hanover NH, USA\\
Email: chris.snyder@dartmouth.edu}

\end{flushleft}

\vfill

\begin{center}
\today\\

\href{https://drive.google.com/file/d/1velnr9hupdiyx7guISHG12wDkyOJcACx/view?usp=sharing}{Click  here for the latest version}
\end{center}

\vfill 

\noindent {\bf Abstract:}  How much competition should a designer introduce when procuring innovation? We study a designer funding multiple technological approaches whose feasibility is learned through experimentation. Between independent contracts and contests, we introduce \emph{interdependent contracts}, which combine approach-specific rewards with continuation decisions that depend on rivals' outcomes. When both breakthroughs and breakdowns are contractable, these contracts implement the optimum by extracting firms' rents and coordinating experimentation across approaches. When only breakthroughs are contractable, we compare simple mechanisms with constant contractual payments. Under technological symmetry, optimal interdependent contracts terminate experimentation at a success cutoff or deadline. Fixed-prize contests can implement the same innovation outcomes, but their relative expenditure depends on the experimentation horizon: for interior success cutoffs, contests are more expensive when the horizon is short and less expensive when it is sufficiently long. Rank-contingent contests can weakly outperform constant-payment interdependent contracts, but technological heterogeneity can reverse this ranking because individualized contracts better tailor incentives to different approaches. We further characterize optimal interdependent contracts with time-varying payments, extending the comparison beyond simple mechanisms. Our results show how outcome contractability, technological heterogeneity, and payment flexibility shape the desirable form of competition in innovation procurement.

\vfill 


\vfill 

\noindent {\bf Journal of Economic Literature codes:} O31 (Innovation and Invention: Processes and Incentives); D82 (Asymmetric and Private Information, Mechanism Design);  D86 (Economics of Contract: Theory); H57 (National Government Expenditures and Related Policies: Procurement)

\vfill

\noindent {\bf Acknowledgments:} The authors thank the Market Shaping Accelerator team for helpful comments and support. The authors are grateful to Bruno Strulovici for his helpful comments and suggestions. The authors gratefully acknowledge funding from the Sloan Foundation under grant G--2025--25183.

\end{titlepage}

\section{Introduction}
\label{s:introduction}

Innovation procurement often involves multiple technological approaches to
the same problem. A buyer seeking to remove atmospheric CO$_2$, for example,
might support direct air capture, enhanced weathering, and biomass carbon
removal, each with uncertain prospects for success. The buyer can reward
each approach separately through individual contracts or have firms compete
for a common prize pool. Which arrangement best balances the value of
innovation against the spending needed to induce it, and how much
competition is optimal?

Both arrangements are widely used. Frontier's advance market commitment
for permanent carbon removal relies on supplier-specific agreements across
distinct technologies
\citep{frontier280earth2024,frontierlithos2023,frontierexergi2024}.
Operation Warp Speed similarly supported vaccine candidates through
separate development and procurement agreements spanning mRNA,
viral-vector, and recombinant-protein platforms
\citep{gao2021,tabarrok2023}. Innovation prizes instead link rewards to
competitors' performance. The Ansari XPRIZE offered \$10 million to the
first team meeting its spaceflight target \citep{xprize2004}, while the
DARPA Grand Challenge rewarded the fastest vehicle completing its course
\citep{darpa2005}. More recently, XPRIZE Carbon Removal has brought
multiple carbon-removal technologies into competition for a common prize
pool.

Economic theory offers extensive guidance on how competition shapes innovation \citep{aghion2005competition,gilbert2006schumpeter}, the design of R\&D contracts \citep{shan2017optimal,poblete2017managing}, and innovation contests \citep{taylor1995digging,che2003optimal,halac2017contests}. Recent work also examines the relative merits and interaction of contracts and prizes in innovation settings \citep{che2021prizes,fan2024contest}. We contribute to this literature by asking a distinct question: \emph{how much should an innovator's rewards and continuation depend on its competitors' performance?} We distinguish two dimensions of competition: interdependence in continuation decisions and interdependence in payments, and we examine their respective roles in coordinating experimentation and providing incentives.

Our model features a principal procuring innovations from firms pursuing
distinct technological approaches. Each approach has uncertain feasibility.
Experimentation can produce a breakthrough if the approach is feasible or
a breakdown that reveals its infeasibility. Only the principal values successful innovations. Each firm's payoff is the received payment net of the experimentation cost.
The principal values additional
innovations net of the payments required to induce experimentation. She
chooses how firms' rewards and continuation depend on outcomes across
approaches. 
We begin by studying parsimonious payment schemes. Contracts
specify fixed payments for breakthroughs and breakdowns,
while contests allocate rewards through a common prize pool
or a rank-contingent prize schedule. Under independent contracts,
neither a firm's payment nor its continuation depends on its
rivals' performance. Contests allow both to depend on relative
performance. Between them lie \emph{interdependent contracts}:
payments depend only on a firm's own technological outcomes,
but its contract can end in response to rivals' developments.
These contracts introduce competition through continuation
decisions while preserving individualized rewards.
This framework allows us to compare the performance of
simple mechanisms and identify the value of introducing
interdependence in continuation decisions and payments.

Our first main result shows that when both breakthroughs and
breakdowns are contractable, simple interdependent contracts
can implement the optimum. Payments for both outcomes allow
the principal to extract all firms' information rents, while
interdependent stopping coordinates experimentation across
approaches. In particular, the principal can terminate an
unresolved approach in response to its competitors' outcomes.
Therefore, optimal procurement can be achieved by requiring interdependence in
continuation, but not in payments.

This arrangement has an institutional analogue in U.S. Department of
Defense procurement. The DoD can fund parallel R\&D efforts while retaining
the right to terminate individual contracts before completion. Under the
Federal Acquisition Regulation's Termination for Convenience provisions,
an agency may end a contract when continuation is no longer in its
interest \citep[sec.~52.249-2(a)]{far522492}. Progress along an alternative technological path
can provide an economic reason to do so.\footnote{The analogy is not exact.
Under FAR Part~49, terminated contractors may recover allowable incurred
costs and related settlement expenses, supported by documentation and
auditing. Our baseline model rules out contracting on effort or incurred
costs. In the pure good-news environment, effort-independent termination
compensation is optimally zero because it weakens effort incentives; see
\cref{prop:zeroC}.}
In this arrangement, compensation remains contract-specific, but one
developer's progress can affect whether another continues to receive
funding. Our analysis explains why this intermediate arrangement can be
valuable even without competition for a common prize.

Our second main result examines the relative performance of
fixed-prize contests and interdependent contracts with constant
breakthrough payments when only breakthroughs are contractable.
For symmetric technologies, we first characterize the optimal
interdependent contracts within this class: unresolved firms
experiment until a deadline or until the number of breakthroughs
reaches a cutoff
(\cref{thm:optimal interdependent contracts absent breakdown}).
We then construct a class of equal-sharing fixed-prize contests
using the same stopping rules. Although the two mechanisms
generate identical innovation outcomes under full effort, their
minimum expected expenditures need not coincide.
For any interior success cutoff, there exists a unique critical
experimentation horizon. Fixed-prize contests require strictly
higher expected expenditure than interdependent contracts when
the horizon is shorter than this threshold, but strictly lower
expenditure when it is longer. At the critical horizon, the two
mechanisms are expenditure-equivalent. For the boundary cases
of stopping at the first success or allowing all approaches
to continue until success or the deadline, the two mechanisms
have the same minimum expected expenditure
(\cref{prop:expenditure_equivalent_contest_ic,cor:fixed_prize_expenditure}).

This comparison highlights a tradeoff between pooling rewards
and preserving effort incentives. Under interdependent contracts,
each successful firm receives a fixed payment regardless of how
many rivals succeed. A fixed-prize contest instead divides a
common pool among successful firms, making individual rewards
depend on the total number of winners. We identify the prize
pool that equates expected expenditures under full effort and
characterize when it is sufficient to induce full effort.
For interior cutoffs, this condition reduces to the initial
incentive constraint. When the horizon is short, the
expenditure-equivalent prize is insufficient to motivate
experimentation, requiring a larger prize pool. When the
horizon is sufficiently long, full effort can be implemented
with a smaller pool, generating expenditure savings.
Thus, even among simple mechanisms, pooling rewards can
either increase or decrease incentive costs.

We also study rank-contingent contests, which condition prizes on the
order of success. These contests can implement the optimal interdependent
contracts and therefore weakly dominate them
(\cref{thm:rank-contingent-dominance}). Their potential cost advantage
comes from targeting incentives by rank rather than pooling rewards
across winner counts. A fixed contractual payment must satisfy the most
demanding incentive constraint across all ranks a firm might occupy;
rank-contingent prizes can instead match each rank's incentive requirement.
This flexibility can reduce spending when requirements differ across
ranks, but yields no additional saving under a cutoff stopping rule,
where they coincide. 

Technological heterogeneity can reverse these comparisons. When approaches
differ in feasibility or development rates, interdependent contracts can
tailor payments to each firm's incentive requirements. An anonymous
rank-contingent contest must instead pay the same prize to whichever firm
occupies a given rank, potentially overcompensating firms that are easier
to motivate. This disadvantage can make even the optimal rank-contingent
contest more expensive than interdependent contracts
(\cref{ex:asymmetric-rank-contest-analytical,ex:asymmetric-rank-contest-numerical}).
Similarly, a fixed-prize contest may require a sufficiently large prize
pool to motivate the hardest-to-incentivize firm that it becomes more
costly than individualized contracts.

Together, these results show how contractability, technological
heterogeneity, and the experimentation horizon shape the desirable
form of competition. When both
breakthroughs and breakdowns are contractable, interdependent
continuation coordinates experimentation while approach-specific
payments extract rents. When only breakthroughs are contractable,
payment interdependence can further reduce incentive costs, but its
advantage depends on the prize structure, stopping rule, and
experimentation horizon. Technological heterogeneity can strengthen
the case for individualized rewards. The central procurement question
is thus not whether to introduce competition, but where to introduce
it: in continuation decisions, in payments, or in both.




\subsection{Literature Review}
\label{s:literature}

Our paper studies how innovation procurement mechanisms coordinate
parallel experimentation. \citet{che2021prizes} examine how cash
prizes and follow-on contracts balance incentives to innovate
against efficient implementation, while \citet{kremer2022designing}
design advance market commitments to protect supplier investment
from ex post hold-up. We focus on decisions during experimentation:
how discoveries on one technological approach should affect the
continuation and rewards of others.

We also contribute to the comparison of contests and individual
contracts. \citet{LazearRosen1981} establish conditions under which
rank-based rewards provide efficient incentives, while
\citet{GreenStokey1983} identify gains from filtering common shocks.
\citet{fan2024contest} show how contests mitigate incentive
distortions arising from learning and renegotiation.
Our comparison holds under commitment, with risk-neutral firms
and independent technological uncertainty.
We compare fixed-prize contests with constant-payment
interdependent contracts implementing the same experimentation
and stopping process, isolating the effect of pooling rewards
across winner counts. Under symmetry and interior success
cutoffs, contests are more expensive over short horizons but
less expensive over sufficiently long ones; expenditure
equivalence holds at the first- and last-success cutoffs.
Heterogeneous technologies can favor individualized contracts
because rewards can be tailored to firms' incentive requirements.

Dynamic contest models study the interaction between information
and rewards. In \citet{halac2017contests} and
\citet{bimpikis2019designing}, rivals' progress informs agents
about a common innovation's feasibility.
Our technologies have independent feasibility, so disclosure
affects incentives through rewards and termination without
changing beliefs about a firm's own feasibility.
\citet{Ely2023contest} optimize feedback, termination, and prize
allocation to maximize effort under a fixed prize budget.
Our principal maximizes the concave value of innovations net
of endogenous expenditure.

Our analysis also relates to the joint design of disclosure and monetary incentives in dynamic moral hazard. \citet{ely2025feedback} show that withholding feedback about an agent's own success can sustain effort but increase incentive costs, leading to an optimal transition from silence to transparency. We share their concern that dynamic incentives may induce agents to delay effort, but study a different role of disclosure. Firms observe their own outcomes, while the principal controls disclosure of rivals' progress. Because technological states are independent, disclosure affects incentives through continuation and payment consequences rather than beliefs about own feasibility.

We further relate to dynamic contracting for experimentation: \citet{bergemann1998venture,bergemann2005financing} study moral hazard and learning to
project termination, and \citet{halac2016optimal} study their interaction with private information about
ability.
We examine how continuation should respond to outcomes across
a portfolio.
\citet{moroni2022experimentation} studies complementary milestones,
where agents can free-ride on others' early successes before
participating in later stages.
In our setting, successes have diminishing marginal value,
so a breakthrough can shorten the remaining experimentation
period.

Our mixed-news learning process is also related to \citet{strulovici2010learning}, who studies collective experimentation with favorable and unfavorable Poisson news. Unlike their collective stopping environment, our principal can terminate approaches individually, and a common success-cutoff rule emerges endogenously under symmetry. Moreover, our mixed-news results identify a distinct role for contractable
failure. \citet{benkert2020designing} implement first-best research
with interim transfers and a final prize, using interim payments
to align the principal's stopping incentives and entry fees to
extract surplus.
We achieve zero-rent implementation through nonnegative payments
for breakthroughs and breakdowns, separating effort incentives
from portfolio selection.
\citet{akcigit2016role} also reward disclosures of technological
dead ends to prevent duplicated research.
Because our approaches' feasibility is independent, breakdown
payments eliminate incentive rents even without this
informational spillover between technologies.

Finally, we contribute to the literature on the optimal degree of competition. Existing work studies competition through entry
\citep{fullerton1999auctioning}, prize profiles and allocation
rules \citep{letina2023optimal}, or prize dispersion under
private information and convex effort costs
\citep{zhang2024optimal}.
We distinguish two dimensions of interdependence:
whether rivals' outcomes affect continuation and whether
they affect rewards.
Interdependent contracts coordinate continuation while retaining
individualized payments; contests add payment interdependence.
This comparison identifies when reward pooling improves
procurement beyond coordinated stopping.
Our time-varying-payment analysis separately evaluates the gain
from payment flexibility for a given stopping rule.

\section{Model}
\label{s:model}


\subsection{Environment}

A principal seeks solutions to a problem $\calP$ from $n$ firms, each pursuing a distinct technological approach $i \in \{1, \dots, n\}$. Each approach has a binary state $\theta_i \in \{G, B\}$, corresponding to being feasible ($G$) or infeasible ($B$). States are independent across approaches, and all parties share a common prior $\delta_i = \prob{\theta_i = G} \in (0,1)$.

Time $t \in [0, \infty)$ is continuous. At each instant $t$, firm $i$ chooses effort $\strategy_i(t) \in [0,1]$ at flow cost $\cost \strategy_i(t)$, where $\cost > 0$. 
Learning follows an exponential bandit process. While firm $i$ exerts effort $\strategy_i(t)$:
 If $\theta_i = G$, a \emph{breakthrough} arrives at Poisson rate $\rate_i \strategy_i(t)$, where $\rate_i > 0$ is the development rate.
 If $\theta_i = B$, a \emph{breakdown} arrives at Poisson rate $\mu_i \strategy_i(t)$, where $\mu_i \geq 0$.
Shirking ($\strategy_i(t) = 0$) yields neither cost nor news. This framework nests \emph{pure good-news learning} ($\mu_i = 0$; \citealt{halac2017contests}), where a breakthrough provides definitive proof of feasibility (e.g., a successful Phase III trial), as well as \emph{mixed-news learning} ($\mu_i > 0$; \citealt{strulovici2010learning,board2013reputation, keller2015breakdowns}), where approaches are also subject to conclusive failures (e.g., a prototype failing a critical safety test).


\subsection{Information and Strategies}

Let $\outcome_{i,t} \in \{\text{unresolved}, \text{success}, \text{breakdown}\}$ denote the status of approach $i$ at time $t$. Both firm $i$ and the principal observe $\outcome_{i,t}$ immediately, whereas rival firms observe it only if disclosed by the principal.

The principal commits to a disclosure policy generating a public message process $(M_t)_{t \geq 0}$, which induces the public filtration $\mathbb{F} = (\mathcal{F}_t)_{t \geq 0}$ with $\mathcal{F}_t = \sigma(M_s : 0 \leq s \leq t)$ satisfying the usual conditions. Firm $i$'s private filtration is $\mathbb{F}^i = (\mathcal{F}_t^i)_{t \geq 0}$, where $\mathcal{F}_t^i = \mathcal{F}_t \vee \sigma(\outcome_{i,s} : 0 \leq s \leq t)$. Given $\mathcal{F}_t^i$, firm $i$'s posterior belief is $\private_t^i = \prob{\theta_i = G \mid \mathcal{F}_t^i}$. 

A strategy for firm $i$ up to deadline $T$ is an $\mathbb{F}^i$-predictable effort process $(\strategy_i(t))_{t \in [0,T]}$ taking values in $[0,1]$. Let $\Strategy_i(t) = \int_0^t \strategy_i(v)\,dv$ denote cumulative effort up to time $t$.


\subsection{Payoffs}

Firms are risk-neutral and protected by limited liability, requiring transfers $\text{Payment}_i \geq 0$. If firm $i$'s project terminates at a (potentially stochastic) stopping time $\tau_i \leq T$, its expected payoff is
\begin{equation}
    U_i = \expect{\text{Payment}_i - \cost \int_0^T\strategy_i(t)dt}.
\end{equation}
Each firm's outside option is normalized to zero, so participation requires $U_i \geq 0$.

The principal derives value $\val(k)$ when $k \in \{0, \dots, n\}$ approaches succeed. We assume $\val(k)$ is weakly increasing and concave, with marginal value $\Delta \val(k) \equiv \val(k+1) - \val(k)$ for $k \leq n-1$. The principal maximizes expected net surplus:
\begin{equation}
    \obj = \expect{\val(k) - \text{Spending}}, \quad \text{where} \quad \text{Spending} = \sum_{i=1}^n \text{Payment}_i.
\end{equation}


\subsection{Mechanisms}

The principal motivates firms up to a fixed horizon $T$, which captures project urgency and serves as a unified benchmark across mechanisms. Let $\projset$, $X$ and $U$ denote the realized sets of approaches that have achieved a breakthrough, a breakdown, or remained unresolved by termination, respectively. We study three classes of mechanisms: independent contracts, interdependent contracts, and contests.

\medskip
\noindent\textbf{Independent contracts ($\mathcal{C}$).}
An independent contract mechanism $\contracts = \{\contract_i\}_{i \in I}$ selects a set of active firms $I \subseteq \{1, \dots, n\}$ and offers each an individual contract $\contract_i = (T, \Payment_i, \payment_i)$. Each firm $i \in I$ operates until the deterministic deadline $T$ and receives $\Payment_i \geq 0$ upon a breakthrough, $\payment_i \geq 0$ upon a breakdown, and $0$ if unresolved. Payments depend solely on firm $i$'s own final outcome, yielding total spending
\[
    \mathrm{Spending}(\contracts) = \sum_{i \in \projset} \Payment_i + \sum_{i \in X} \payment_i.
\]

\medskip
\noindent\textbf{Interdependent contracts ($\Hybrid$).}
An interdependent contract mechanism $\hybrids = \{\hybrid_i\}_{i \in I}$ allows each firm's deadline to depend stochastically on the outcomes of rival firms. Formally, each contract is a quintuple $\hybrid_i = \big(\Payment_i, \payment_i, \{\compensation_i(t)\}_{t \in [0,T]}, \tau_i, \disclosure_i\big)$, where:
 $\Payment_i \geq 0$ and $\payment_i \geq 0$ are fixed payments for a breakthrough and breakdown occurring at any $t \leq \tau_i$;
 $\tau_i \in [0,T]$ is an endogenous stopping time adapted to the joint outcome history of all firms;
 $\disclosure_i \in \{0,1\}$ governs whether termination at $\tau_i$ is disclosed to firm $i$ immediately ($\disclosure_i = 1$) or only at $T$ ($\disclosure_i = 0$);
  $\compensation_i(t) \geq 0$ is a continuous compensation schedule paid if firm $i$ is unresolved when terminated at $\tau_i = t \leq T$ and termination is disclosed ($\disclosure_i = 1$).

Independent contracts correspond to the special case where $\tau_i \equiv T$ and $\compensation_i(t) \equiv 0$. The total spending is
\[
    \mathrm{Spending}(\hybrids) = \sum_{i \in \projset} \Payment_i + \sum_{i \in X} \payment_i + \sum_{i \in U} \compensation_i(\tau_i)\disclosure_i.
\]
Alternatively, interdependent contracts can be interpreted as a contest with approach-specific prizes $\Payment_i$, eliminating prize-sharing uncertainty conditional on success while keeping the principal's total outlay contingent on aggregate outcomes.

We restrict attention to the set of \emph{admissible symmetric} stopping profiles, $\taus = (\tau_1, \dots, \tau_n) \in \stopset$, defined by two conditions: (i) conditional on firm $i$ remaining unresolved at time $t$, the distribution of $\tau_i$ is independent of firm $i$'s effort history; and (ii) unresolved firms facing identical public histories face identical stopping rules, regardless of identity.\footnote{In our analysis, stopping policies depend solely on calendar time $t$ and the aggregate public count of successes, which satisfies both conditions.} 

\medskip
\noindent\textbf{Contests ($\mech$).}
A contest rewards firms based on relative performance. Let $\successtime_i \in [0,T] \cup \{\nosuccess\}$ denote firm $i$'s success time ($\successtime_i = \nosuccess$ if no success occurs before termination), and let $\rank_i(\successtimes)$ denote firm $i$'s chronological rank among successful firms under profile $\successtimes = (\successtime_1, \dots, \successtime_n)$.

A contest $\mech = (\Payment, \shares, \drule)$ specifies a maximum prize pool $\Payment \geq 0$, a symmetric public disclosure rule $\drule = (M_t, \disclosure_t)_{t \in [0,T]}$, and a prize-sharing rule $\shares = (\share_i(\successtimes))_{i=1}^n \in \mathbb{R}_+^n$ satisfying:
\begin{enumerate}
    \item[(i)] \textit{Anonymity}: $\share_i(\successtimes) = \share(\successtime_i, \sigma(\successtimes_{-i}))$ for any permutation $\sigma$;
    \item[(ii)] \textit{No reward for failure}: $\share_i(\nosuccess, \successtimes_{-i}) = 0$;
    \item[(iii)] \textit{Weak budget balance}: $\sum_{i=1}^n \share_i(\successtimes) \leq 1$ for all $\successtimes$.
\end{enumerate}
Firm $i$ receives $\Payment \share_i(\successtimes)$, and the principal retains any unallocated surplus, yielding $\mathrm{Spending}(\mech) = \Payment \sum_{i=1}^n \share_i(\successtimes)$. This structure nests two canonical specifications:

\noindent
\textit{Fixed-prize contests ($\jointcontest$):} Following \citet{halac2017contests}, condition (iii) is strengthened to \emph{strict budget balance} whenever at least one firm succeeds ($\sum_{i=1}^n \share_i(\successtimes) = 1$ if $\projset \neq \varnothing$). The entire pool $\Payment$ is disbursed upon any success:
    \[
        \mathrm{Spending}(\jointcontest) = \indicate{\projset \neq \varnothing} \Payment.
    \]

    \noindent
 \textit{Rank-contingent contests ($\rankcontest$):} Specified by $\rankcontest = (\boldsymbol{\Payment}, \drule)$, where $\boldsymbol{\Payment} = (\Payment_1, \dots, \Payment_n) \in \mathbb{R}_+^n$ is a vector of prizes awarded strictly by order of success. Setting the pool to $\Payment = \sum_{r=1}^n \Payment_r$, firm $i$'s share is $\share_i(\successtimes) = \Payment_{\rank_i(\successtimes)} / \Payment$ if $\successtime_i \neq \nosuccess$ and $0$ otherwise. When $k $ firms succeed, total spending is
    \[
        \mathrm{Spending}(\rankcontest) = \sum_{r=1}^{k} \Payment_r.
    \]

\section{Mixed-news: Separation of Incentives and Portfolio Selection}\label{sec:mixed-news}

The main goal of this section is to establish that in mixed-news learning environment, the novel interdependent contracts introduced by our paper allows the principal to separate effort incentives from portfolio selection. When both breakthroughs and conclusive breakdowns are contractable, appropriately chosen payments implement full effort while leaving each firm zero expected rent. We begin with independent contracts to explain how rewarding evidence of infeasibility makes this possible. We then introduce interdependent contracts, which allow the principal to use outcomes from one approach to determine whether experimentation on another should continue. The same payments sustain full effort under these contingent termination decisions, so the principal’s remaining problem is to choose a continuation policy that maximizes expected innovation value net of effort costs. Payments therefore provide effort incentives, while interdependent continuation coordinates experimentation across the portfolio. Under symmetric technologies, downward-drifting beliefs, and a concave innovation value function, we characterize the optimal policy through deadlines that are weakly decreasing in the number of successes.

\subsection{Primer: Independent Contracts}\label{sec:contracts}
We begin with independent contracts, which provide a natural benchmark for the procurement problem. Under an independent contract, each firm's payment and deadline depend only on the outcome of its own technological approach, so the principal cannot use information generated by one approach to alter the incentives or continuation of another. This benchmark isolates how technological learning shapes the cost of inducing experimentation. Under pure good-news learning, unsuccessful experimentation makes a firm increasingly pessimistic about its own technology, requiring the principal to leave information rents in order to sustain effort. By contrast, when both successes and breakdowns are contractable, appropriately chosen payments for the two outcomes can extract these rents completely. The latter observation will provide a useful building block for our analysis of interdependent contracts.

Let $\private_t^i$ denote the firm's posterior belief at time $t$ that its approach $i$ is feasible ($\theta_i = G$), conditional on having worked continuously up to time $t$ without experiencing a success or a breakdown. 
If the approach is feasible, the probability of reaching time $t$ without success is $e^{-\rate_i t}$. If the approach is infeasible, the probability of reaching time $t$ without a breakdown is $e^{-\mu_i t}$. By Bayes' rule, the posterior belief $\private_t^i$ is:
\begin{equation}\label{eq:posterior-mixed}
    \private_t^i  = \frac{\delta_i e^{-\rate_i t}}{\delta_i e^{-\rate_i t} + (1-\delta_i)e^{-\mu_i t}}
\end{equation}
Taking the derivative of $\private_t^i$ with respect to time yields the law of motion for the firm's belief:
\begin{equation}\label{eq:belief dynamics}
    \diff{\private_t^i}{t} = -\private_t^i(1-\private_t^i)(\rate_i - \mu_i).
\end{equation}

As different belief dynamics would have different implications for the firm's incentive, we distinguish the following two regimes: When  $\rate_i > \mu_i$, successes are expected to arrive faster than breakdowns, and the absence of news is a pessimistic signal. Consequently, $\private_t^i$ strictly decreases over time. We call this a \emph{downward drift} environment. One special case of a downward drift environment is $\mu_i=0$, where breakdown news is infeasible. We call this a \emph{pure good news learning} environment. In contrast, when $\rate_i < \mu_i$, breakdowns are expected to arrive faster than successes, and surviving without an event is an optimistic signal. Consequently, $\private_t^i$ strictly increases over time. We call this an \emph{upward drift} environment.

Despite the different implications for the belief dynamics in different learning environments, the next result establishes that whether the principal can extract full rent from a firm is determined by the availability of breakdown news. 
\begin{proposition}[Zero-rent contract]\label{prop:zero-rent-contracts}
\singlespace
Fix $T>0$. An independent contract can implement full effort
until resolution or $T$, while leaving firm $i$ zero expected
rent, if and only if $\mu_i>0$.
\end{proposition}

\begin{proof}[Proof of \cref{prop:zero-rent-contracts}]
To prove this proposition, we will first establish the contrapositive of the `only if' part by showing that in a pure good news learning environment ($\mu_i=0$), it is impossible to extract full rent. Then we proceed to establish the `if' part by constructing the first-best contract when breakdown news is feasible. 

    \paragraph{Step 1: Pure Good News Learning}
Suppose $\mu_i=0$. Then firm $i$ is in a pure good news learning environment.
Given that breakdown news is infeasible, the principal cannot pay upon breakdown. Thus, she is restricted to the following class of independent contracts: $\contract_i = (T, \Payment_i)$, where $\Payment_i \ge 0$ is the payment upon success. 
Since the firm learns nothing if it shirks, once it stops working, it will never become optimal to restart later. The firm's strategy reduces to choosing a single optimal stopping time $\tau \in [0, T]$, at which point it permanently stops working if a success has not already occurred.

Let $S_i(t)$ be the unconditional probability that the firm survives to time $t$ (no success occurs before time $t$):
$    S_i(t) = (1- \delta_i)e^{-\mu_i t}+ \delta_i e^{-\rate_i t} = (1- \delta_i)+ \delta_i e^{-\rate_i t}$.
The firm chooses $\tau$ to maximize its expected payoff:
\begin{equation}
    \max_{\tau \in [0, T]} U_i(\tau) = \int_{0}^{\tau} S_i(t) \left[ \private_t^i \rate_i \Payment_i - \cost \right] dt.
\end{equation}
Taking the derivative of the objective function with respect to $\tau$ yields the marginal expected payoff at time $t$: $ U_i'(t) = S_i(t) \left[ \private_t^i \rate_i \Payment_i - \cost \right]$.
Because $S_i(t) > 0$, the firm's optimal stopping decision is entirely governed by whether the instantaneous expected profit, $\private_t^i \rate_i \Payment_i - \cost$, is positive.

For the independent contract $\contract_i = (T, \Payment_i)$ to be incentive compatible, the firm must be willing to work continuously up to the deadline $T$. 
 Since $\private_t^i$ strictly decreases over time when $\mu_i=0$,  firm $i$'s marginal expected payoff strictly decreases. To ensure firm $i$ works until the deadline, the local flow payoff must be weakly positive at the terminal time $T$:
$\private_T^i \rate_i \Payment_i \ge \cost$.
Thus, the minimum budget requirement for incentive compatibility is $ \Payment_i \ge \frac{\cost}{\rate_i \private_T^i}$.

Let $\totalcost_i(T)$ denote the total effort cost of a firm working continuously until $T$:
\begin{equation}
    \totalcost_i(T) = \int_0^T \cost S_i(t) dt.
\end{equation}
The principal's total expected payment is $W_i(T) = \delta_i (1 - e^{-\rate_i T}) \Payment_i$. Because the firm's net expected payoff is $U_i(T) = W_i(T) - \totalcost_i(T) \ge 0$, the total expected payment the principal can ever achieve is bounded below by $\totalcost_i(T)$. 

Notice that $\private_t^i$ strictly decreases over time, and any incentive-compatible contract such that firm $i$ works until $T$ in a history of no success satisfies $U_i'(T)\geq 0$. These together imply that $U_i'(t) > 0$ for all $t < T$, granting the firm strictly positive information rent ($U_i(T) > 0$). Consequently, the principal's expected payment strictly exceeded the total cost of effort ($W_i(T) > \totalcost_i(T)$).
Thus, any incentive-compatible contract in a pure good news learning environment cannot extract full rent from the firm.


    \paragraph{Step 2: Mixed News Learning.}
    
Suppose now  $\mu_i>0$.
Since breakdown evidence is feasible,  the principal can use a larger class of contracts $\contract_i = (T, \Payment_i, \payment_i)$, where $\Payment_i \ge 0$ is the payment upon breakthrough and $\payment_i \ge 0$ is the payment upon a breakdown. Then the firm's expected payoff from stopping at $\tau \in [0, T]$ is
\begin{equation*}
    U_i(\tau) = \int_{0}^{\tau} S_i(t) \left[ \private_t^i \rate_i \Payment_i + (1-\private_t^i)\mu_i \payment_i - \cost \right] dt.
\end{equation*}
Thus, the marginal payoff is $U_i'(t) = S_i(t) \left[ \private_t^i \rate_i \Payment_i + (1-\private_t^i)\mu_i \payment_i - \cost \right]$.


Being able to verify failed attempts allows the principal to provide insurance against the scenario in which the firm works but eventually fails. Consider a contract that sets $\Payment_i = \frac{\cost}{\rate_i}$, and $\payment_i = \frac{\cost}{\mu_i}$.
Under this contract, each firm's instantaneous flow payoff is perfectly flat and identically zero for all $t$:
\begin{equation*}
    \private_t^i \rate_i \Payment_i + (1-\private_t^i)\mu_i \payment_i = \cost, \quad \forall t\in[0,T]
\end{equation*}

Thus, $U_i'(t) = 0$ for all $t$. This implies that each firm is perfectly indifferent between stopping and continuing at all times, and has zero information rent as $U_i(T) = 0$. 

Recall that $\totalcost_i(T)$ is the total effort of a firm working continuously until $T$.
And the principal's total expected payment is $W_i(T) = \delta_i (1 - e^{-\rate_i T}) \Payment_i + (1-\delta_i)(1 - e^{-\mu_i T}) \payment_i$. Because the firm's net expected payoff is $U_i(T) = W_i(T) - \totalcost_i(T) \ge 0$, the lowest total expected payment the principal can ever achieve is bounded below by $\totalcost_i(T)$. Under the current contract, the principal's total expected payment achieves the lower bound: $W_i(T) = \totalcost_i(T)$.

\end{proof}

It is worth noting that the zero-rent contract is not unique in the \emph{upward drift} environment. When $\rate_i < \mu_i$, some of the restricted contracts that only pay for success can also achieve the first best. 
This is because in a upward drift environment, $\private_t^i$ strictly increases over time, and the firm's marginal expected payoff crosses zero from below at most once when $\payment_i=0$. The optimal stopping time is a corner solution: $\tau = 0$ (not start at all) or $\tau = T$ (work until the end once start). Thus, the incentive compatibility constraint is the global condition that the ex-ante expected benefit exceeds the unconditional expected total cost, $U_i(T) \ge 0$.  By setting $\payment_i=0$ and $\Payment_i=\frac{\cost}{\rate_i} + \frac{\cost (1-\delta_i)(1 - e^{-\mu_i T})}{\delta_i \mu_i (1 - e^{-\rate_i T})}$ such that $U_i(T) = 0$, the principal can also extract full rent from firm $i$.

Essentially, paying for breakdown is providing insurance for working. It de-risks the innovation processes and could potentially reduce procurement spending.
We say the insurance has a value if allowing $\payment_i>0$ leaves the principal strictly better off than setting $\payment_i=0$. Thus, there is a value of insurance if and only if $\rate_i >  \mu_i$.

Under independent contracts, the principal can utilize outcome-contingent payments to implement full rent extraction. However, given the uncertainty of the feasibility of each technological path, once the principal sets up the contracts, she cannot control the number of successful technological paths to procure. We highlight that this is the main limitation of independent contracts. In the next section, we will propose interdependent contracts, where the principal can utilize outcome-contingent payments to extract full rent, and can additionally use the information generated by one approach to determine whether experimentation on another should continue. We will show that the interdependent termination decisions allow the principal to balance the value of successful innovations and the spending.

\subsection{Optimal Interdependent Contracts}\label{sec:hybrid-mixed-news}
We now allow the principal to make the continuation of one technological approach contingent on outcomes realized by others.  Under independent contracts, the principal commits ex ante to developing each funded approach until its own deadline, even when subsequent discoveries make continued experimentation undesirable. Interdependent contracts instead allow her to respond dynamically to information generated across the portfolio, terminating some unresolved approaches while continuing others. Under mixed-news learning, where contractable successes and breakdowns permit full rent extraction. Once firms' rents are eliminated, mechanism design reduces to an optimal stopping problem: the principal chooses how long to continue experimentation as innovations arrive, and we characterize the resulting stopping policy under symmetric technologies and a concave value function. 

First, we establish that when the principal can contract on both conclusive good news (success) and conclusive bad news (breakdown), she can extract the full information rent from each firm.

\begin{proposition}[Full rent extraction]\label{prop:first best stochastic contract}
\singlespace
Suppose $\mu_i>0$ for all $1\leq i\leq n$. There exists a class of interdependent contracts under which the principal can extract the full information rent from each firm.
\end{proposition}

 Consider a stochastic contract for approach $i$, denoted by $\hybrid_i^\dagger = \big(\Payment_i^\dagger, \payment_i^\dagger, \{\compensation_i^\dagger(t)\}_{t\in[0,T]}, \tau_i, \disclosure_i^\dagger\big)$, with breakthrough payment $\Payment_i^\dagger = \frac{\cost}{\rate_i}$, breakdown payment $\payment_i^\dagger = \frac{\cost}{\mu_i}$, zero termination compensation $\compensation_i^\dagger(t) = 0$ for all $t \in [0,T]$, and full disclosure $\disclosure_i^\dagger = 1$ such that the realization of $\tau_i$ is publicly announced. 
Under the constructed contract and an arbitrary cumulative effort path $\Strategy(t)$, each firm's instantaneous flow payoff is
\begin{align}
     p_{\Strategy(t)}^i \rate_i \left( \frac{\cost}{\rate_i} \right) + (1-p_{\Strategy(t)}^i)\mu_i \left( \frac{\cost}{\mu_i} \right) - \cost \nonumber =0, \nonumber
\end{align}
where $p_{\Strategy(t)}^i$ is the firm's posterior belief on the approach being feasible under the cumulative effort path $\Strategy(t)$.

Thus, each firm's marginal payoff of effort is exactly zero everywhere, and each firm is locally indifferent between working and shirking at every instant $t$, regardless of its past effort history. By the standard tie-breaking assumption, continuous effort ($\strategy_i(t) = 1$ for all $t$) is dynamically incentive compatible.
Moreover, since each firm has zero total expected payoff under full effort, the principal extracts the full information rent from each firm.
Similar to the analysis for contracts, the ability to contract on breakdown is what enables full rent extraction.\footnote{Similarly, in a mixed news learning environment with upward drift ($\mu_i>\rate_i$), it is possible to use a success-only contract ($\payment_i = 0$) to extract full rent.}\footnote{Notice that the zero-rent contract is prior-free, requiring no knowledge of $\delta_i$. However, the required breakdown payment $\payment_i = \cost / \mu_i$ grows inversely with the breakdown rate. If breakdowns are highly infrequent ($\mu_i \to 0$), full rent extraction requires arbitrarily large payments, which may not be implementable if the principal has a budget constraint.}

\bigskip

We introduce the following symmetric assumption that will help to tighten the characterization of the optimal mechanisms.
\begin{assumption}[Symmetric technologies]\label{assump:symmetry}
    Suppose approaches are symmetric: $\rate_i=\rate$, $\mu_i=\mu$, and $\delta_i=\delta$ for all $i$,
where $0<\delta<1$ and $\rate>0$.
\end{assumption}

Let $\tau^\dagger=\inf\left\{t\in[0,T]:t\geq \rho(\eta_t)\text{ or } t=T\text{ or } a(t)=0\right\},$
where the endogenous deadline following $k$ successes is
\begin{equation}
\label{eq:mixed-stopping-time}
\rho(k)=
\begin{cases}
\displaystyle
\inf \left\{ t\geq0: p_t\rate\Delta\val(k)\leq\cost \right\}
&
\text{if }\delta\rate\Delta\val(k)>\cost,
\\[1.2em]
0,
&
\text{if }\delta\rate\Delta\val(k)\leq\cost.
\end{cases}
\end{equation}

\begin{theorem}\label{thm:optimal interdependent contracts}
If $\mu_i>0$ for all $i$, then there exist interdependent contracts that are optimal among mechanisms that
implement full effort until each approach is resolved
or permanently terminated.  

    Under Assumption \ref{assump:symmetry} and $\rate>\mu>0$, the interdependent contracts $\{\hybrid_i^\dagger,i=1,\dots,n\}$, where $\tau_i=\tau^\dagger$ maximize the principal's objective. Moreover, the endogenous deadline following $k$ successes $\rho(k)$ is weakly decreasing in $k$.
%
\end{theorem}

\begin{proof}[Proof of \cref{thm:optimal interdependent contracts}]
The proof consists of three main steps. First, we argue that the interdependent contracts $\{\hybrid_i^\dagger,i=1,\dots,n\}$ achieve the lower bound of the principal's expected spending to implement full effort until $\tau_i\wedge T$ for each firm $i$. This implies that using interdependent contracts can achieve the optimum among all mechanisms. We highlight that this observation does not depend on whether the approaches are symmetric.
Second, we show that the principal's maximization problem can be reduced to an equivalent problem of choosing the optimal stopping policy among the full rent extraction contracts. Lastly, we further impose symmetry and characterize the optimal stopping policy.

\textbf{Step 1: full rent extraction.} 
Consider any individually rational and incentive-compatible mechanism $M$ that implements full effort until $\tau_i^M = \tau_i\wedge T$ for each firm $i$. For each firm $i$, individual rationality implies that 
$U_i(M)=\expect{\mathrm{Payment}_i(M)}-\cost\expect{\Strategy_i^M(\tau_i^M)}\geq 0$.
Therefore,
$\expect{\mathrm{Payment}_i(M)}
\geq
\cost\expect{\Strategy_i^M(\tau_i^M)}$,
and summing across firms yields the lower bound of the principal's expected spending to implement full effort until $\tau_i^M\wedge T$ for each firm $i$: 
$\expect{\mathrm{Spending}(M)}
\geq
\sum_i \cost\expect{\Strategy_i^M(\tau_i^M)}.$

Under the proposed interdependent contracts $\{\hybrid_i^\dagger,i=1,\dots,n\}$, $\expect{\mathrm{Payment}_i}=\cost\expect{\Strategy_i(\tau_i^M)}$ for each $\hybrid_i$. Thus, 
$\expect{\mathrm{Spending}}=\sum_i \cost\expect{\Strategy_i(\tau_i^M)}$, which achieves the lower bound of the principal's expected spending to implement full effort until $\tau_i^M$ for each firm $i$.

\textbf{Step 2: Separation of stopping and spending.}
From the above analysis, we know that under the interdependent contracts $\{\hybrid_i^\dagger,i=1,\dots,n\}$, continuous effort is weakly optimal for the firm and generates zero expected rent. Throughout the analysis, we adopt the convention that a firm works when indifferent.

Let $\boldsymbol{\tau}=(\tau_1,\ldots,\tau_n)$ denote an arbitrary profile of stopping rules, where $\tau_i$ is the time at which approach $i$ is terminated if it has not previously succeeded or broken down. 
Let $a(t)$ denote the number of unresolved approaches that remain active at time $t$, and let $\eta_{\boldsymbol{\tau}}$ denote the total number of successes generated before the corresponding approaches are resolved or terminated.

We claim that the principal's mechanism-design problem reduces to
\begin{equation}
\label{eq:planner_stopping_problem}
\max_{\boldsymbol{\tau}\in\mathcal{T}}
\left\{
\mathbb{E}\left[\val(\eta_{\boldsymbol{\tau}})\right] - \mathbb{E}\left[
\int_0^\infty \cost\,a(t)\,dt
\right]
\right\}.
\end{equation}

This is because under continuous effort, the principal's expected payment to firm $i$ is exactly equal to the firm's expected effort cost. Summing across firms yields an expected total payment of $\mathbb{E}\left[ \int_0^\infty \cost\,a(t)\,dt \right]$. Therefore, the principal's expected payoff coincides with expected social surplus. Maximizing the principal's payoff over admissible interdependent contracts is consequently equivalent to choosing the stopping policy $\boldsymbol{\tau}$ that solves \eqref{eq:planner_stopping_problem}.

\textbf{Step 3: Optimal stopping policy.}
We omit the characterization of the optimal stopping policy under symmetric technologies provide it in \cref{apx:optimal_stopping_good_news}.
\end{proof}

The first two steps establish that interdependent contracts with the specified payments, compensation schedule, and disclosure decision can achieve optimality.
We highlight that we do not need symmetry for the first two steps of the proof. Thus, although we don't have a characterization of the optimal stopping policy under asymmetric technologies, we know that there exist some interdependent contracts that can achieve optimality.

Here we sketch the intuition for the optimal stopping policy. Conditional on $k$ successes having occurred, every unresolved approach faces a common deadline $\rho(k)$. Before this deadline, all unresolved firms continue to work; once calendar time reaches the deadline, all remaining unresolved approaches are terminated. 
Notice that  $\rho(k)$ is the calendar time such that each firm's posterior belief at $\rho(k)$ equals $\frac{c}{\rate \Delta\val(k)}$.
 At time $t$, an unresolved approach succeeds at instantaneous rate $p_t\rate$, so its instantaneous expected contribution from another success is $p_t\rate\Delta\val(k)$. Experimentation continues precisely while this marginal benefit exceeds the flow cost $\cost$.

The stopping rule creates a dynamic competitive externality across firms. Because $\rate>\mu$, the absence of a resolution is bad news about an approach's feasibility, so $p_t$ declines with calendar time. Concavity introduces a second force: as more innovations are obtained, the marginal value of another innovation falls. Consequently, $\Delta\val(k+1)\leq\Delta\val(k)$, which implies $\rho(k+1)\leq \rho(k)$. Therefore, each success weakly ratchets the common deadline inward. A firm that remains unresolved may suddenly face a shorter remaining horizon when another firm succeeds. Unlike a conventional rank-order contest, firms are not competing for shares of a fixed prize pool: each firm receives the same fixed success payment specified by its stochastic contract. Competition instead operates through the stopping rule. Firms are effectively racing against a clock that becomes more demanding as the principal accumulates successful innovations. When $\val$ is strictly concave, this ratchet is strict whenever the deadline remains interior. When $\val$ is linear, by contrast, every additional innovation has the same marginal value, so successes do not alter the deadline and the stochastic stopping rule collapses to a common deterministic deadline.

The simplicity of the rule is less immediate than the marginal condition might suggest. If $p_t\rate\Delta\val(k)\leq\cost$, then stopping is locally optimal: the expected instantaneous benefit from experimenting with an unresolved approach is no greater than its flow cost. Concavity makes this local comparison globally valid. If no resolution occurs, the posterior $p_t$ falls, making experimentation still less attractive. A breakdown leaves the number of successes unchanged. A success raises $k$, but concavity lowers $\Delta\val(k)$. Thus, once the process enters the region where $p\rate\Delta\val(k)\leq\cost$, every possible subsequent evolution weakly reinforces the case for stopping. The stopping region is therefore absorbing. There is no reason to incur a current experimentation loss in the hope of reaching a future state in which experimentation becomes more attractive.

The more subtle part of the result concerns what happens outside the stopping region. The inequality $p_t\rate\Delta\val(k)>\cost$ immediately shows that terminating \emph{all} unresolved approaches cannot be optimal, but symmetry alone does not imply that all of them should remain active. Even when firms are identical, the principal could in principle terminate some unresolved approaches and continue with an arbitrary subset. Symmetry only implies that the principal is indifferent about \emph{which} firms to retain; it does not imply that the optimal number of retained firms must be either zero or all of them.

The proof therefore has to rule out such selective termination. The key step is to compare the value generated by the candidate stopping policy with the unresolved approaches $a+1$ to that generated with only $a$. Adding an approach weakly increases the distribution of the number of innovations obtained before stopping. More importantly, after accounting both for the additional approach's own effort cost and for the fact that its success can shorten the deadline, and thereby reduce the effort subsequently supplied by other firms, its incremental expected surplus remains weakly positive. Formally, the difference in continuation values can be represented as a sum of nonnegative terms proportional to $\rate-\mu$. Hence, the principal benefits from having an additional active approach in the continuation region. It follows recursively that immediately discarding any subset of unresolved approaches cannot improve the principal's payoff. Whenever continuation is optimal, there exists an optimal policy that retains all unresolved approaches. Together with the absorbing stopping-region argument, this verifies the all-or-nothing policy stated in the result.

The condition $\rate>\mu$ helps to explain this result economically. Since a feasible approach resolves through success more quickly than an infeasible approach resolves through breakdown, remaining unresolved is bad news for feasibility. The principal therefore benefits from exploiting available approaches while their posterior prospects remain sufficiently favorable. Parallel experimentation creates additional opportunities to obtain innovations before this no-news-is-bad-news learning drives the posterior to the stopping threshold. Choosing to discard identical approaches sacrifices these opportunities without generating any benefit. The proof makes this intuition precise while also accounting for the endogenous shortening of the deadline after successes.

Concavity is the crucial condition that makes the stopping decision myopic. When innovations are substitutes, a success reduces the marginal value of subsequent innovations. Hence, if experimentation is already unprofitable at the current state, neither waiting nor obtaining another success can move the principal into a state in which experimentation becomes more attractive. This is what allows the optimal stopping rule to be characterized entirely by comparison between the expected marginal benefit of another innovation and the cost of continued experimentation.

This logic generally breaks down when $\val$ is convex. With convex value, innovations are dynamically complementary: obtaining one success raises the value of obtaining subsequent successes.  Although experimentation has negative instantaneous surplus at the current state, a success can move the principal into a future state in which continued experimentation is substantially more valuable. As a result, the principal may optimally continue experimentation even when $p_t\rate\Delta\val(k)<\cost$, because the relevant benefit of a success includes the option value of reaching more valuable continuation states.
For example, suppose $\val(0)=\val(1)=0$ and $\val(2)>0$ is sufficiently large. At zero successes, the marginal benefit of another innovation is smaller than the cost of continued experimentation. However, when multiple unresolved approaches remain, experimentation can be socially valuable because the first success creates the possibility of obtaining the highly valuable second success. Moreover, this option value depends on how many unresolved approaches remain available. The optimal stopping boundary under convex values therefore generally depends on the number of successes $k$, the shape of the value function, and the number of unresolved approaches $a$. In this sense, concavity is what collapses an intrinsically dynamic experimentation problem into the simple success-dependent deadline characterized above.

The stopping rule provides a useful comparison with
\citet{strulovici2010learning}, whose mixed-news specification
in Section~5.2 yields the same posterior dynamics as ours.
In his baseline utilitarian benchmark, each additional
success reveals another beneficiary of the common risky
action, increasing the social return to continuing that
action. The utilitarian continuation deadline therefore
moves later as successes accumulate
\citep[Theorem~2]{strulovici2010learning}.
In our setting, completed innovations retain their value
after experimentation ends, while concavity of $\val$
makes further innovations less valuable at the margin.
Consequently, $\rho(k)$ is weakly decreasing in $k$:
additional breakthroughs move the continuation deadline
weakly earlier. The opposite comparative statics reflect
the complementarity between success and continued
collective action in his setting, and the substitutability
of innovations in ours.

The scope for selective continuation also differs.
In his collective-action setting, all individuals must
take the same action, so continuation is all-or-nothing
by construction. Our principal can instead terminate
approaches individually. Nevertheless, under the symmetry
and concavity assumptions of Theorem~\ref{thm:optimal interdependent contracts}, an optimal policy
keeps all remaining unresolved approaches active until
a common stopping time. Common termination thus emerges
from optimal portfolio selection despite the principal's
ability to discontinue approaches selectively.
\section{Pure Good News: Simple Mechanisms}\label{sec:good-news-constant-payment}
Under pure good-news learning, the absence of contractable breakdowns prevents the principal from eliminating firms’ information rents, making the design of rewards central to procurement costs. We focus on simple mechanisms: interdependent contracts with constant breakthrough payments, fixed-prize contests with a constant total prize pool, and rank-contingent contests with a fixed payment for each success rank.
We first characterize optimal interdependent contracts that implements full effort until a common exogenous deadline $T$.
Then we ask whether contests can improve on them by linking rewards to relative performance. Comparing mechanisms that implement the same experimentation process allows us to isolate differences in incentive costs. Under symmetric technologies and interior success cutoffs, fixed-prize contests are more costly over short horizons but can lower expected expenditure over sufficiently long ones. Rank-contingent contests, in turn, can reproduce the optimal interdependent-contract outcome. These results show how pooling rewards can improve procurement even when the principal already coordinates continuation across approaches, while identifying the circumstances in which individualized payments remain less expensive.
Throughout this section and next, we assume that $\mu_i=0$, so that breakdown news is not contractable.

\subsection{Optimal Interdependent Contracts under Constant Payments}\label{sec:hybrid-good-constant}
We now illustrate the relationship between the DoD's Federal
Acquisition Regulation (FAR) mechanism and our interdependent contracts.
Parallel development awards act as approach-specific breakthrough
payments, $\Payment_i$, while the government's option to halt an
ongoing project upon a rival's success corresponds to the endogenous
termination time, $\tau_i$. Issuing a Termination for Convenience
notice is equivalent to disclosing the realized termination
($\disclosure_i=1$), as the contractor immediately learns of the
decision and ceases work. Finally, the Part~49 termination settlement
is analogous to the compensation schedule, $\compensation_i(t)$.

This comparison raises a natural question. From a theory perspective,
absent breakdown payments in the pure good-news learning environment,
can the principal utilize the compensation schedule to reduce total
spending? Put it differently, from the comparison's perspective, can
we exactly reproduce DoD's contract arrangement in our framework?

In this subsection, first, we show that the analogy is not exact.
In \cref{prop:prize-lower-bound}, we characterize the limitations of
termination compensation as an incentive device: it is not a substitute
for breakthrough payments to incentivize effort in a moral hazard
setting without any monitoring device.

\begin{proposition}[Success-Prize Lower Bound]
\label{prop:prize-lower-bound}
\singlespace
Fix any constant success prize
$\Payment_i<\cost/(\rate_i p_T^i)$, any non-negative compensation
schedule $\compensation_i(\cdot)\geq0$ and either disclosure rule.
Define the effort threshold
\[
\hat e_i
\equiv
\inf\left\{
e\geq0:\rate_i p_e^i\Payment_i<\cost
\right\}.
\]
Then in every equilibrium,
$\Strategy_i(T)\leq\hat e_i<T$ almost surely.
Moreover, if $\Payment_i\leq\cost/(\rate_i p_0^i)$, then unconditional
shirking ($\strategy_i(t)\equiv0$) is a strictly dominant strategy.
\end{proposition}

Consequently, to incentivize
firm $i$ that remains unresolved 
at dates arbitrarily close to $T$ to keep working, it
requires $\Payment_i\geq\cost/(\rate_i p_T^i)$.
This necessary condition holds under any non-negative compensation schedule
and either disclosure rule.
The core intuition is that termination compensation cannot substitute
for the breakthrough payment. As a firm exerts cumulative effort
$\Strategy_i(t)$ without a breakthrough, its posterior belief
$p_{\Strategy_i(t)}^i$ that its approach remains feasible strictly
deteriorates. Consequently, once
$\rate_i p_{\Strategy_i(t)}^i\Payment_i<\cost$, the breakthrough
payment no longer covers the marginal cost of effort, even if the
contract is certain to remain active. Stochastic termination can
only further depress the return to effort: when termination is
concealed, the firm risks bearing effort costs for a contract that
has already been terminated, while under disclosure, both the
benefit and cost of effort are realized only while the contract
remains active. Thus, no choice of stopping or disclosure rule can
restore incentives once the success prize falls below this
viability threshold.

Termination compensation cannot overturn this conclusion because it
is paid only if the firm remains unresolved when termination occurs.
Because reducing effort makes a breakthrough less likely, shirking
actually weakly increases the probability of remaining eligible for
compensation. Hence, offering a non-negative compensation schedule
makes stopping effort weakly more attractive, rather than restoring
the incentive to continue. Truncating any strategy $\strategy_i$
once cumulative effort exceeds $\hat e_i$ unambiguously improves
the firm's payoff by saving effort costs while weakly increasing
expected termination compensation.
This establishes the bound on
cumulative effort under any admissible stopping rule, compensation
schedule, and disclosure rule.

\cref{prop:prize-lower-bound} establishes an important limitation of
termination compensation. Because compensation is paid only if the
firm remains unresolved, it cannot rescue a breakthrough payment
that is too small to motivate late-stage experimentation. Indeed,
the proof suggests a stronger result: absent an auditing technology
that makes compensation contingent on incurred effort, termination
compensation actually works against effort incentives.
We exploit this observation to characterize optimal
interdependent contracts under the following assumption. Moreover, we restrict attention to
interdependent contracts with constant breakthrough payments that
implement full effort for any given deadline $T$, whenever a funded firm remains unresolved
and its contract remains active. This is because we are interested in the comparison between contests and
interdependent contracts that implement full effort whenever
a firm remains unresolved and active for any given deadline $T$.

\begin{assumption}[Continuation requirement]
\label{assump:pure-good-news-benchmark}
For every funded firm $i\in I$,  for every $t\in[0,T)$, 
$$Q_i(t)
=
\Pr\!\left(
\tau_i>t
\mid
\text{firm $i$ remains unresolved at $t$}
\right)>0.$$
\end{assumption}

This restriction permits early termination on some histories while
requiring a positive probability of continued experimentation
arbitrarily close to $T$ for each funded firm.
Under Assumption~\ref{assump:symmetry}, write
$\Payment^*=\cost/(\rate p_T)$ and define
$m^*
=
\min\left\{
k\in\{0,\ldots,n-1\}:
\Delta\val(k)\leq\Payment^*
\right\}$,
with the convention that $m^*=n$ if the set is empty.
If $m^*\geq1$, consider the contracts
$\hybrid_i^*
=
\bigl(
\Payment_i^*,
\{\compensation_i^*(t)\}_{t\in[0,T]},
\tau_i^*,
\disclosure_i^*
\bigr)$, 
where $\Payment_i^*=\Payment^*$, $\disclosure_i^*=1$,
$\compensation_i^*(t)=0$, and $\tau_i^*=\tau^*
=
\inf\left\{
t\in[0,T]:\eta_t\geq m^*
\right\}\wedge T$,
with $\inf\varnothing=+\infty$.
If $m^*=0$, the proposed mechanism instead offers no contracts.

\begin{theorem}
\label{thm:optimal interdependent contracts absent breakdown}
Under Assumption~\ref{assump:symmetry}, \ref{assump:pure-good-news-benchmark} and $\mu=0$, among the interdependent contracts implementing full-effort and with constant payments, the principal's
objective is maximized 
by offering $\{\hybrid_i^*:i=1,\ldots,n\}$ when $m^*\geq1$.
When $m^*=0$, it is optimal to fund no firms.
\end{theorem}

For $m^*\geq1$, the proposed stopping rule satisfies the continuation
requirement because there is positive probability of no
breakthroughs before any date $t<T$.
When $m^*=0$, no firm is funded, so the continuation requirement
does not apply to any firm and is vacuously satisfied.
We prove \cref{thm:optimal interdependent contracts absent breakdown}
in two steps. First, in \cref{prop:zeroC}, we fix an admissible
stopping rule satisfying the continuation requirement in
Assumption~\ref{assump:pure-good-news-benchmark} and show that, to implement
full effort while a firm remains unresolved and active,
(1) immediate disclosure of termination is weakly preferred to
concealed termination;
(2) zero compensation is weakly preferred to any non-negative
compensation schedule; and
(3) the minimum constant breakthrough payment is
$\cost/(\rate_i p_T^i)$.
Second, in \cref{prop:optimal_stopping_pure_good_news}, we show
that the specified stopping policy is optimal within this benchmark
given these payment, compensation, and disclosure choices.

\begin{lemma}[Optimality of Zero Compensation and Full Disclosure]
\label{prop:zeroC}
\singlespace
Fix an admissible stopping rule $\tau_i$ satisfying the continuation
requirement in Assumption~\ref{assump:pure-good-news-benchmark}.
Among contracts with constant breakthrough payments that implement
full effort whenever firm $i$ remains unresolved and its contract
remains active, it is optimal to set
$\disclosure_i^*=1$, $\compensation_i^*(t)=0$, and
$\Payment_i^*=\cost/(\rate_i p_T^i)$.
\end{lemma}

The proof is provided in
\cref{apx:pure-good-news-optimal-stochastic}.
This result follows from three observations. First, immediate
disclosure of termination weakly dominates concealed termination.
When termination is hidden, the firm does not know whether its
contract remains active: it bears the cost of effort for sure,
while a breakthrough earns the success prize only if the contract
has not already been terminated. Disclosure eliminates this wedge
by informing the firm immediately when its contract ends, thereby
relaxing its effort incentive constraint and preventing socially
wasteful effort after termination.

Second, once termination is disclosed, positive termination
compensation is unnecessary and, in fact, weakens effort incentives.
For any strategy $\strategy_i$, the probability of remaining
unresolved at a given termination date is decreasing in cumulative
effort $\Strategy_i(t)$. Since compensation is paid precisely when
the firm is terminated while unresolved, a deviation that reduces
effort weakly increases its expected compensation. Removing
$\compensation_i(t)$ therefore makes every low-effort deviation
weakly less attractive relative to full effort, while also reducing
the principal's spending; participation remains satisfied because
the firm can always shirk and obtain zero.

Finally, with immediate disclosure and zero compensation, the
success prize is the only instrument to motivate effort.
By Assumption~\ref{assump:pure-good-news-benchmark}, an unresolved firm can
remain active with positive probability arbitrarily close to $T$.
Since $p_t^i$ declines with cumulative effort, to implement full effort, the binding
incentive constraint is the one at $T$, yielding
the minimum constant prize
$\Payment_i^*=\cost/(\rate_i p_T^i)$.

This zero-compensation result clarifies the model's relationship
with FAR Part~49. Under FAR Part~49, termination settlements are
based on documented allowable costs and work already performed,
rather than on the termination date alone. In our model, incurred
effort is not observable or contractable, so $\compensation_i(t)$
can depend on the timing of termination but not on verified
expenditure.
As \cref{prop:zeroC} shows, this distinction is consequential:
effort-independent termination compensation cannot substitute
for breakthrough payment and, in fact, weakens effort incentives.
Thus, actual termination settlements differ from the pure
compensation instrument in our model because the government can
audit documented costs and work performed.

Given the optimal payment and disclosure choices characterized in
\cref{prop:zeroC}, the principal's remaining task within the
benchmark is to choose when to terminate experimentation.
Under symmetric technologies and a concave value function,
the stopping problem admits a particularly simple solution.
More generally, the same stopping rule is optimal when the
net marginal value of successive innovations satisfies the
single-crossing condition described below.

\begin{lemma}[Optimal Stopping Time for Symmetric
Interdependent Contracts]
\label{prop:optimal_stopping_pure_good_news}
\singlespace
Under Assumption~\ref{assump:symmetry} and \ref{assump:pure-good-news-benchmark}, consider the  payment 
$\Payment_i^*=\Payment^*=\cost/(\rate p_T)$,
the 
disclosure rule $\disclosure_i^*=1$, and $\compensation_i^*(t)=0$
for every funded firm.
Define $m^*
=
\min\left\{
k\in\{0,\ldots,n-1\}:
\Delta\val(k)\leq\Payment^*
\right\}$,
with $m^*=n$ if the set is empty.
If $m^*\geq1$, an optimal stopping policy is
\[
\tau^*_{m^*}
=
\inf\left\{
t\in[0,T]:\eta_t\geq m^*
\right\}\wedge T.
\]
If $m^*=0$, it is optimal to fund no firms.
\end{lemma}

This result follows from a simple comparison between the value
and the cost of successive innovations.
Under \cref{prop:zeroC}, every successful innovation costs the
principal the same prize $\Payment^*$.
Hence, if experimentation terminates after $k$ successes, the
principal's net payoff is
$G(k)\equiv\val(k)-k\Payment^*$.
Obtaining one additional innovation changes this payoff by
$G(k+1)-G(k)=\Delta\val(k)-\Payment^*$.
\cref{prop:optimal_stopping_pure_good_news}
can be generalized to all value functions satisfying the following
single-crossing condition: Suppose that the net marginal value
$\Delta\val(k)-\Payment^*$ single-crosses zero from above in $k$:
if $\Delta\val(k)-\Payment^*\leq0$ for some $k$, then
$\Delta\val(\ell)-\Payment^*\leq0$ for every $\ell>k$.
The single-crossing condition says that, once an additional
innovation ceases to be worthwhile, no subsequent innovation
becomes worthwhile again. Before this point, the principal
prefers one more success; after this point, the principal weakly
prefers to stop. This gives rise to a fixed target number of
innovations $m^*$.

The single-crossing condition is weaker than concavity of $\val$.
Concavity requires $\Delta\val(k)$ itself to be weakly decreasing
in $k$, whereas
the single-crossing condition
only requires its position relative to the constant cost
$\Payment^*$ to change sign at most once.
Thus, marginal values may be non-monotone before or after the
crossing. What matters for the stopping problem is only that,
after the marginal value first falls weakly below the cost of an
additional success, it never rises above that cost again.
Concavity of $\val$ is therefore a sufficient, but not necessary,
condition for the result.

Once the constant prize $\Payment^*$ is
fixed, the optimal continuation decision before $T$ depends
only on the number of successes.
Independence of feasibility across firms implies that rivals'
successes do not change the posterior feasibility of a given
unresolved approach.
As time passes without success, the arrival rate of a future
success declines, but the principal's net payoff conditional on
that success remains $\Delta\val(k)-\Payment^*$.
The principal therefore has no reason to abandon an innovation
with positive net value merely because it has not yet arrived.
For $m^*\geq1$, experimentation continues until $m^*$ successes
have been obtained or $T$ is reached.

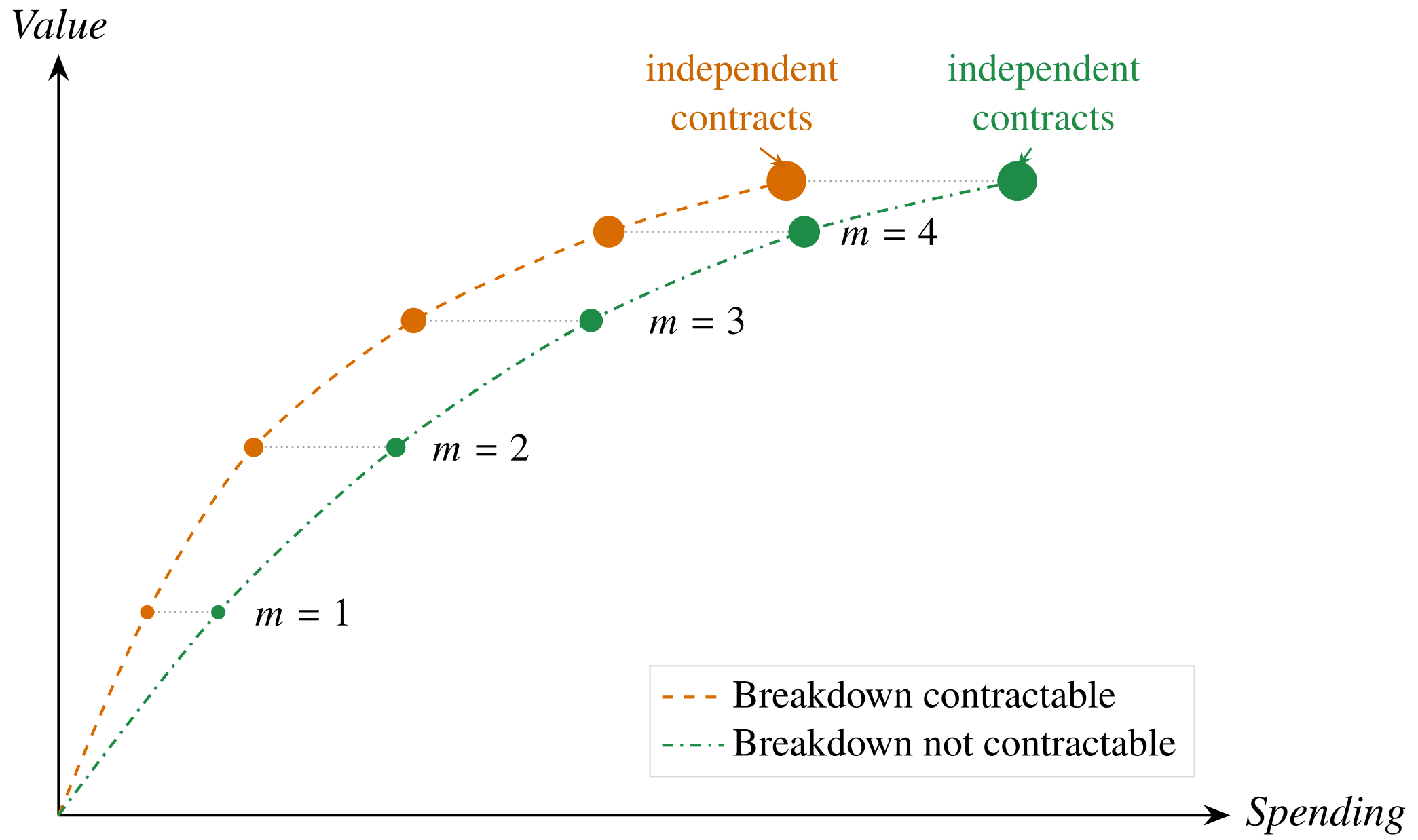
\begin{figure}[t]
\centering
\begin{tikzpicture}
\begin{axis}[
    axis lines=left,
    axis line style={-{Stealth[length=2.6mm,width=2mm]}, line width=0.7pt},
    xmin=0, xmax=6.6, ymin=0, ymax=6.0,
    xtick=\empty, ytick=\empty,
    clip=false,
    width=13cm, height=9.0cm,
    every axis x label/.style={at={(ticklabel* cs:1.0)}, anchor=west, font=\itshape},
    every axis y label/.style={at={(ticklabel* cs:1.0)}, anchor=south, font=\itshape},
    xlabel={Spending},
    ylabel={Value},
    legend style={
        at={(0.97,0.05)}, 
        anchor=south east, 
        font=\small, 
        draw=gray!30, 
        fill=white, 
        fill opacity=0.9, 
        text opacity=1
    },
    legend cell align={left},
]

\addplot[thick, smooth, dashed, color=orange!85!black] coordinates {
    (0,0) (0.5,1.6) (1.1,2.9) (2.0,3.9) (3.1,4.6) (4.1,5.0)
};
\addlegendentry{Breakdown contractable }

\addplot[thick, smooth, dashdotted, color=plaincol] coordinates {
    (0,0) (0.9,1.6) (1.9,2.9) (3.0,3.9) (4.2,4.6) (5.4,5.0)
};
\addlegendentry{Breakdown not contractable }

\draw[gray!70, densely dotted, line width=0.5pt] (axis cs:0.5,1.6) -- (axis cs:0.9,1.6);
\draw[gray!70, densely dotted, line width=0.5pt] (axis cs:1.1,2.9) -- (axis cs:1.9,2.9);
\draw[gray!70, densely dotted, line width=0.5pt] (axis cs:2.0,3.9) -- (axis cs:3.0,3.9);
\draw[gray!70, densely dotted, line width=0.5pt] (axis cs:3.1,4.6) -- (axis cs:4.2,4.6);
\draw[gray!70, densely dotted, line width=0.5pt] (axis cs:4.1,5.0) -- (axis cs:5.4,5.0);

\node[circle, fill=orange!85!black, inner sep=1.4pt] at (axis cs:0.5,1.6) {};
\node[circle, fill=orange!85!black, inner sep=1.9pt] at (axis cs:1.1,2.9) {};
\node[circle, fill=orange!85!black, inner sep=2.5pt] at (axis cs:2.0,3.9) {};
\node[circle, fill=orange!85!black, inner sep=3.1pt] at (axis cs:3.1,4.6) {};
\node[circle, fill=orange!85!black, inner sep=3.9pt] at (axis cs:4.1,5.0) {};

\node[circle, fill=plaincol, inner sep=1.4pt] at (axis cs:0.9,1.6) {};
\node[circle, fill=plaincol, inner sep=1.9pt] at (axis cs:1.9,2.9) {};
\node[circle, fill=plaincol, inner sep=2.3pt] at (axis cs:3.0,3.9) {};
\node[circle, fill=plaincol, inner sep=3.1pt] at (axis cs:4.2,4.6) {};
\node[circle, fill=plaincol, inner sep=3.9pt] at (axis cs:5.4,5.0) {};

\node[anchor=west, font=\small] at (axis cs:1.02,1.6) {$m=1$};
\node[anchor=west, font=\small] at (axis cs:2.02,2.9) {$m=2$};
\node[anchor=west, font=\small] at (axis cs:3.24,3.9) {$m=3$};
\node[anchor=west, font=\small] at (axis cs:4.32,4.6) {$m=4$};

\node[color=orange!80!black, font=\small, anchor=south, align=center] at (axis cs:3.85,5.28)
    {independent \\contracts};
\draw[orange!80!black, line width=0.6pt, -{Stealth[length=1.6mm]}]
    (axis cs:3.95,5.26) -- (axis cs:4.10,5.10);
\node[color=plaincol, font=\small, anchor=south, align=center] at (axis cs:5.55,5.28)
    {independent \\contracts};
\draw[plaincol, line width=0.6pt, -{Stealth[length=1.6mm]}]
    (axis cs:5.48,5.26) -- (axis cs:5.40,5.10);

\end{axis}
\end{tikzpicture}
\caption{Interdependent contracts: breakdown contractable vs not contractable.}
\label{fig:hybrids}
\end{figure}

To conclude this section, we summarize the relationship between
independent and interdependent contracts in different learning
environments in \cref{fig:hybrids}.
Independent contracts are a special case of interdependent
contracts where no stochastic deadline is imposed.
Among contracts with constant payment and implementing full effort, allowing
interdependent termination weakly improves the principal's
optimal payoff relative to independent contracts implementing
full effort through the same horizon $T$.
The additional stopping flexibility allows the principal to
respond to successes on other approaches while retaining
approach-specific breakthrough payments.
Moreover, when breakdown news is contractable, both independent
and interdependent contracts can extract full rent from all
firms, which weakly reduces the principal's expected spending.

\subsection{Contests}\label{sec:contests}
Having characterized optimal interdependent contracts, we now ask whether contests can further improve incentive provision. Since interdependent contracts already allow continuation decisions to depend on rivals' outcomes, any advantage of contests must arise from how rewards depend on relative performance. We focus on simple mechanisms with time-invariant payments: interdependent contracts offer fixed outcome-contingent payments, fixed-prize contests commit to a constant prize pool, and rank-contingent contests offer fixed prizes for each success rank. Under symmetric technologies, we show that fixed-prize contests can outperform interdependent contracts over sufficiently long horizons, while the expenditure ranking reverses for short horizons. We then show that rank-contingent contests can implement optimal interdependent contracts under symmetry. These results isolate the role of payment interdependence in shaping the cost of innovation incentives.

\subsubsection{Fixed-Prize Contests}\label{sec:optimal-contest}

We now compare fixed-prize contests with interdependent contracts under symmetric technologies and pure good-news learning.
Recall that, by \cref{thm:optimal interdependent contracts absent breakdown}, optimal interdependent contracts offer a constant breakthrough payment
$    \Payment^*=\frac{\cost}{\rate\private_T}$,
and terminate experimentation after $m^*$ successes or at $T$, whichever occurs first.
We construct a class of fixed-prize contests that use the same stopping rule but distribute a common prize among successful firms. We first compare expected expenditures under full effort and then examine whether the contests can implement full effort.

\textbf{Fixed-prize contests with success cutoffs.}
For each $m\in\{1,\ldots,n\}$, consider a fixed-prize contest $\jointcontest_m$ with the following features:
(i) the principal commits to a total prize $\Payment(m)$;
(ii) the prize is divided equally among successful firms, with no payment if no firm succeeds; and
(iii) the principal discloses no intermediate successes and announces termination when $m$ firms have succeeded.
Thus, the contest terminates at
\[
    \tau_m
    =\inf\{t\in[0,T]:\eta_t\ge m\}\wedge T,
\]
where $\eta_t$ counts the number of realized successes by time $t$ under the mechanism.
Let $\mathcal J=\{\jointcontest_m:1\le m\le n\}$ denote this class.
To compare outcomes, let $\bar\eta_t$ denote the number of potential breakthroughs by time $t$ under uninterrupted full effort by all firms.\footnote{Unlike $\eta_t$, which counts breakthroughs actually realized under the mechanism and therefore stops increasing after termination, $\bar\eta_t$ is defined under hypothetical uninterrupted full effort through $t$, including after the mechanism would have terminated. The two processes need not coincide after early termination or under deviations from full effort.} Under Assumption~\ref{assump:symmetry} and $\mu=0$, $\bar\eta_T\sim\operatorname{Binomial}(n,q)$, where $q=\delta(1-e^{-\rate T})$.

Under full effort, both $\jointcontest_m$ and interdependent contracts using $\tau_m$ produce the same number of successful innovations, i.e., $\eta_{\tau_m}=\min\{\bar\eta_T,m\}$. 
Therefore, they generate the same expected innovation value: $\expect{\val(\min\{\bar\eta_T,m\})}$.

The two mechanisms differ in their payments. Interdependent contracts offer $\Payment^*$ to each successful firm, whereas the fixed-prize contest pays the entire pool whenever at least one firm succeeds. Consequently, the expected  spending in a contest $\jointcontest_m$ is $ \prob{\bar\eta_T>0}\Payment(m)$, while the expected expenditure under constant-payment interdependent contracts using the same stopping rule $\tau_m$ is $\expect{\min\{\bar\eta_T,m\}}\Payment^*$.
The second expression uses the constant-payment implementation of full effort under interdependent contracts established in \cref{prop:zeroC}. For this comparison, the stopping rule $\tau_m$ is fixed; it need not be optimal unless $m=m^*$.
This identifies the fixed prize that equates expected expenditures under full effort. Whether this prize is sufficient to induce full effort is a separate question.

\begin{theorem}
[Expenditure equivalence and full-effort implementation]
\label{prop:expenditure_equivalent_contest_ic}

Suppose Assumption~\ref{assump:symmetry} holds, with $n\ge2$ and $\mu=0$. For each $m\in\{1,\ldots,n\}$, consider $\jointcontest_m$ with total prize
\begin{equation}
\label{eq:expenditure_equivalent_pool}
    \Payment(m)
    =
    \Payment^*
    \frac{\expect{\min\{\bar\eta_T,m\}}}
    {\prob{\bar\eta_T>0}}.
\end{equation}
Then:

\begin{enumerate}
\item[(i)] \textit{Expenditure equivalence.}
Under full effort, $\jointcontest_m$ and interdependent contracts offering $\Payment^*$ and using $\tau_m$ achieve the same expected innovation value and expected expenditure.

\item[(ii)] \textit{Full-effort implementation.}
For $m=1$ and $m=n$, the prize in \eqref{eq:expenditure_equivalent_pool} is the minimum fixed prize that implements full effort.
For $1<m<n$, the expenditure-equivalent prize implements full effort if and only if
\begin{equation}
\label{eq:expenditure_equivalent_ic}
    \frac{\expect{\min\{\bar\eta_T,m\}}}
    {\prob{\bar\eta_T>0}}
    \ge
    m\frac{\private_T}{\delta}.
\end{equation}

\item[(iii)] \textit{Expenditure comparison.}
For each $1<m<n$, there exists a unique critical horizon $T_m^*>0$ such that \eqref{eq:expenditure_equivalent_ic} holds if and only if $T\ge T_m^*$.
\end{enumerate}
\end{theorem}

The proof is provided in \cref{apx:optimal-contest}. Here we only provide intuition for the proof.
The dynamic incentive constraint reflects two competing effects of delaying experimentation. Waiting preserves the possibility of earning a prize later, potentially with a larger share if few competitors succeed, but also risks losing the opportunity to win if competitors reach the stopping cutoff. Under equal sharing, these effects are closely linked: the expected prize share can be decomposed into a baseline share proportional to contest survival probability and a deadline-dependent residual. Moreover, the expected prize share and contest survival probability satisfy a simple linear relationship.

The equal-sharing structure of the contest allows us to simplify the whole set of dynamic incentive compatibility constraints to the initial incentive constraint.
Under equal sharing,  the expected prize share and contest survival probability satisfy a simple linear relationship. At the expenditure-equivalent prize, this allows every intermediate incentive constraint to be expressed as a nonnegative weighted sum of the initial and terminal constraints. Since the terminal constraint is always strictly satisfied for interior cutoffs, the initial constraint alone determines whether full effort is implementable.

This immediately implies the following comparison:
\begin{corollary}[Expenditure Comparison]
\label{cor:fixed_prize_expenditure}
Under Assumption~\ref{assump:symmetry} and $\mu=0$. Fix a stopping rule $\tau_m$.
For each $1<m<n$, the minimum expected expenditure required to implement full effort under the contest $\jointcontest_m$ is strictly higher than (strictly lower than, or equal to)  that under constant-payment interdependent contracts when $T<T_m^*$ ($T>T_m^*$, or $T=T_m^*$).
When $m=1$ and $m=n$, the two mechanisms have the same minimum expected expenditure.
\end{corollary}

The corollary shows that neither mechanism uniformly minimizes expenditure for a given stopping rule. For intermediate success cutoffs, fixed-prize contests may require greater expenditure than interdependent contracts when the horizon is short, but can become strictly less expensive when the horizon is sufficiently long. Thus, the relative cost of the two mechanisms depends not only on the number of successful innovations required, but also on the duration of experimentation.

To compare the two mechanisms, we construct an expenditure-equivalent prize pool that equates the expected expenditure of a fixed-prize contest and constant-payment interdependent contracts under full effort. This prize equals the contractual breakthrough payment multiplied by the expected number of winners conditional on at least one success. By construction, both mechanisms yield the same innovation outcomes and expected expenditure if the contest implements full effort. The remaining question is therefore whether the expenditure-equivalent prize provides sufficient incentives. If it is insufficient, the contest requires a larger prize pool and is more expensive than interdependent contracts. If it provides strictly more incentives than necessary, the principal can reduce the prize pool while preserving full effort, making the contest less expensive. This approach allows us to compare minimum implementation costs without explicitly solving for the optimal fixed prize in every environment.

The expenditure comparison reflects a tradeoff between pooling rewards and preserving effort incentives. An expenditure-equivalent fixed-prize contest redistributes rewards from histories with many successful firms to those with few successes. This redistribution strengthens incentives for late experimentation, when a firm can expect a relatively large share of the pool if the contest has not yet terminated. Indeed, for interior cutoffs, the terminal incentive constraint is always slack at the expenditure-equivalent prize. However, prize sharing can weaken incentives to exert effort initially, as a firm's eventual reward may be diluted by subsequent successes. When the experimentation horizon is short, the expenditure-equivalent prize fails to provide sufficient initial incentives, requiring a larger pool than the expenditure-equivalent benchmark. As the horizon increases, the contractual breakthrough payment rises to compensate for declining posterior feasibility, making the expenditure-equivalent pool increasingly generous. Beyond a critical horizon, the contest implements full effort with a strictly smaller prize pool, generating expenditure savings relative to interdependent contracts.

\subsubsection{Rank-contingent Contests}\label{sec:rank-contingent contests}
 
 Unlike under interdependent contracts, where the principal directly specifies a stopping rule, stopping in a contest is induced by firms' optimal effort responses to the publicly disclosed information and the rank-contingent payment rule.

 \begin{theorem}[Dominance of rank-contingent contests]
\label{thm:rank-contingent-dominance}
\singlespace
Under Assumption~\ref{assump:symmetry} and $\mu=0$, the optimal constant-payment interdependent contracts characterized in \cref{thm:optimal interdependent contracts absent breakdown} can be implemented by a rank-contingent contest. Thus, there exists at least one rank-contingent contest that weakly dominates the optimal constant-payment interdependent contracts characterized in \cref{thm:optimal interdependent contracts absent breakdown}.
\end{theorem}

The proof is provided in \cref{apx:rank-contingent contests}. The main idea of the proof is to construct a rank-contingent contest that implements the optimal constant-payment interdependent contracts characterized in \cref{thm:optimal interdependent contracts absent breakdown}. 
Notice that optimal interdependent contracts admit a symmetric stopping rule that is a rank-specific cutoff rule. Fix such a stopping profile $\taus\in\stopset$, with cutoffs $\{\gamma_r(\taus)\}_r$. Suppose every firm can occupy every attainable success rank with positive probability. Consider a collection of interdependent contracts that implements full effort before $\taus$. By \cref{prop:zeroC},  without loss of optimality, we can restrict to contracts with zero termination compensation and deterministic identity-contingent success prizes $\Payment_i$. For every attainable success rank $r$, define the rank-specific incentive requirement \[ \Payment_r(\taus) := \frac{\cost}{\rate p_{\gamma_r(\taus)}}. \]
This is the success prize required to induce an unresolved firm to exert effort up to the cutoff for the $r$-th success.

\begin{lemma}[Rank-specific incentive targeting]
\label{lem:rank-specific-incentive-targeting}
\singlespace
Suppose $\gamma_r$ is decreasing. Then every interdependent contract implementing full effort under $\taus$ must satisfy
\begin{equation}
    \Payment_i\geq \max_r \Payment_r(\taus)
    \qquad
    \text{for every }i,
    \label{eq:contract-common-lower-bound}
\end{equation}
where the equality is obtained when $\gamma_r$ is constant.
Moreover, there exists a rank-contingent contest with rank-specific payments $\Payment_r(\taus)$ and a symmetric public disclosure rule $\drule\in\dset$ that induces the same rank-specific cutoff rule $\taus$ and implements full effort before stopping.
\end{lemma}

In general, rank-specific cutoff rules allow the continuation horizon to vary with the success rank being pursued. When these cutoffs differ, the incentive requirements differ across ranks. An interdependent contract must nevertheless offer each firm a single success prize satisfying $\Payment_i \geq \max_{r} \cost / (\lambda p_{\gamma_r})$, because ex ante the firm may occupy any attainable rank. A rank-contingent contest instead conditions the payment on the realized rank, $\Payment_r(\taus) = \cost / (\lambda p_{\gamma_r})$. Thus, whenever a rank with a strictly lower incentive requirement is realized with positive probability, the contest avoids the overpayment inherent in the identity-contingent contract and the dominance is strict.
However, under a concave value function, \cref{thm:optimal interdependent contracts absent breakdown} shows that the optimal stopping rule is \emph{fixed-count stopping}. As a result, the proposed rank-contingent contest is payoff-equivalent to the optimal constant-payment interdependent contracts characterized in \cref{thm:optimal interdependent contracts absent breakdown}.

\begin{proof}[Proof of \cref{thm:rank-contingent-dominance}]
Let $\hybrids^*$ be an optimal interdependent contract mechanism. By \cref{thm:optimal interdependent contracts absent breakdown}(or \cref{prop:optimal_stopping_pure_good_news}), there exists an optimal interdependent contract mechanism whose stopping profile $\taus^*$ admits a fixed-count stopping rule. Apply \cref{lem:rank-specific-incentive-targeting} to $\taus^*$. There exists a rank-contingent contest that induces the same stopping rule and full-effort profile. Hence, the two mechanisms generate the same success process and the same realized set of successful innovations. Moreover, on every realized path with $k$ successes, $\sum_{r=1}^{k}\Payment_r(\taus^*) = k\max_r \Payment_r(\taus^*)$. 
Thus, the constructed rank-contingent contest generates the same value as the optimal interdependent contract mechanism while making the same payments path by path. It therefore yields the principal the same expected payoff. Since the optimal rank-contingent contest can do at least as well as this constructed contest, the result follows. 
\end{proof}

Moreover, the dominance is strict if, under an optimal interdependent-contract mechanism, there exists an attainable success rank $r$ such that $\Payment_r(\taus^*) < \max_r \Payment_r(\taus^*)$ and the $r$-th success occurs with positive probability.

\section{Pure Good News: Optimal Mechanism}\label{sec:good-time-varying}

In this section, we allow breakthrough payments to vary with calendar time.
We also admit stopping rules whose latest possible continuation date
is earlier than $T$, relaxing the continuation requirement
$Q_i(t)>0$ for every $t<T$ used in \cref{sec:hybrid-good-constant}.
We first characterize the least costly payment schedule for a given
termination rule, which isolates the value of payment flexibility. One implication is that it is without loss of optimality to restrict attention to constant payments in independent contracts.
We then choose payments and stopping jointly. We show that among contracts that implement full effort
whenever a firm is unresolved and active, with all experimentation
ending by $T$, relaxing the constant payments restriction in interdependent contracts improves the principal's objective.  All proofs are provided in the appendix.

\paragraph{Payments to implement a given termination rule.}
Consider a firm whose experimentation is terminated at a random
time $\sigma\in[0,T]$ that is independent of its feasibility,
effort, and breakthrough process.
The firm receives no information about $\sigma$ beyond whether
termination has occurred. Termination is announced immediately,
and no compensation is paid upon termination.
Under pure good-news learning, an infeasible approach never
generates news, whereas a feasible approach experiences a
breakthrough at rate $\rate$ per unit of effort.
After $t$ units of effort, the probability of remaining unresolved
and the posterior probability of feasibility conditional on
remaining unresolved are, respectively,
\[
S(t)=1-\delta+\delta e^{-\rate t},
\qquad
p_t=\frac{\delta e^{-\rate t}}{S(t)}.
\]
The following lemma characterizes the minimum cost of sustaining
full effort under this termination risk.

\begin{lemma}[Optimal Payments under Independent Stochastic Termination]
\label{lem:independent_termination_payments}
Suppose $\mu=0$ and fix $T>0$.
Among all nonnegative measurable breakthrough-payment schedules
with finite expected expenditure that implement full effort
until success or termination, the unique minimum-cost
schedule, up to Lebesgue-null sets on active dates, is
\begin{equation}
\label{eq:independent_horizon_optimal_payment}
\Payment^\sigma(t)
=
\expect{\left.
    \frac{\cost}{\rate p_\sigma}
    \,\right|\,\sigma>t},
\qquad
t<T \text{ with } \prob{\sigma>t}>0.
\end{equation}
The minimum expected expenditure is
\begin{equation}
\label{eq:independent_horizon_minimum_cost}
\totalcost^{\min}(\sigma)
=
\expect{
    \frac{\cost S(\sigma)}{\rate}
    \left(e^{\rate\sigma}-1\right)
}.
\end{equation}
\end{lemma}

The payment formula averages the incentive requirements associated
with the termination dates that remain possible.
For a deterministic horizon $s>0$, the minimum constant
breakthrough payment is $\cost/(\rate p_s)$.
At an active date $t$, the optimal reward is the conditional
average of these payments over horizons satisfying $\sigma>t$.
Later termination dates require larger rewards because
unsuccessful experimentation lowers the posterior probability
of feasibility. As the contract survives, earlier termination
dates are ruled out, so $\Payment^\sigma(t)$ is weakly increasing.
It can remain constant over intervals on which termination
cannot occur.

The expenditure formula has a parallel interpretation.
Let
$\mathcal C(s)\equiv
\cost S(s)(e^{\rate s}-1)/\rate$
denote the minimum expected payment under a deterministic
horizon $s$.
Then \eqref{eq:independent_horizon_minimum_cost} states that
$\totalcost^{\min}(\sigma)=\expect{\mathcal C(\sigma)}$:
minimum expected expenditure is the average of the
deterministic-horizon implementation costs evaluated at the
possible termination dates.

Two implications clarify when payment flexibility matters.
First, if termination occurs at a fixed date, the conditional
terminal incentive requirement does not change over time.
Given that  independent contracts use a deterministic deadline, the following result implies that restricting to constant payment in independent contracts is without loss of optimality.

\begin{proposition}[Optimality of Constant Breakthrough Payments under a Deterministic Deadline]
\label{prop:constant_payment_optimal}
Suppose $\mu=0$ and fix a deterministic deadline $T>0$.
Among all nonnegative, absolutely continuous breakthrough-payment
schedules that implement full effort until
success or the deadline $T$, the principal's expected payment is
uniquely minimized by the constant schedule
\begin{equation}
    \Payment^*(t)=\frac{\cost}{\rate p_T},
    \qquad t\in[0,T].
\end{equation}
\end{proposition}

A deterministic deadline thus leaves no expenditure gain from
allowing time-varying rewards. The relevant incentive problem
is dynamic: a firm can postpone effort, preserve a more
optimistic posterior, and then pursue a larger reward later.
Consequently, a schedule that merely covers the contemporaneous
effort cost, $\cost/(\rate p_t)$, need not induce full effort
from the outset. The optimal constant schedule satisfies the
terminal incentive requirement while avoiding an increase in
rewards that would encourage postponement.

To derive the second implication of \cref{lem:independent_termination_payments}, consider an exogenous termination time $\tau$ that is
independent of the firm's feasibility, effort, and breakthrough
process, with no information about $\tau$ disclosed before
termination. Applying
\cref{lem:independent_termination_payments} to
$\sigma=\tau\wedge T$ gives the following characterization.

\begin{proposition}[Optimal Time-varying Payment under an Exogenous Stochastic Stopping Time]
\label{prop:stochastic_termination_optimal_payment}
Suppose $\mu=0$ and fix a deterministic deadline $T>0$.
Among all nonnegative measurable breakthrough-payment schedules
that implement full effort until
success or $\tau\wedge T$ and have finite expected
expenditure, the principal's expected payment is uniquely
minimized, up to Lebesgue-null sets on active dates, by
\begin{equation}
\label{eq:optimal_payment_stochastic_termination}
\Payment^*(t)
=\expect{\left.
\frac{\cost}{\rate p_{\tau\wedge T}}
\,\right|\,\tau>t},
\qquad
t<T\ \text{with}\ \prob{\tau>t}>0.
\end{equation}
\end{proposition}

Stochastic termination changes the cost of postponing effort:
waiting can cause the firm to lose its opportunity to experiment.
This permits lower early rewards than a constant prize tied
to the latest possible continuation date.
To make the comparison precise, let
$L=\operatorname*{ess\,sup}(\tau\wedge T)>0$.
For the same termination rule, the minimum constant prize is
$\cost/(\rate p_L)$, whereas the optimal time-varying schedule
never exceeds this amount.
Payment flexibility therefore weakly reduces expected expenditure,
with strict savings whenever
$\prob{0<\tau\wedge T<L}>0$.
If termination occurs only at $0$ or $L$, the optimal payment
remains constant on all active dates, and there is no strict gain.
Thus, it is the possibility of termination at a positive date
before the latest continuation date that generates the savings.

\paragraph{Optimal Interdependent Contracts: Jointly choosing payments and termination.}

We now characterize the optimal interdependent contracts that allow for time-varying payments.
Firms observe their own outcomes and whether their contracts
remain active, while intermediate successes by other firms are
concealed. Termination is announced immediately, and no
compensation is paid upon termination.
These information and compensation rules are maintained in
the following characterization.

For any symmetric stopping rule $\tau$,
based on the principal's observed breakthroughs,
define $\sigma_i^\tau$ as the termination time generated by
that rule if firm $i$ were kept unresolved, while the other
firms exerted full effort.
 Before firm $i$ succeeds,
its contract remains active precisely until $\sigma_i^\tau$,
which is independent of its own feasibility, effort, and
breakthrough process. Because intermediate rival successes
are concealed, the firm learns about this horizon only through
its contract's survival. Thus, \cref{lem:independent_termination_payments}
applies even though the principal's stopping rule
is endogenous to the portfolio's outcomes.

The optimal rule compares the marginal value of another
success with a time-dependent implementation-cost threshold.
Define
\[
\kappa(t)
\equiv
\frac{\cost}{\rate}
\left(1+\frac{1-\delta}{\delta}e^{2\rate t}\right)
\]
and let
\begin{equation}
\label{eq:optimal_interdependent_stopping}
\tau^*\equiv
\inf\big\{t\in[0,T]:\eta_t=n\ \text{or}\
\Delta\val(\eta_t)\le\kappa(t)\big\}\wedge T.
\end{equation}
The marginal-value condition is evaluated only when
$\eta_t<n$, and the infimum of the empty set is $+\infty$.
The associated breakthrough-payment schedule is

\begin{equation}
\label{eq:optimal_interdependent_timevarying_payment}
\Payment_i^*(t)
=
\expect{\left.
\frac{\cost}{\rate p_{\sigma_i^{\tau^*}}}
\,\right|\,\sigma_i^{\tau^*}>t},
\qquad
t<T\ \text{with}\ \prob{\sigma_i^{\tau^*}>t}>0.
\end{equation}
Symmetry implies that this payment schedule is common
to all firms.

\begin{theorem}[Optimal Interdependent Contracts with Time-Varying Breakthrough Payments]
\label{thm:optimal_interdependent_timevarying}
Under Assumption \ref{assump:symmetry}, suppose $\mu_i=0$
for every firm $i$. 
Among all symmetric stopping rules and nonnegative measurable
breakthrough-payment schedules with finite expected expenditure
that implement full effort whenever a firm is unresolved and
active, the principal's objective is maximized by the stopping
time $\tau^*$ in \eqref{eq:optimal_interdependent_stopping}
and the payments $\Payment_i^*(t)$ in
\eqref{eq:optimal_interdependent_timevarying_payment}.
A success that triggers termination receives its breakthrough
payment.
\end{theorem}

The threshold $\kappa(t)$ has a useful economic interpretation.
Extending a deterministic horizon increases the probability
of a breakthrough at rate $\delta\rate e^{-\rate t}$ and
increases minimum expected expenditure at rate $\mathcal C'(t)$.
Their ratio is
\[
\kappa(t)
=
\frac{\mathcal C'(t)}{\delta\rate e^{-\rate t}}.
\]
Thus, $\kappa(t)$ measures the marginal procurement expenditure
per additional expected success, including the incentive cost
of sustaining effort.
As unsuccessful experimentation continues, the posterior
probability of feasibility falls and the reward required to
sustain effort rises. These effects are reflected in an increasing
procurement cost per additional expected success.
After minimizing transfers, the number of unresolved firms
scales both the expected gain and the cost of continuing,
so the continuation decision compares
$\Delta\val(\eta_t)$ directly with $\kappa(t)$.

The stopping rule responds to two sources of diminishing returns.
The passage of time raises $\kappa(t)$, while additional successes
lower $\Delta\val(\eta_t)$ by concavity.
Either can make further experimentation unprofitable.
Moreover, once the marginal-value condition calls for stopping,
neither further waiting nor additional successes can restore
the case for continuation.
This explains why the optimal rule is characterized by a single
continuation cutoff for each success count.

For $k=0,\ldots,n-1$, let $\gamma_k$ denote the calendar-time
deadline for continuing after $k$ successes.
These deadlines provide an equivalent description of the
principal's stopping rule:
\begin{equation}
\gamma_k =
\begin{cases}
0,
& \displaystyle
\Delta\val(k)\le \frac{\cost}{\rate\delta},
\\[10pt]
\displaystyle
\frac{1}{2\rate}
\log\left[
\frac{\delta}{1-\delta}
\left(
\frac{\rate\Delta\val(k)}{\cost}-1
\right)
\right],
& \displaystyle
\frac{\cost}{\rate\delta}
<\Delta\val(k)<\kappa(T),
\\[12pt]
T,
& \displaystyle
\Delta\val(k)\ge\kappa(T).
\end{cases}
\label{eq:optimal_rank_cutoffs}
\end{equation}
Because $\val$ is concave,
$\gamma_0\ge\gamma_1\ge\cdots\ge\gamma_{n-1}$.

The three cases distinguish immediate termination, an endogenous
deadline before $T$, and continuation until the exogenous deadline.
In the interior case, the deadline equates the marginal value
of another success with its marginal implementation cost.
When $\gamma_k=T$, experimentation continues to $T$ unless
another success changes the relevant cutoff.
The ordering of the deadlines also means that a new success
can shorten the remaining experimentation period or trigger
immediate termination.
If $\gamma_0=0$, no experimentation takes place, and the payment
schedule need not be specified on any active date.

This result clarifies the role of calendar time in the principal's
decision. With a constant prize held fixed, the net gain from
another success depends on the number of successes already
obtained, leading to the fixed-count stopping rule in the
constant-payment benchmark.
When payments and termination are chosen jointly in the broader
class considered here, the minimized cost of inducing another
success also depends on how long experimentation has continued.
The optimal contract therefore combines payments that respond
to the remaining termination risk with deadlines that depend
on the number of successes already achieved.

\section{Robustness: Asymmetric Technologies}
The symmetry assumption plays an important role in the dominance of
rank-contingent contests. Under symmetry, a firm's incentive requirement at a
given history depends only on the success rank being pursued, so rank is
sufficient for targeting payments. With asymmetric technologies, however, the
required payment can depend on both the firm's identity and the success rank.
Interdependent contracts can accommodate this heterogeneity through
identity-specific success prizes, whereas an anonymous rank-contingent contest
must offer the same prize to whichever firm occupies a given rank. As a result,
rank-contingent payments may overcompensate easier-to-incentivize firms, and the
dominance result can be reversed. The following example first illustrates this
force holding the induced effort and stopping behavior fixed.

\begin{example}
\label{ex:asymmetric-rank-contest-analytical}
The symmetry assumption is important for the dominance of rank-contingent contests over interdependent contracts. To see this, consider two firms and suppose $\mu_i=0$ for $i=1,2$. The mechanism stops at the first success or at the terminal deadline $T$, so that $\tau=\inf\{t:\eta_t=1\}\wedge T$. Hence there is only one attainable success rank.

Suppose that, at the binding continuation history, the two firms have different incentive requirements: $\cost / (\lambda_1 p_{1,T}) = 1$ and $\cost / (\lambda_2 p_{2,T}) = 10$. Thus, firm $2$ is substantially more costly to incentivize than firm $1$.

Under interdependent contracts, the principal can condition the success prize on firm identity. The minimum success prizes that induce full effort are therefore $\Payment_1=1$ and $\Payment_2=10$. Consequently, on a realized path in which firm $1$ produces the first success, the principal pays only $\Payment_1=1$, whereas on a path in which firm $2$ produces the first success, she pays $\Payment_2=10$.

Now consider an anonymous rank-contingent contest that induces the same stopping rule and effort profile. Since there is only one attainable rank, let $\Payment_1^{RC}$ denote the prize paid to the first successful firm. Because either firm can be the first successful firm with positive probability, the common rank-one prize must satisfy both firms' incentive constraints:
\[
    \Payment_1^{RC}
    \geq
    \max\left\{
        \frac{\cost}{\lambda_1p_{1,T}},
        \frac{\cost}{\lambda_2p_{2,T}}
    \right\}
    = 10.
\]
Hence, the least costly rank-contingent contest sets $\Payment_1^{RC}=10$.

It follows that, on every path in which firm $1$ succeeds first, $\Payment_1^{RC} = 10 > 1 = \Payment_1$. On paths in which firm $2$ succeeds first, the two mechanisms make the same payment. Since firm $1$ produces the first success with positive probability, the rank-contingent contest has strictly greater expected spending than the interdependent contracts while inducing exactly the same effort and stopping behavior.

The source of the failure is that rank alone is no longer sufficient to determine the incentive requirement. Interdependent contracts can condition payments jointly on the firm's identity and its own outcome, whereas an anonymous rank-contingent contest must offer a single prize to whichever firm occupies a given rank. The rank-contingent prize must therefore satisfy the incentive constraint of the most costly firm that can occupy that rank, which can lead to overpayment when a less costly firm succeeds. Under symmetric firms, this cross-firm heterogeneity disappears, and rank becomes sufficient for targeting incentive payments.
\end{example}

The preceding example isolates the source of the inefficiency by holding the
effort and stopping policy fixed. It shows that, once incentive requirements
differ across firms, an anonymous rank-contingent contest may require strictly
greater expenditure to implement the same innovation process. This comparison
alone, however, does not establish that interdependent contracts outperform the
\emph{optimal} rank-contingent contest: the latter may respond to the payment
inefficiency by choosing a different prize and thereby inducing different effort
and stopping behavior. The following numerical example allows both mechanism
classes to optimize and shows that the reversal can persist even after this
adjustment.

The following example shows that technological asymmetry can reverse the
comparison between interdependent contracts and anonymous
rank-contingent contests, even when each mechanism is optimized.
We keep payments constant over time and allow firms' experimentation
horizons to end before $T$.
\begin{example}[Failure of rank-contingent dominance under asymmetry]
\label{ex:asymmetric-rank-contest-numerical}

Consider two firms with $\cost=1$ and $T=4$. The principal values
only the first success:
$    \val(0)=0$, and $ \val(k)=80$ for all $k\geq1.$
The technologies are
$    (\delta_1,\lambda_1, \mu_1)=(0.6,2,0)$,
   $(\delta_2,\lambda_2, \mu_2)=(0.3,0.4,0).$
Both mechanisms terminate experimentation after the first success
or at $T$, while allowing individual firms to stop earlier.

After $t$ units of unsuccessful experimentation, firm $i$'s
posterior probability of feasibility is
$    p_{i,t}
    =
    \frac{\delta_i e^{-\lambda_i t}}
         {1-\delta_i+\delta_i e^{-\lambda_i t}}.$
For a constant success prize $\Payment_i$, an interior effort
cutoff $\gamma_i\in(0,T)$ satisfies
$
    \lambda_i p_{i,\gamma_i}\Payment_i=\cost.
$
A firm does not start experimenting if
$\lambda_i\delta_i\Payment_i\leq\cost$; it works until $T$,
absent an earlier success, if
$\lambda_i p_{i,T}\Payment_i\geq\cost$.

\medskip
\noindent
\textbf{Interdependent contracts.}
The principal can offer different prizes to the two firms and
therefore choose their effort cutoffs separately. To implement
a positive cutoff $\gamma_i\leq T$, the minimum constant prize is
$
    \Payment_i(\gamma_i)
    =
    \frac{\cost}{\lambda_i p_{i,\gamma_i}}.
$
A cutoff of $\gamma_i=0$ means that firm $i$ does not experiment,
and its prize can be set to zero.

Let $\pi_i(\gamma_1,\gamma_2)$ denote the probability that firm $i$
produces the first success when each firm experiments until its
cutoff or until either firm succeeds. Since at most one prize
is paid, the principal's expected payoff is
$
    \Pi^H(\gamma_1,\gamma_2)
    =
    \sum_{i=1}^{2}
    \bigl[80-\Payment_i(\gamma_i)\bigr]
    \pi_i(\gamma_1,\gamma_2).
$
The principal chooses $(\gamma_1,\gamma_2)\in[0,T]^2$
to maximize this payoff.

\medskip
\noindent
\textbf{Anonymous rank-contingent contests.}
Because experimentation ends at the first success, the contest
has a single relevant prize: a common payment $\Payment^{RC}$
to whichever firm succeeds first. Before that success, there is
no realized success information to disclose. The prize therefore
determines each firm's effort cutoff through its own posterior:
$
    \lambda_i
    p_{i,\gamma_i(\Payment^{RC})}
    \Payment^{RC}
    =
    \cost
$
whenever the induced cutoff is interior, with the boundary
conventions described above.

Let $\pi^{RC}(\Payment^{RC})$ denote the probability of at least
one success before the induced cutoffs. The principal chooses
the common prize to maximize
$
    \Pi^{RC}(\Payment^{RC})
    =
    \bigl(80-\Payment^{RC}\bigr)
    \pi^{RC}(\Payment^{RC}).
$
Thus, the interdependent contracts allow the principal to choose
two prizes, whereas the anonymous contest requires one prize
to serve both firms' incentive needs.

\medskip
\noindent
\textbf{Comparison.}
Numerical maximization gives the following results; all reported
effort cutoffs are interior.
\begin{center}
\begin{tabular}{lcc}
\hline
 & Interdependent contracts & Rank-contingent contest \\
\hline
Prize if firm 1 succeeds first & $5.257$  & $12.068$ \\
Prize if firm 2 succeeds first & $22.093$ & $12.068$ \\
Firm 1's effort cutoff        & $1.329$  & $1.773$  \\
Firm 2's effort cutoff        & $3.029$  & $1.237$  \\
Probability of success        & $0.6511$ & $0.6316$ \\
Expected expenditure          & $5.387$  & $7.622$  \\
Principal's expected payoff   & $46.700$ & $42.904$ \\
\hline
\end{tabular}
\end{center}
Hence,
$
    \Pi^{H,*}
    \approx46.700
    >
    42.904
    \approx\Pi^{RC,*},
$
implying that the optimal anonymous rank-contingent contest is 
strictly dominated by interdependent contracts in this example.

\end{example}

The numerical comparison illustrates the incentive tradeoff.
Relative to the optimal interdependent contracts, the common
contest prize raises firm 1's reward and extends its effort
horizon, while reducing firm 2's reward and shortening its
effort horizon. The principal consequently spends more while
obtaining a lower probability of success.
Individualized prizes avoid this tradeoff by allowing the
principal to choose each technology's experimentation horizon
separately. An anonymous contest cannot increase the reward
for firm 2 without also increasing the payment when firm 1
succeeds first. Under symmetry, the firms' incentive
requirements coincide, so this conflict does not arise.


\bigskip

\section{Conclusion}

How much competition should a designer introduce when procuring innovation? We argue that the choice is not simply between independent contracts and contests. Interdependent contracts offer an intermediate mechanism that preserves approach-specific rewards while allowing firms' continuation to depend on competitors' outcomes. This distinction separates two dimensions of competition: interdependence in continuation decisions and interdependence in payments.

Our results show that the value of each dimension depends critically on which technological outcomes are contractable. When both breakthroughs and breakdowns are contractable, interdependent contracts can fully extract firms' rents and coordinate experimentation through optimal stopping. This provides an economic rationale for procurement arrangements, such as those used by the U.S. Department of Defense, that fund parallel technological approaches while retaining the flexibility to terminate individual projects.

When only breakthroughs are contractable, competition over rewards can offer additional benefits, but its advantage is not universal. Under symmetric technologies, we show that fixed-prize contests are more expensive than interdependent contracts with constant payments over short horizons, but less expensive over sufficiently long horizons, holding experimentation and stopping fixed. The two mechanisms are expenditure-equivalent at the first- and last-success cutoffs. Rank-contingent prizes and time-varying payments provide further flexibility, while technological heterogeneity can favor individualized contracts over anonymous contests.

Taken together, our findings suggest that competition in innovation procurement is not inherently beneficial or costly. Interdependence in continuation can improve the coordination of experimentation, while interdependence in payments has more nuanced effects on incentive costs. The central design question is therefore not simply whether to introduce competition, but how and along which dimensions technological approaches should be made interdependent.

\clearpage

\end{spacing}

\appendix
\section{Appendix: Omitted Proof in \cref{sec:mixed-news}}\label{apx:optimal_stopping_good_news}

\begin{proof}[Remaining proof of \cref{thm:optimal interdependent contracts}]
We solve the induced social planner's stopping problem:
retained unresolved approaches exert full effort until resolution
or permanent termination, and all experimentation ends no later
than $T$.
We verify the candidate policy stated in the theorem.
Recall that $\val$ is weakly increasing and concave:
$\Delta\val(k)\geq0$ and
$\Delta\val(k+1)\leq\Delta\val(k)$.
Throughout this verification, $\rate>\mu>0$.

\medskip
\noindent
\textbf{Step 1: Candidate stopping boundary.}
Consider a candidate continuation region
\[
\continuation
=
\left\{
(k,t):
k\in\{0,\ldots,n-1\},\quad 0\leq t<T,\quad
p_t\rate\Delta\val(k)>\cost
\right\}.
\]
The candidate policy requires that to continue at a history with
$k$ successes, the instantaneous expected social gain from
operating one unresolved approach for an additional instant and
then stopping is positive:
$p_t\rate\Delta\val(k)>\cost$.
Moreover, the candidate policy retains all unresolved approaches
in $\continuation$ and terminates all unresolved approaches
outside $\continuation$.
In particular, all remaining approaches are terminated
at $T$, even if their instantaneous expected social gain is
still positive.

For each $k=0,\ldots,n-1$, define
\[
\rho(k)
=
\inf\left\{
t\geq0:
p_t\rate\Delta\val(k)\leq\cost
\right\}.
\]
If $\delta\rate\Delta\val(k)\leq\cost$, then $\rho(k)=0$.
Otherwise, $\rho(k)>0$ is uniquely determined by
$p_{\rho(k)}\rate\Delta\val(k)=\cost$.
Using \eqref{eq:posterior-mixed} gives the explicit expression
of $\rho(k)$.
Define the horizon-capped deadline
\[
\bar{\rho}(k)\equiv\min\{\rho(k),T\}.
\]
For $k<n$ and $0\leq t\leq T$, the continuation condition is
equivalently $t<\bar{\rho}(k)$.

Because $\Delta\val(k)$ is weakly decreasing in $k$,
$\rho(k+1)\leq\rho(k)$
and $\bar{\rho}(k+1)\leq\bar{\rho}(k)$
for $k=0,\ldots,n-2$.
If $\val$ is strictly concave and both
uncapped deadlines are positive, the inequality
between the uncapped deadlines is strict.
Capping at $T$ can turn this strict inequality into
equality when both capped deadlines equal $T$.
If $\val$ is linear, $\Delta\val(k)$ is constant and hence
$\rho(k)$ \proofedit{and $\bar{\rho}(k)$ are}
independent of $k$.

\medskip
\noindent
\textbf{Step 2: The candidate stopping region is optimal.}
At $t=T$, termination is mandatory, so there is no
continuation decision to verify.
Fix a state \proofedit{at $t<T$} with $k$ successes and
$a\geq1$ unresolved approaches such that
$p_t\rate\Delta\val(k)\leq\cost$.
Suppose the planner continues any $b\leq a$ unresolved approaches
for an additional instant.
Applied to the stopping payoff $\val(k)$, the
instantaneous expected net-surplus rate, including effort costs,
is
\[
(\mathcal L^b\val)(k,t)-b\cost
=
b\left[p_t\rate\Delta\val(k)-\cost\right]
\leq0.
\]

Moreover, the stopping region is absorbing.
If no news occurs, $p_t$ decreases by
\eqref{eq:belief dynamics}, making experimentation weakly less
attractive.
Following a breakdown, the number of successes remains unchanged.
Following a success, the state moves from $k$ to $k+1$,
and, if an unresolved approach remains,
concavity implies
$\Delta\val(k+1)\leq\Delta\val(k)$.
Hence,
$p_t\rate\Delta\val(k+1)
\leq p_t\rate\Delta\val(k)\leq\cost$.
If no unresolved approach remains, experimentation
ends. Reaching $T$ also ends experimentation.
Thus, no possible transition can return the state to the
candidate continuation region.

Consequently, starting from the stopping region, the instantaneous
expected net contribution of experimentation remains nonpositive
under every admissible continuation policy.
By Dynkin's formula,
\[
\mathbb E\left[
\val(\eta_\sigma)
-
\cost\int_t^\sigma a(s)\,ds
\right]
\leq\val(k)
\]
for every admissible future stopping time
$\sigma\in[t,T]$.
Immediate termination of all unresolved approaches is therefore
optimal outside $\continuation$.

\medskip
\noindent
\textbf{Step 3: Candidate continuation value and verification
within the continuation region.}

Let $W(k,a,t)$ denote expected social surplus under the candidate
policy starting at time $t$ with $k$ successes and $a$ unresolved
approaches.
By construction, $W(k,0,t)=\val(k)$, and
$W(k,a,t)=\val(k)$ whenever
$t\geq\proofedit{\bar{\rho}(k)}$
\proofedit{and $k<n$}.
In particular, the finite-horizon terminal condition is
\[
W(k,a,T)=\val(k)
\]
for every feasible state.
If $t<\proofedit{\bar{\rho}(k)}$ and $a\geq1$, the candidate
policy operates all $a$ unresolved approaches.
Hence $W$ satisfies
\begin{align}
0
={}&
W_t(k,a,t)-a\cost
\nonumber\\
&+
a p_t\rate
\left[W(k+1,a-1,t)-W(k,a,t)\right]
\nonumber\\
&+
a(1-p_t)\mu
\left[W(k,a-1,t)-W(k,a,t)\right].
\label{eq:candidate_HJB}
\end{align}
\proofedit{The candidate value is continuous in time and
piecewise continuously differentiable between the finitely many
capped deadlines. The time-dependent Dynkin calculation below
can therefore be applied interval by interval.}
The only remaining issue is whether the planner could profitably
deviate by terminating only some unresolved approaches while
continuing the others.

Consider an arbitrary admissible stopping policy.
Recall that $\eta_s$ denotes the number of successes at time $s$
and $a_s$ the number of unresolved approaches the policy keeps
active.
Consider the candidate value process $W(\eta_s,a_s,s)$.
Between endogenous resolution events and deliberate termination
events, all retained approaches work continuously.
\proofedit{Write
\[
Y_s
\equiv
W(\eta_s,a_s,s)-\int_t^s\cost a_v\,dv.
\]
}

If the state lies in the candidate continuation region,
\eqref{eq:candidate_HJB} implies that
$\proofedit{Y_s}$ has zero drift.
If the state lies in the candidate stopping region
\proofedit{and $s<T$}, then
$W(\eta_s,a_s,s)=\val(\eta_s)$, and the drift is
$a_s\left[
\proofedit{p_s}\rate\Delta\val(\eta_s)-\cost
\right]\leq0$
\proofedit{whenever $a_s>0$; it is zero if $a_s=0$}.
Finally, suppose the policy deliberately terminates some
unresolved approaches, reducing their number from $a$ to $b<a$.
We use the following lemma, proved below.

\begin{lemma}[Value of an Additional Unresolved Approach]
\label{lem:value_monotone_active}
\proofedit{For every feasible comparison with
$k,a\in\mathbb Z_+$, $k+a+1\leq n$, and $0\leq t\leq T$,}
\begin{equation}
\label{eq:value_monotone_active}
W(k,a+1,t)\geq W(k,a,t).
\end{equation}
\end{lemma}

By \cref{lem:value_monotone_active},
$W(k,a,t)\geq W(k,b,t)$ for every $b<a$.
Thus, immediately terminating any strict subset of unresolved
approaches cannot increase continuation surplus.
\proofedit{The same argument includes termination of all active
approaches and interventions at the initial date.
In particular, iterating the lemma down to zero active approaches
gives
\[
W(k,a,t)\geq W(k,0,t)=\val(k).
\]
}

It follows that under any admissible policy,
$\proofedit{Y_s}$ is a supermartingale
\proofedit{up to the policy's terminal time}.
Under the candidate policy it is a martingale until termination.
\proofedit{Indeed, the candidate policy continues only in
$\continuation$, where the drift is zero, and terminates only
where its continuation value equals $\val$.
Thus its deliberate termination decisions create no downward
jump in $Y_s$, including at the deadline $T$.}

Let $\sigma$ denote the terminal time induced by an arbitrary
admissible policy.
\proofedit{We have $\sigma\leq T$.
Since $a_s\leq n$, $\val$ is finite on $\{0,\ldots,n\}$,
and accumulated effort costs are bounded by $n\cost T$,
the stopped process $Y_{s\wedge\sigma}$ is bounded.
Optional sampling therefore applies directly, without
localization.
Using
$W(\eta_\sigma,a_\sigma,\sigma)\geq\val(\eta_\sigma)$
from \cref{lem:value_monotone_active} gives}
\begin{align*}
\mathbb E\left[
\val(\eta_\sigma)
-
\int_t^\sigma\cost a_s\,ds
\right]
&\leq
\mathbb E\left[
W(\eta_\sigma,a_\sigma,\sigma)
-
\int_t^\sigma\cost a_s\,ds
\right]\\
&\leq W(k,a,t).
\end{align*}
\proofedit{Under the candidate policy, the stopped process is a
martingale and
$W(\eta_{\tau^*},a_{\tau^*},\tau^*)=\val(\eta_{\tau^*})$.
Consequently,
\[
\mathbb E\left[
\val(\eta_{\tau^*})
-
\int_t^{\tau^*}\cost a_s\,ds
\right]
=
W(k,a,t).
\]
}
Hence, no admissible stopping policy yields greater expected
surplus than the candidate policy.

Therefore, the optimal policy retains all unresolved approaches
whenever \proofedit{$t<T$ and}
$p_t\rate\Delta\val(k)>\cost$, and terminates all unresolved
approaches whenever
$p_t\rate\Delta\val(k)\leq\cost$
\proofedit{or $t=T$}.
Equivalently,
\proofedit{
starting from the current date $t$,
\[
\tau^*
=
\inf\left\{
s\in[t,T):
a_s=0
\ \text{or}\
(\eta_s,s)\notin\continuation
\right\}
\wedge T,
\]
with $\inf\varnothing=+\infty$.
While an unresolved approach remains and $s<T$, the condition
$(\eta_s,s)\notin\continuation$ is equivalent to
$s\geq\rho(\eta_s)$.
At the initial date, set $t=0$.
This completes the verification of the optimal stopping policy
in \cref{thm:optimal interdependent contracts}.}
\end{proof}

\begin{proof}[Proof of \cref{lem:value_monotone_active}]
Recall that the horizon-capped deadline is $\bar{\rho}(\ell)\equiv\min\{\rho(\ell),T\}$, for $\ell=0,\ldots,n-1.$
Consider a feasible comparison with $k+a+1\leq n$ and $0\leq t\leq T$.
The result is immediate if $t\geq\proofedit{\bar{\rho}(k)}$,
since the candidate policy terminates all unresolved approaches and
$W(k,a,t)=W(k,a+1,t)=\val(k)$.
Hence, consider a continuation state with
$t<\proofedit{\bar{\rho}(k)}$.

\medskip
\noindent
\textbf{Step 1: Characterize the distribution of future successes under the candidate policy.}
Fix $k$ and $t<\proofedit{\bar{\rho}(k)}$, and measure elapsed time
from the current state by $u\geq0$.
\proofedit{All success and resolution times used below are potential
times under uninterrupted full effort from the current state.
The candidate stopping policy realizes these outcomes only while
experimentation remains active.}
Conditional on an approach being unresolved at time $t$, let
\[
q_u(u)=p_te^{-\rate u}+(1-p_t)e^{-\mu u}
\]
denote the probability that it remains unresolved for an additional
$u$ units of time. Its posterior probability of being feasible
conditional on remaining unresolved until $t+u$ is
\[
\pi_t(u)=
\frac{p_te^{-\rate u}}
     {p_te^{-\rate u}+(1-p_t)e^{-\mu u}}.
\]
Finally, let $q_s(u)=p_t(1-e^{-\rate u})$ denote the probability,
conditional on being unresolved at time $t$, that the approach
succeeds by time $t+u$. Then
\begin{equation}\label{eq:sprime_identity}
q_s'(u)
=p_t\rate e^{-\rate u}
=\rate\pi_t(u)q_u(u).
\end{equation}

For each \proofedit{$j=0,\ldots,a$}, define the remaining deadline
after $j$ additional successes by
\proofedit{
\[
d_j\equiv\max\{\bar{\rho}(k+j)-t,0\}.
\]
}
Because $\val$ is concave, $\proofedit{\bar{\rho}(k)}$ is weakly
decreasing in $k$, so $d_0\geq d_1\geq\cdots\geq d_a$.
\proofedit{Moreover, $0\leq d_j\leq T-t$.}
Let $N_a$ denote the number of additional successes generated under
the candidate policy when the state begins with $a$ unresolved
approaches.

We first characterize the distribution of $N_a$.
For every \proofedit{$1\leq j\leq a+1$} such that $d_{j-1}>0$,
\begin{equation}
\label{eq:success_tail_characterization}
\{N_a\geq j\}
=
\left\{
\text{at least $j$ of the $a$ approaches
\proofedit{potentially} succeed by time }t+d_{j-1}
\right\}.
\end{equation}
Indeed, if the candidate policy obtains at least $j$ additional
successes, then the $j$-th success must occur no later than
$\proofedit{\bar{\rho}(k+j-1)}=t+d_{j-1}$.
Conversely, suppose at least $j$ approaches have
\proofedit{potential} success times
$t_{(1)}\leq\cdots\leq t_{(j)}
\leq\proofedit{\bar{\rho}(k+j-1)}$.
For every $r\leq j$, monotonicity of the deadlines implies
$\proofedit{\bar{\rho}(k+r-1)\geq\bar{\rho}(k+j-1)}$, and therefore
\[
t_{(r)}
\leq t_{(j)}
\leq\proofedit{\bar{\rho}(k+j-1)}
\leq\proofedit{\bar{\rho}(k+r-1)}.
\]
Hence, each of these $j$ successes occurs before the deadline
applicable at its rank, so all $j$ are realized by the candidate
policy.
\proofedit{These event identities hold almost surely, since potential
success times have no atoms at the deterministic deadlines.
If $d_{j-1}=0$, then $\Pr(N_a\geq j)=0$.}

Since approaches are independent, the number that
\proofedit{potentially} succeed by $t+d_{j-1}$ is binomial with
success probability $q_s(d_{j-1})$. Therefore,
\begin{equation}
\label{eq:binomial_tail}
\Pr(N_a\geq j)
=
\sum_{\ell=j}^{a}
\binom{a}{\ell}
q_s(d_{j-1})^\ell
\bigl[1-q_s(d_{j-1})\bigr]^{a-\ell}.
\end{equation}

\medskip
\noindent
To establish that the candidate value is weakly increasing in the
number of unresolved approaches, we compare the candidate policy
starting with $a+1$ unresolved approaches to that starting with $a$.
We decompose the resulting change in expected surplus into two
components: the change in expected gross innovation value, denoted
by $\Delta\mathcal V_a$, and the change in expected effort cost,
denoted by $C_{a+1}-C_a$. Thus,
\begin{equation}
\label{eq:incremental_surplus}
W(k,a+1,t)-W(k,a,t)
=
\Delta\mathcal V_a-\bigl(C_{a+1}-C_a\bigr).
\end{equation}
We define and compute $\Delta\mathcal V_a$ precisely in Step 2 and
$C_{a+1}-C_a$ in Step 3. We then combine these two expressions in
Step 4 to show that the incremental surplus in
\eqref{eq:incremental_surplus} is nonnegative.

\medskip
\noindent
\textbf{Step 2: Compute the incremental expected innovation value from one additional approach.}
We next compare the candidate policy starting with $a+1$ unresolved
approaches to that starting with $a$. For any nonnegative
integer-valued random variable $N$ \proofedit{with $k+N\leq n$},
\[
\val(k+N)
=
\val(k)+
\sum_{j=1}^{\proofedit{n-k}}\Delta\val(k+j-1)\mathbf 1\{N\geq j\}.
\]
Thus,
\begin{equation}
\label{eq:value_tail_representation}
\mathbb E[\val(k+N_a)]
=
\val(k)+
\sum_{j=1}^{a}\Delta\val(k+j-1)\Pr(N_a\geq j).
\end{equation}

Let $Q_j^a\equiv\Pr(N_a\geq j)$. Using the binomial recursion,
\begin{align}
Q_j^{a+1}-Q_j^a
&=
q_s(d_{j-1})
\Pr\left(
\operatorname{Binomial}
\bigl(a,q_s(d_{j-1})\bigr)=j-1
\right)
\nonumber\\
&=
\binom{a}{j-1}
q_s(d_{j-1})^j
\bigl[1-q_s(d_{j-1})\bigr]^{a-j+1}.
\label{eq:incremental_tail}
\end{align}
For $j=0,\ldots,a$, define
\begin{equation}
\label{eq:Hj_def}
H_j(u)
\equiv
\binom{a}{j}
q_s(u)^{j+1}[1-q_s(u)]^{a-j}.
\end{equation}

Let
\[
\Delta\mathcal V_a
\equiv
\mathbb E[\val(k+N_{a+1})]
-
\mathbb E[\val(k+N_a)]
\]
denote the change in expected gross innovation value from starting
with $a+1$ rather than $a$ unresolved approaches.
Using the tail representation above and
\eqref{eq:incremental_tail}, we obtain
\begin{align}
\Delta\mathcal V_a
&=
\sum_{j=1}^{a+1}
\Delta\val(k+j-1)\bigl(Q_j^{a+1}-Q_j^a\bigr)
\nonumber\\
&=
\sum_{j=0}^{a}\Delta\val(k+j)H_j(d_j),
\label{eq:incremental_innovation_value}
\end{align}
where $Q_{a+1}^{a}\equiv0$.
Since $q_s(0)=0$, we have $H_j(0)=0$;
\proofedit{in particular, terms with $d_j=0$ vanish.}
Hence,
\begin{equation}
\label{eq:incremental_value_integral}
\Delta\mathcal V_a
=
\sum_{\substack{j=0,\ldots,a\\d_j>0}}
\int_0^{d_j}
\Delta\val(k+j)H_j'(u)\,du.
\end{equation}

\medskip
\noindent
\textbf{Step 3: Compute the incremental expected effort cost.}
We now compute the corresponding change in expected effort cost.
\proofedit{Set $C_0=0$. For $a\geq1$,}
consider a tagged approach among $a$ initially unresolved approaches.
At elapsed time $u$, the tagged approach remains active if two
conditions hold: it is itself unresolved, and if exactly $j$ of the
other $a-1$ approaches have \proofedit{potentially} succeeded, the
deadline associated with $k+j$ successes has not yet been reached.
\proofedit{Because the deadlines are nonincreasing in the success
count, these conditions also ensure that no earlier stopping
boundary has been crossed.}
Hence, its probability of being active at elapsed time $u$ is
\[
q_u(u)\sum_{j=0}^{a-1}
\mathbf 1\{u<d_j\}\binom{a-1}{j}
q_s(u)^j[1-q_s(u)]^{a-1-j}.
\]
By symmetry, the expected total effort cost starting with $a$
unresolved approaches is
\begin{equation}
\label{eq:cost_a}
C_a
=
\cost\int_0^{\proofedit{T-t}}
a q_u(u)
\sum_{j=0}^{a-1}
\mathbf 1\{u<d_j\}\binom{a-1}{j}
q_s(u)^j[1-q_s(u)]^{a-1-j}\,du.
\end{equation}

Subtracting the corresponding expressions for $a+1$ and $a$ gives
\begin{align}
C_{a+1}-C_a
={}&
\cost
\sum_{\substack{j=0,\ldots,a\\d_j>0}}
\int_0^{d_j}q_u(u)
\Bigg[
(a+1)\binom{a}{j}
q_s(u)^j[1-q_s(u)]^{a-j}
\nonumber\\
&\hspace{2.8cm}
-
a\binom{a-1}{j}
q_s(u)^j[1-q_s(u)]^{a-1-j}
\Bigg]du,
\label{eq:cost_difference_raw}
\end{align}
where the second term is interpreted as zero when $j=a$
\proofedit{(including the comparison $a=0$)}.

Using $a\binom{a-1}{j}=(a-j)\binom{a}{j}$, the expression in square
brackets equals
\[
\binom{a}{j}
q_s(u)^j[1-q_s(u)]^{a-j-1}
\bigl[(j+1)-(a+1)q_s(u)\bigr].
\]
On the other hand, differentiating \eqref{eq:Hj_def} gives
\begin{align*}
H_j'(u)
&=
\binom{a}{j}q_s(u)^j[1-q_s(u)]^{a-j-1}\\
&\qquad{}\times
\bigl[(j+1)-(a+1)q_s(u)\bigr]\proofedit{q_s'(u)}.
\end{align*}
Combining this expression with \eqref{eq:sprime_identity} yields
\[
q_u(u)
\Bigg[
(a+1)\binom{a}{j}q_s(u)^j[1-q_s(u)]^{a-j}
-
a\binom{a-1}{j}q_s(u)^j[1-q_s(u)]^{a-1-j}
\Bigg]
=
\frac{H_j'(u)}{\rate\pi_t(u)}.
\]
Therefore,
\begin{equation}
\label{eq:incremental_cost_integral}
C_{a+1}-C_a
=
\sum_{\substack{j=0,\ldots,a\\d_j>0}}
\int_0^{d_j}
\frac{\cost}{\rate\pi_t(u)}H_j'(u)\,du.
\end{equation}

\medskip
\noindent
\textbf{Step 4: Combining incremental value and incremental cost.}
Combining \eqref{eq:incremental_value_integral} and
\eqref{eq:incremental_cost_integral}, we obtain
\begin{equation}
\label{eq:value_difference_pre_ibp}
W(k,a+1,t)-W(k,a,t)
=
\sum_{\substack{j=0,\ldots,a\\d_j>0}}
\int_0^{d_j}
\left[
\Delta\val(k+j)-\frac{\cost}{\rate\pi_t(u)}
\right]H_j'(u)\,du.
\end{equation}

For each $j$ with $d_j>0$, define
\[
B_j(u)
\equiv
\Delta\val(k+j)-\frac{\cost}{\rate\pi_t(u)}.
\]
\proofedit{
Since $\pi_t(u)=p_{t+u}$ and
$t+d_j=\bar{\rho}(k+j)=\min\{\rho(k+j),T\}$,
the boundary value satisfies
\[
B_j(d_j)
=
\begin{cases}
0,
& \rho(k+j)\leq T,\\[4pt]
\displaystyle
\Delta\val(k+j)-\frac{\cost}{\rate p_T},
& T<\rho(k+j).
\end{cases}
\]
In the first case, $d_j>0$ implies $\rho(k+j)>t\geq0$, so the
endogenous stopping condition gives
$\rate\pi_t(d_j)\Delta\val(k+j)=\cost$.
In the second case, the horizon arrives before that stopping
condition binds, so
$\rate p_T\Delta\val(k+j)>\cost$.
Consequently, $B_j(d_j)\geq0$, with strict inequality when
$T<\rho(k+j)$.
}
Moreover,
\[
B_j'(u)
=
\frac{\cost}{\rate}
\frac{\proofedit{\pi_t'(u)}}{\pi_t(u)^2}
=
-\frac{\cost(\rate-\mu)}{\rate}
\frac{1-\pi_t(u)}{\pi_t(u)}
<0.
\]
Since $H_j(0)=0$, integration by parts gives
\begin{align}
\int_0^{d_j}B_j(u)H_j'(u)\,du
&=
\proofedit{B_j(d_j)H_j(d_j)}
-
\int_0^{d_j}B_j'(u)H_j(u)\,du
\nonumber\\
&=
\proofedit{B_j(d_j)H_j(d_j)}
+
\frac{\cost(\rate-\mu)}{\rate}
\int_0^{d_j}
\frac{1-\pi_t(u)}{\pi_t(u)}H_j(u)\,du
\geq0.
\end{align}
Summing across $j$ yields the exact representation
\begin{align}
\label{eq:value_monotonicity_representation}
W(k,a+1,t)-W(k,a,t)
={}&
\proofedit{
\sum_{\substack{j=0,\ldots,a\\d_j>0}}
B_j(d_j)H_j(d_j)
}
\nonumber\\
&+
\frac{\cost(\rate-\mu)}{\rate}
\sum_{\substack{j=0,\ldots,a\\d_j>0}}
\int_0^{d_j}
\frac{1-\pi_t(u)}{\pi_t(u)}H_j(u)\,du
\geq0.
\end{align}
\proofedit{The first sum is the nonnegative boundary contribution
from the finite horizon; its terms vanish whenever the
corresponding endogenous deadline is reached by $T$.}
This establishes \eqref{eq:value_monotone_active}.
Iterating the inequality gives $W(k,a,t)\geq W(k,b,t)$
for every $b<a$.
\end{proof}

\begin{proof}[Proof of \cref{prop:first best stochastic contract}]

\textbf{Step 1: The Firm's Maximization Problem.}
Conditional on surviving without a success or breakdown up to cumulative effort $\Strategy(t)$, the firm's posterior belief that its approach is feasible ($\theta_i = G$) is given by Bayes' rule:
\begin{equation}
    p_{\Strategy(t)} = \frac{\delta_i e^{-\rate_i \Strategy(t)}}{\delta_i e^{-\rate_i \Strategy(t)} + (1-\delta_i)e^{-\mu_i \Strategy(t)}}
\end{equation}
The unconditional probability of reaching cumulative effort $\Strategy(t)$ without a terminal event (success or breakdown) is:
\begin{equation}
    S(\Strategy(t)) = \delta_i e^{-\rate_i \Strategy(t)} + (1-\delta_i)e^{-\mu_i \Strategy(t)}
\end{equation}
The contract remains active at chronological time $t$ with probability $Q_i(t) = \Pr(\tau > t)$. By assumption, this distribution is independent of the firm's effort process conditional on the firm not having succeeded. The firm's expected payoff for choosing an arbitrary measurable effort path $\strategy_i$ is governed by the integral:
\begin{equation}
    U_i(\strategy_i) = \int_{0}^{T} \strategy_i(t) Q_i(t) S(\Strategy(t)) F(\Strategy(t)) dt
\end{equation}
where the instantaneous flow payoff is $F(\Strategy(t)) = p_{\Strategy(t)} \rate_i \Payment_i + (1-p_{\Strategy(t)})\mu_i \payment_i - \cost$.

\textbf{Step 2: Incentive Compatibility Constraints and Zero Rent.}
To establish dynamic incentive compatibility, we must show that at any arbitrary time $t \in [0, T]$, given any history of cumulative effort $\Strategy(t) \in [0, t]$, the firm has no strict incentive to shirk ($\strategy_i(t) < 1$). 

Under the constructed contract, each firm's instantaneous flow payoff is
\begin{align}
     & p_{\Strategy(t)} \rate_i \left( \frac{\cost}{\rate_i} \right) + (1-p_{\Strategy(t)})\mu_i \left( \frac{\cost}{\mu_i} \right) - \cost \nonumber \\
    &= p_{\Strategy(t)}\cost + (1-p_{\Strategy(t)})\cost - \cost = 0. \nonumber
\end{align}

This implies that the marginal payoff of effort is exactly zero everywhere. 
Therefore, the firm is locally indifferent between working and shirking at every instant $t$, regardless of its past effort history. By the standard tie-breaking assumption, continuous effort ($\strategy_i(t) = 1$ for all $t$) is dynamically incentive compatible.
Furthermore, because the instantaneous flow payoff is identically zero everywhere, each firm's total expected payoff is zero under full effort.
\begin{equation}
    U_i(\strategy_i) = \int_{0}^{T} \strategy_i(t) Q_i(t) S(\Strategy(t)) [0] dt = 0
\end{equation}
Thus, the principal's total expected payment to firm $i$ is
\begin{equation}
    W_i(T) = \int_0^T Q_i(t) S(t) \left[ p_t \rate_i \frac{\cost}{\rate_i} + (1-p_t) \mu_i \frac{\cost}{\mu_i} \right] dt = \int_0^T \cost Q_i(t) S(t) dt =\totalcost_i(T),
\end{equation}
which coincides with the maximum expected cost of a firm that works continuously ($\strategy_i(t) = 1$, such that $\Strategy(t) = t$) until $T$. 
Thus, the principal extracts full rent from each firm.
\end{proof}

\section{Appendix: Omitted Proof in \cref{sec:hybrid-good-constant}}\label{apx:pure-good-news-optimal-stochastic}
\subsection{Proof of \cref{prop:prize-lower-bound}}
Under pure good-news learning, after cumulative effort $e$ without success, firm $i$'s posterior probability that its approach remains feasible is
\[
p_e^i = \frac{\delta_i e^{-\rate_i e}}{1-\delta_i+\delta_i e^{-\rate_i e}},
\]
which satisfies $dp_e^i/de = -\rate_i p_e^i(1-p_e^i) < 0$. Let $S_i(e) = 1-\delta_i+\delta_i e^{-\rate_i e}$ denote the unconditional probability that approach $i$ remains unresolved after cumulative effort $e$.

Consider an arbitrary strategy $\strategy_i$ with cumulative effort $\Strategy_i(t) = \int_0^t\strategy_i(v)\,dv$. Recall that $Q_i(t) $ is the probability that the contract remains active. Define the weight on effort cost by
\[
W_i(t) = 
\begin{cases}
Q_i(t), & \disclosure_i=1,\\
1, & \disclosure_i=0.
\end{cases}
\]
Under disclosed termination, the firm ceases working once termination occurs, whereas under concealed termination it may continue incurring effort costs even after its contract is inactive. Thus, $Q_i(t)\leq W_i(t)$.

At cumulative effort $\Strategy_i(t)$, the instantaneous success rate conditional on remaining unresolved is $\rate_i p_{\Strategy_i(t)}^i$. Hence, excluding termination compensation, the expected contribution of effort at date $t$ is bounded above by $S_i(\Strategy_i(t)) \strategy_i(t) W_i(t) \left[ \rate_i p_{\Strategy_i(t)}^i\Payment_i-\cost \right]$.

Now consider any strategy $\strategy_i$ that exerts effort on a positive-measure set $\Lambda = \left\{ t: \strategy_i(t)>0,\; \Strategy_i(t)>\hat e_i \right\}$. Construct a truncated strategy $\tilde{\strategy}_i$ that coincides with $\strategy_i$ until cumulative effort reaches $\hat e_i$, and exerts zero effort thereafter. Because $\Strategy_i(t)$ is continuous and non-decreasing, its cumulative effort is $\tilde{\Strategy}_i(t) = \min\{\Strategy_i(t),\hat e_i\}$.

For every $t\in\Lambda$, the definition of $\hat e_i$ implies $\rate_i p_{\Strategy_i(t)}^i\Payment_i-\cost < 0$. Removing such effort strictly increases the firm's payoff from the success-prize and effort-cost components. Moreover, since $\tilde{\Strategy}_i(t)\leq\Strategy_i(t)$ and $S_i(e)$ is decreasing, we have $S_i(\tilde{\Strategy}_i(t)) \geq S_i(\Strategy_i(t))$. Because termination compensation is non-negative and is paid only when the firm remains unresolved at termination, reducing effort weakly increases the firm's expected termination compensation. 

It follows that $U_i(\tilde{\strategy}_i) > U_i(\strategy_i)$, meaning any strategy exerting effort beyond $\hat e_i$ is strictly dominated. If $\Payment_i < \cost / (\rate_i p_T^i)$, the strict monotonicity of $p_e^i$ implies $\hat e_i < T$, proving part~(i). If $\Payment_i \leq \cost / (\rate_i p_0^i)$, then $\rate_i p_e^i\Payment_i < \cost$ for all $e>0$, yielding $\hat e_i=0$. Any positive effort is strictly dominated by shirking, establishing part~(ii). If this holds for all firms, universal shirking is the unique equilibrium. Part~(iii) follows immediately.


\subsection{Proof of \cref{prop:zeroC}}
We proceed in three steps.

\medskip
\noindent
\textbf{Step 1: Full disclosure weakly dominates concealed termination.}
Suppose first that termination is concealed ($\disclosure_i=0$). 
By the continuation requirement in
Assumption~\ref{assump:pure-good-news-benchmark},
$Q_i(t)>0$ for every $t<T$.
Thus, firm $i$ can remain unresolved and active with positive
probability arbitrarily close to $T$.  Because the firm does not observe whether its contract has been terminated, it incurs the cost of any effort it exerts regardless of whether $\tau_i>t$. A success, however, generates the prize $\Payment_i$ only if the contract remains active. Thus, along the full-effort path, a necessary condition for effort at date $t$ is $Q_i(t)\rate_i p_t^i\Payment_i \geq \cost$.

Now disclose termination immediately ($\disclosure_i=1$), holding fixed the success prize, compensation schedule, and stopping rule. Whenever the firm is permitted to continue working, it knows its contract is active. The corresponding incentive condition becomes $\rate_i p_t^i\Payment_i \geq \cost$. Since $Q_i(t)\leq 1$, disclosure weakly relaxes the firm's effort incentive constraint. Furthermore, disclosure prevents the firm from expending effort after the contract has ended. This replacement is strictly beneficial whenever concealed termination causes post-termination effort with positive probability, meaning an optimal contract can always be chosen with $\disclosure_i^*=1$.

\medskip
\noindent
\textbf{Step 2: Zero termination compensation is optimal.}
Restricting attention to disclosed termination, consider an arbitrary non-negative compensation schedule $\compensation_i(t)$ that implements full effort. Let $\strategy_i^F$ denote the full-effort strategy: $\strategy_i^F(t)=1$ whenever firm $i$ remains unresolved and its contract is active. Prior to termination, its cumulative effort is $\Strategy_i^F(t) = \int_0^t\strategy_i^F(v)\,dv = t$. 

Consider any alternative strategy $\strategy_i$ with cumulative effort $\Strategy_i(t) = \int_0^t\strategy_i(v)\,dv$. Since $\strategy_i(t)\in[0,1]$, we have $\Strategy_i(t) \leq \Strategy_i^F(t) = t$ prior to termination. Let $K_i(\strategy_i)$ denote the expected termination compensation generated by strategy $\strategy_i$. By the admissibility of the stopping rule, the distribution of $\tau_i$ conditional on firm $i$ remaining unresolved does not depend on firm $i$'s own effort strategy. Therefore,
\[
K_i(\strategy_i) = \int_{[0,T]} S_i(\Strategy_i(t)) \compensation_i(t)\,dF_{\tau_i}(t).
\]
Because $S_i(e)$ is decreasing and $\Strategy_i(t) \leq \Strategy_i^F(t)$, we have $S_i(\Strategy_i(t)) \geq S_i(\Strategy_i^F(t))$. Given $\compensation_i(t)\geq 0$, it follows that $K_i(\strategy_i) \geq K_i(\strategy_i^F)$ for every deviation $\strategy_i$.

Now, construct a new contract identical to the original except that termination compensation is set to zero. Denoting payoffs under the original and modified contracts by $U_i$ and $U_i^0$, respectively, we see for every deviation $\strategy_i$:
\begin{align*}
U_i^0(\strategy_i^F)-U_i^0(\strategy_i) 
&= U_i(\strategy_i^F)-U_i(\strategy_i) + K_i(\strategy_i)-K_i(\strategy_i^F)\\
&\geq U_i(\strategy_i^F)-U_i(\strategy_i) \geq 0.
\end{align*}
Thus, removing termination compensation weakly strengthens every incentive constraint. Individual rationality is also preserved: unconditional shirking yields $U_i^0(\strategy_i\equiv0)=0$, and since full effort remains incentive compatible, $U_i^0(\strategy_i^F) \geq 0$. Removing compensation preserves implementation and participation while weakly reducing expected spending, so we can optimally set $\compensation_i^*(t)=0$ for $dF_{\tau_i}$-almost every $t$.

\medskip
\noindent
\textbf{Step 3: The optimal success prize.}
With full disclosure and zero termination compensation, consider firm $i$ at a history where it remains unresolved and its contract is active. Given cumulative effort $\Strategy_i(t)$, the marginal return from effort is $\rate_i p_{\Strategy_i(t)}^i\Payment_i-\cost$. Hence, full effort requires $\rate_i p_{\Strategy_i(t)}^i\Payment_i \geq \cost$.

Along the full-effort path, $\strategy_i^F(t)=1$ implies $ \Strategy_i^F(t)=t$. The incentive condition becomes $\rate_i p_t^i\Payment_i \geq \cost$ for every $t<T$ at which the firm can remain active. Because Assumption~\ref{assump:pure-good-news-benchmark} requires $Q_i(t)>0$ for every $t<T$, the firm can remain active arbitrarily close to $T$. Since $p_t^i$ is strictly decreasing, the incentive constraint at $T$ is the tightest:
$\rate_i p_T^i\Payment_i \geq \cost$.

By Proposition~\ref{prop:prize-lower-bound}, this condition is necessary for full effort under any non-negative compensation schedule. Under full disclosure and zero compensation, it is also sufficient. The principal therefore chooses the smallest success prize satisfying this constraint: $\Payment_i^* = \cost / (\rate_i p_T^i)$. At this prize, $\rate_i p_{\Strategy_i^F(t)}^i\Payment_i^* = \rate_i p_t^i\Payment_i^* > \cost$ for every $t<T$, meaning full effort is strictly optimal at every pre-termination history.

\subsection{Proof of \cref{prop:optimal_stopping_pure_good_news}}\label{apx:optimal_stopping_pure_good_news}

Fix the optimal interdependent contracts characterized in \cref{prop:zeroC}. Under symmetry, every successful firm receives the same success prize $\Payment^* = \cost / (\lambda p_T)$, and termination compensation is zero. Hence, conditional on the mechanism terminating after $k$ successful innovations, the principal's realized payoff is $G(k) \equiv \val(k)-k\Payment^*$. Thus, once the contract terms are fixed, the principal's stopping problem is equivalent to choosing when to halt the success process to maximize the expected value of $G(\eta_\tau)$.

We verify that the proposed cutoff rule, $\tau^* = \inf\{t\in[0,T]:\eta_t\ge m^*\}\wedge T$, is optimal. Recall that $m^*$ is defined as the smallest $k$ such that the marginal value of an innovation, $\Delta\val(k) \equiv \val(k+1)-\val(k)$, falls weakly below $\Payment^*$. For $k=0,\ldots,n-1$, the incremental net payoff from one additional success is $G(k+1)-G(k) = \Delta\val(k)-\Payment^*$. Because $\val$ is concave, $\Delta\val(k)$ is weakly decreasing in $k$. By the definition of $m^*$, we have $G(k+1)-G(k)>0$ for all $k<m^*$, whereas $G(k+1)-G(k)\le 0$ for all $k\ge m^*$. Therefore, $G(k)$ is strictly increasing up to $m^*$ and weakly decreasing thereafter.

To see that this cutoff rule is optimal, consider the potential success process that would obtain if all firms continued to work until $T$. Let $\eta_T$ denote the total number of successes on this potential path. Any admissible stopping rule can only truncate this process. Hence, for every realization, if a stopping rule $\tau$ terminates after $\eta_{\tau}$ successes, it must be that $\eta_{\tau}\le \eta_T$.

Under our proposed cutoff rule, the number of successes at termination is $\eta_{\tau^*} = \min\{\eta_T,m^*\}$. We can compare this outcome realization-by-realization with that generated by an arbitrary admissible stopping rule $\tau$. If $\eta_T<m^*$, then $G(k)$ is strictly increasing over all $k\le \eta_T$. Since $\eta_{\tau}\le\eta_T$, it follows that $G(\eta_{\tau}) \le G(\eta_T) = G(\eta_{\tau^*})$. If instead $\eta_T\ge m^*$, then $m^*$ maximizes $G(k)$ over all $k\le\eta_T$, implying $G(\eta_{\tau}) \le G(m^*) = G(\eta_{\tau^*})$.

Thus, for every realization of the underlying success process, $G(\eta_{\tau^*})\ge G(\eta_{\tau})$ for any admissible stopping rule $\tau$. Taking expectations yields
\[
    \mathbb{E}\big[ \val(\eta_{\tau^*}) - \Payment^*\eta_{\tau^*} \big] \ge \mathbb{E}\big[ \val(\eta_{\tau}) - \Payment^*\eta_{\tau} \big].
\]
Hence, $\tau^*$ is optimal.

Finally, notice why no calendar-time cutoff is needed prior to $T$. Conditional on an unresolved firm remaining active at time $t$, its posterior probability of feasibility is $p_t = \delta e^{-\lambda t} / (1-\delta+\delta e^{-\lambda t})$, which dictates the rate at which future successes arrive. However, this belief does not affect the principal's net payoff from an additional realized success, which is always $\Delta\val(k)-\Payment^*$. Moreover, by the independence of feasibility across approaches, the successes of other firms convey no information about the feasibility of a given unresolved approach. Calendar time therefore affects only how likely it is that another success arrives before $T$, not whether such a success is valuable conditional on arriving. This explains why the optimal stopping region depends only on the number of successes and is fully characterized by the static threshold $m^*$.

\section{Appendix: Omitted Proof in \cref{sec:optimal-contest}}\label{apx:optimal-contest}

\bigskip
\begin{proof}[Proof of \cref{prop:expenditure_equivalent_contest_ic}]
Part (i) follows directly from
the expected spending comparison
and \eqref{eq:expenditure_equivalent_pool}.
We establish part (ii) by characterizing a firm's incentives to exert full effort when its rivals do so.

\medskip
\noindent\textit{Step 1. Dynamic incentive compatibility.}
Fix a firm $i$ and suppose its rivals exert full effort.
Let $\bar\eta_t^{-i}$ denote the number of potential
breakthroughs among its $n-1$ rivals by $t$, defined from
the uncensored full-effort paths even if the contest
would have terminated earlier.
Define
\[
    Q(t)=\prob{\bar\eta_t^{-i}<m},
    \qquad
    H(t)=\expect{
        \frac{\indicate{\bar\eta_t^{-i}<m}}
        {\min\{m,1+\bar\eta_T^{-i}\}}
    }.
\]
Under cutoff disclosure, the function $Q(t)$ measures
contest survival: if firm $i$ remains unresolved, it
loses the opportunity to win when rivals reach the cutoff. The function $H(t)$ combines this
survival probability with the expected prize share from
an immediate breakthrough. Thus $H(t)/Q(t)$ is the
expected share conditional on survival, incorporating
prize dilution by other successful firms.

Let $W_G(t)$ denote firm $i$'s continuation payoff from
exerting full effort, conditional on its approach being
feasible, remaining unresolved, and the contest being
active at time $t$.
Conditional on the contest surviving until $t$, the
probability that it remains active at a future date
$s\ge t$ is $Q(s)/Q(t)$. Given feasibility, the
probability that firm $i$ has not achieved a breakthrough
between $t$ and $s$ is $e^{-\rate(s-t)}$.
If a breakthrough occurs at $s$, the expected prize is
$\Payment(m)H(s)/Q(s)$, yielding an expected continuation
benefit of $\frac{1}{Q(t)}
\int_t^T \rate e^{-\rate(s-t)}
\Payment(m)H(s)\,ds.$
Meanwhile, the firm incurs effort cost $\cost$ at
date $s$ only if it remains unresolved and the
contest is still active, giving an expected
continuation cost of $\frac{\cost}{Q(t)}
\int_t^T e^{-\rate(s-t)}Q(s)\,ds.$
Thus, 
\begin{equation}
\label{eq:contest_feasible_continuation}
W_G(t)=\frac{1}{Q(t)}
\int_t^T e^{-\rate(s-t)}
\left[
\rate\Payment(m)H(s)-\cost Q(s)
\right]ds.
\end{equation}

    To derive the dynamic incentive constraint, consider a deviation in which firm $i$ shirks over an infinitesimal interval $[t,t+dt]$ before returning to full effort. Conditional on feasibility, exerting effort generates a breakthrough with probability $\rate\,dt$, yielding the expected prize $\Payment(m)H(t)/Q(t)$ but eliminating the continuation payoff $W_G(t)$. Conditional on infeasibility, effort generates no breakthrough and only incurs the flow cost $\cost$. Thus, the expected incremental payoff from exerting effort rather than shirking is
$\left[
\rate\private_t
\left(\Payment(m)\frac{H(t)}{Q(t)}-W_G(t)\right)
-\cost
\right]dt+o(dt).$
Full effort therefore requires
\begin{equation}
\label{eq:contest_dynamic_ic}
    \rate\private_t
    \left[
    \Payment(m)\frac{H(t)}{Q(t)}
    -W_G(t)
    \right]
    \ge\cost,
    \qquad t<T,
\end{equation}
where $\private_t=
\frac{\delta e^{-\rate t}}
{1-\delta+\delta e^{-\rate t}}$ is the posterior of an
unresolved firm under full effort. The bracketed term
is the immediate success reward minus the continuation
value forgone upon success: shirking preserves this
future opportunity without incurring current effort cost.

This dynamic incentive constraint not only discourages shirking at time \(t\) when the firm has exerted full effort in the past, but also discourages further shirking if the firm has already shirked earlier. Indeed, previous shirking leaves an unresolved firm with a posterior weakly above $\private_t$, and the expression in brackets is strictly positive. To see the latter, note that $H(t)$ is non-increasing and
\[
\begin{aligned}
&\Payment(m)H(t)-Q(t)W_G(t)\\
={}&
\Payment(m)\left[
H(t)-\int_t^T
\rate e^{-\rate(s-t)}H(s)\,ds
\right]+\cost\int_t^T
e^{-\rate(s-t)}Q(s)\,ds
>0.
\end{aligned}
\]
Since cumulative effort cannot exceed calendar time,
an unresolved firm's posterior following any effort
history is weakly above $\private_t$. Moreover,
$Q(t)$, $H(t)$, and $W_G(t)$ are independent of
the firm's own past effort.
The instantaneous gain from effort is increasing
in the posterior, as established above. Hence,
verifying \eqref{eq:contest_dynamic_ic} at
$\private_t$ for every $t<T$ also verifies the
full-effort continuation policy at every posterior
reachable through a deviation. The finite-horizon
dynamic-programming verification argument then
establishes that these conditions are necessary
and sufficient for full-effort implementation.

\medskip
\noindent\textit{Step 2. Change of Variables.}
For the remainder of the proof, let $x=e^{-\rate t}$,
$z=e^{-\rate T}$, $a=(1-\delta)/\delta$, and
$q=\delta(1-z)$. With a slight abuse of notation, write
$Q(x)$ and $H(x)$ for the corresponding functions in
transformed time. We hold the deadline parameter $z$
fixed as $x$ varies over $[z,1]$. In particular,
$H(x)$ also depends on $z$ through the distribution
of rivals' eventual successes; we suppress this fixed
parameter for readability.

\begin{claim}
\label{lem:contest_phi_equivalence}
The dynamic incentive constraint
\eqref{eq:contest_dynamic_ic} is equivalent to
\begin{equation}
\label{eq:expenditure_equivalent_phi}
    \Phi(x)\ge0,
    \qquad x\in[z,1],
\end{equation}
where
\begin{equation}
\label{eq:contest_phi_definition}
\begin{aligned}
\Phi(x)
={}&
\frac{\rate\Payment(m)}{\cost}
\left[
xH(x)-\int_z^x H(u)\,du
\right]-
\left[
(x+a)Q(x)-\int_z^x Q(u)\,du
\right].
\end{aligned}
\end{equation}
\end{claim}

The function $\Phi(x)$ is a normalized measure of the
instantaneous incentive to work rather than shirk:
\[
\frac{(x+a)Q(x)}{\cost}
\left\{
\rate\private_t\left[
\Payment(m)\frac{H(t)}{Q(t)}-W_G(t)
\right]-\cost
\right\},
\qquad x=e^{-\rate t}.
\]
Since the normalization is positive, $\Phi(x)\ge0$
is precisely the dynamic incentive constraint.

\begin{claim}
\label{lem:contest_H_decomposition}
Under the equal-sharing rule of $\jointcontest_m$,
the expected prize share satisfies
\begin{equation}
\label{eq:contest_H_decomposition}
H(x)
=
\frac{Q(x)}{m}
+
\sum_{k=0}^{m-2}
\left(\frac{1}{k+1}-\frac{1}{m}\right)
\prob{\bar\eta_T^{-i}=k},
\end{equation}
where the sum is zero when $m=1$.
\end{claim}

The first term is the baseline share $1/m$ weighted by
contest survival; the second is the extra share when
fewer than $m$ firms ultimately succeed. Since this
second term is independent of $x$, equal sharing ties
the evolution of expected rewards to contest survival:
$H'(x)=Q'(x)/m$. Differentiating $\Phi$ therefore gives
\begin{equation}
\label{eq:expenditure_equivalent_phi_derivative}
    \Phi'(x)
    =
    Q'(x)
    \left[
    \left(
    \frac{\rate\Payment(m)}{m\cost}-1
    \right)x-a
    \right].
\end{equation}

\medskip
\noindent\textit{Step 3. Incentive Compatibility.}
If $m=1$, then $\Payment(1)=\Payment^*$ and $H(x)=Q(x)$. Since $ \frac{\rate\Payment^*}{\cost}
    =\frac{z+a}{z}$,
we obtain $ \Phi(x)
    =
    \frac{a}{z}
    \int_z^x[Q(x)-Q(u)]\,du
    \ge0.$
Thus, full effort is implemented. Moreover, $\Phi(z)=0$, so the terminal incentive constraint binds and $\Payment(1)$ is minimal.

If $m=n$, an unresolved firm cannot be terminated by its rivals, so $Q(x)=1$. Its expected prize share conditional on success is constant: $ H(x)
    =
    \expect{\frac{1}{1+\bar\eta_T^{-i}}}
    =
    \frac{\prob{\bar\eta_T>0}}{nq}.$
By \eqref{eq:expenditure_equivalent_pool}, $\Payment(n)H(x)=\Payment^*.$
Hence, the contest is payoff-equivalent to an independent contract offering $\Payment^*$ upon success. The expenditure-equivalent prize implements full effort and is minimal.

\medskip
Now we consider the interior cutoffs.
\begin{lemma}[Dynamic Incentive]\label{lmm:dynamic-incentive}
    Fix $1<m<n$. At the expenditure-equivalent prize, all dynamic incentive constraints hold if and only if the initial constraint $\Phi(1)\ge0$ holds.
\end{lemma}

\medskip
\noindent\textit{Step 4. Critical horizon and expenditure comparison.}
For an interior cutoff, the initial incentive condition
\eqref{eq:expenditure_equivalent_simple_ic} requires
$\rate\Payment(m)/(m\cost)\ge1/\delta$.
Substituting the expenditure-equivalent prize
\eqref{eq:expenditure_equivalent_pool} and using
$p_T=(\delta-q)/(1-q)$, where $q=\delta(1-z)$, gives
\[
    \expect{\min\{\bar\eta_T,m\}
        \mid\bar\eta_T>0}
    \ge
    m\frac{\delta-q}{\delta(1-q)}.
\]
Thus the threshold compares the expected number of
rewarded successes, conditional on at least one success,
with the winner cap adjusted for the decline from the
prior $\delta$ to the terminal posterior $p_T$.
The left-hand side is continuous and strictly increasing in $q\in(0,\delta)$, whereas the right-hand side is continuous and strictly decreasing. 
As $q\downarrow0$, the two sides converge to $1$ and $m>1$, respectively. As $q\uparrow\delta$, the right-hand side converges to zero.
Consequently, there exists a unique $q$, and hence a unique $T_m^*>0$, at which equality holds. Full effort is implemented by the expenditure-equivalent prize if and only if $T\ge T_m^*$.

Finally, the dynamic incentive constraints are increasing in the total prize.
When $T<T_m^*$, the expenditure-equivalent prize violates the initial incentive constraint. Any prize implementing full effort must therefore be strictly larger.
At $T=T_m^*$, the expenditure-equivalent prize implements full effort, and the initial constraint binds, making the prize minimal.
When $T>T_m^*$, the expenditure-equivalent prize satisfies every dynamic incentive constraint strictly. By continuity, a slightly smaller prize also implements full effort, yielding strictly lower expected expenditure than the interdependent contracts.
This proves part (iii).
\end{proof}

\begin{proof}[Proof of Claim~\ref{lem:contest_phi_equivalence}]
Substituting \eqref{eq:contest_feasible_continuation} into
\eqref{eq:contest_dynamic_ic} gives
\[
\rate\private_t
\left\{
\frac{\Payment(m)H(t)}{Q(t)}
-
\frac{1}{Q(t)}
\int_t^T e^{-\rate(s-t)}
\left[
\rate\Payment(m)H(s)-\cost Q(s)
\right]ds
\right\}
\ge\cost.
\]

Multiplying both sides by $Q(t)/(\cost\private_t)$
and rearranging yields
\begin{equation}
\label{eq:contest_ic_intermediate}
\begin{aligned}
&\frac{\rate\Payment(m)}{\cost}
\left[
H(t)-\rate\int_t^T
e^{-\rate(s-t)}H(s)\,ds
\right]
+\rate\int_t^T
e^{-\rate(s-t)}Q(s)\,ds
\ge\frac{Q(t)}{\private_t}.
\end{aligned}
\end{equation}

Now apply the change of variables.
Since $e^{-\rate(s-t)}=\frac{u}{x}$ and 
$ds=-\frac{du}{\rate u}$,
we have
\[
\rate\int_t^T
e^{-\rate(s-t)}H(s)\,ds
=
\frac{1}{x}\int_z^x H(u)\,du,
\]
and similarly,
\[
\rate\int_t^T
e^{-\rate(s-t)}Q(s)\,ds
=
\frac{1}{x}\int_z^x Q(u)\,du.
\]

Moreover, Bayes' rule implies
$\private_t
=
\frac{\delta e^{-\rate t}}
{1-\delta+\delta e^{-\rate t}}
=
\frac{x}{x+a}$.
Substituting these identities into
\eqref{eq:contest_ic_intermediate} gives
\[
\begin{aligned}
&\frac{\rate\Payment(m)}{\cost}
\left[
H(x)-\frac{1}{x}\int_z^x H(u)\,du
\right]
+\frac{1}{x}\int_z^x Q(u)\,du
\ge
\frac{x+a}{x}Q(x),\\
\Leftrightarrow & \frac{\rate\Payment(m)}{\cost}
\left[
xH(x)-\int_z^x H(u)\,du
\right]
-
\left[
(x+a)Q(x)-\int_z^x Q(u)\,du
\right]
\ge0,
\end{aligned}
\]
which is precisely $\Phi(x)\ge0$.
The equivalence holds for $x\in(z,1]$.
Continuity of $Q$ and $H$ extends the condition
to $x=z$ by taking limits.
\end{proof}

\begin{proof}[Proof of Claim~\ref{lem:contest_H_decomposition}]
Write $t=-\log(x)/\rate$ when referring to the
calendar date corresponding to transformed time $x$.
A breakthrough by firm $i$
at date $t$ earns a positive prize only if the
contest has not already terminated, i.e.,
$\bar\eta_t^{-i}<m$.

Conditional on a breakthrough at $t$, the firm's
eventual prize share is
$\frac{1}{\min\{m,1+\bar\eta_T^{-i}\}}.$
Therefore, the probability-weighted expected share is
$
H(x)
=
\expect{
\frac{
\indicate{\bar\eta_t^{-i}<m}
}{
\min\{m,1+\bar\eta_T^{-i}\}
}
}$.
Using the identity
\[
\frac{1}{\min\{m,1+j\}}
=
\frac{1}{m}
+
\sum_{k=0}^{m-2}
\left(\frac{1}{k+1}-\frac{1}{m}\right)
\indicate{j=k},
\]
we obtain
\[
\begin{aligned}
H(x)
={}&\frac{Q(x)}{m}+\sum_{k=0}^{m-2}
\left(\frac{1}{k+1}-\frac{1}{m}\right)
\prob{
\bar\eta_t^{-i}<m,\,
\bar\eta_T^{-i}=k
}.
\end{aligned}
\]

Since $\bar\eta_T^{-i}=k<m-1$ implies
$\bar\eta_t^{-i}<m$ for every $t\le T$,
$\prob{
\bar\eta_t^{-i}<m,\,
\bar\eta_T^{-i}=k
}
=
\prob{\bar\eta_T^{-i}=k}$.
Substitution establishes the result.
\end{proof}

\begin{proof}[Proof of \cref{lmm:dynamic-incentive}]
    First, consider the terminal constraint.
\begin{claim}[Terminal Expected Prize Share]
\label{lem:contest_terminal_share}
Let $q=\delta(1-z)$, where $z=e^{-\rate T}$.
Under the equal-sharing rule of $\jointcontest_m$,
the expected prize share conditional on contest survival
at the terminal date satisfies
\begin{equation}
\label{eq:contest_terminal_share}
    \frac{H(z)}{Q(z)}
    =
    \frac{1}{
    \expect{\bar\eta_T
    \mid 1\le\bar\eta_T\le m}
    }.
\end{equation}
At the expenditure-equivalent prize
\eqref{eq:expenditure_equivalent_pool},
the terminal incentive constraint satisfies
\begin{equation}
\label{eq:contest_terminal_ic}
\begin{aligned}
\Phi(z)
={}&
(z+a)Q(z)
\left[
\frac{
\expect{\min\{\bar\eta_T,m\}
    \mid\bar\eta_T>0}
}{
\expect{\bar\eta_T
    \mid1\le\bar\eta_T\le m}
}
-1
\right]
\ge0,
\end{aligned}
\end{equation}
with strict inequality if $1<m<n$ and equality
if $m=1$ or $m=n$.
\end{claim}

\begin{claim}
\label{lem:contest_integral_identities}
Under symmetric technologies and full effort, the
contest survival probability $Q$ and expected prize
share $H$ satisfy
\begin{align}
\int_z^1 Q(u)\,du
&=
\frac{\expect{\min\{\bar\eta_T,m\}}}{n\delta},
\label{eq:contest_Q_integral}\\
\int_z^1 H(u)\,du
&=
\frac{\prob{\bar\eta_T>0}}{n\delta},
\label{eq:contest_H_integral}\\
H(1)
&=\frac{\prob{\bar\eta_T>0}}{nq}
+\frac{\expect{(\bar\eta_T-m)^+}}{mnq}.
\label{eq:H1}
\end{align}
\end{claim}

Next, consider the initial constraint. Define
\[
d\equiv\frac{\rate\Payment(m)}{m\cost}
-\frac{1}{\delta}.
\]
Using Claim~\ref{lem:contest_integral_identities},
\eqref{eq:expenditure_equivalent_pool}, and
$\expect{\bar\eta_T}=nq$, we obtain
\begin{equation}
\label{eq:expenditure_equivalent_initial_ic}
\Phi(1)=
\frac{\expect{(\bar\eta_T-m)^+}}
{\expect{\bar\eta_T}}\,d.
\end{equation}
Note that $(\bar\eta_T-m)^+:=\bar\eta_T-\min\{\bar\eta_T,m\}$.
The first factor is the expected number of potential
breakthroughs beyond the winner cap, relative to the
expected total number of potential breakthroughs.
It is strictly positive when $m<n$ (and $T>0$).
The second factor measures the baseline prize incentive
relative to the initial cost of effort, since
$
\delta d=
\frac{\rate\delta[\Payment(m)/m]}{\cost}-1.$
Thus $d$ is the normalized surplus of the initial
expected breakthrough reward at the baseline share
$1/m$ over effort cost; it is not itself the firm's
full continuation payoff. Since the first factor is
positive, the initial incentive constraint is equivalent to
\begin{equation}
\label{eq:expenditure_equivalent_simple_ic}
    \frac{\rate\Payment(m)}{m\cost}
    \ge\frac{1}{\delta}.
\end{equation}
Substituting the expenditure-equivalent prize from
\eqref{eq:expenditure_equivalent_pool}
shows that this is exactly
\eqref{eq:expenditure_equivalent_ic}.

It remains to prove that the initial constraint is also
sufficient. Since $1/\delta=1+a$,
\eqref{eq:expenditure_equivalent_phi_derivative}
becomes
\[
\Phi'(x)=Q'(x)[dx-a(1-x)].
\]
To obtain a useful decomposition, integration by parts
and \eqref{eq:contest_Q_integral} give
\[
\begin{aligned}
\int_z^1(u-z)Q'(u)\,du
&=(1-z)-\int_z^1Q(u)\,du=(1-z)\frac{\expect{(\bar\eta_T-m)^+}}
{\expect{\bar\eta_T}}.
\end{aligned}
\]
Hence \eqref{eq:expenditure_equivalent_initial_ic}
can also be written as
\[
\Phi(1)=\frac{d}{1-z}
\int_z^1(u-z)Q'(u)\,du.
\]
Combining this boundary value with the expression for
$\Phi'(x)$ yields
\begin{equation}
\label{eq:expenditure_equivalent_positive_decomposition}
\begin{aligned}
\Phi(x)
={}&
\left(a-\frac{zd}{1-z}\right)
\int_x^1(1-u)Q'(u)\,du+\frac{d}{1-z}
\int_z^x(u-z)Q'(u)\,du.
\end{aligned}
\end{equation}
Notice that the right-hand side has derivative
$Q'(x)[dx-a(1-x)]$ and equals $\Phi(1)$ at
$x=1$.

Because $Q'(x)>0$ on $(z,1)$ and
$\Phi(z)>0$ by Claim~\ref{lem:contest_terminal_share},
evaluating \eqref{eq:expenditure_equivalent_positive_decomposition}
at $x=z$ implies
$a-zd/(1-z)>0$. Thus, if the initial constraint
holds ($d\ge0$), both terms in the decomposition
are nonnegative at every $x\in[z,1]$.
Equivalently, the decomposition expresses each
intermediate incentive slack as a nonnegative
linear combination of the terminal and initial slacks:
\[
\Phi(x)=
\frac{\int_x^1(1-u)Q'(u)\,du}
     {\int_z^1(1-u)Q'(u)\,du}\,\Phi(z)
+\frac{\int_z^x(u-z)Q'(u)\,du}
      {\int_z^1(u-z)Q'(u)\,du}\,\Phi(1).
\]
This explains why the initial constraint is decisive
even though the incentive slack need not be monotone
over time: the terminal constraint is already strictly
satisfied, and the initial constraint ensures all
remaining constraints. Conversely, if $d<0$, then
$\Phi(1)<0$ by \eqref{eq:expenditure_equivalent_initial_ic},
so full effort cannot be implemented.
This establishes part (ii).

\end{proof}

\begin{proof}[Proof of Claim~\ref{lem:contest_terminal_share}]
At the terminal date, if $k<m$ rivals have succeeded,
a breakthrough by firm $i$ earns a share $1/(k+1)$.
Thus, $H(z)
=
\sum_{k=0}^{m-1}
\frac{1}{k+1}
\prob{\bar\eta_T^{-i}=k}$ and 
$
Q(z)
=
\sum_{k=0}^{m-1}
\prob{\bar\eta_T^{-i}=k}$.

By symmetry,
$\bar\eta_T\sim\mathrm{Bin}(n,q)$ and
$\bar\eta_T^{-i}\sim\mathrm{Bin}(n-1,q)$.
The binomial identity
$
k\prob{\bar\eta_T=k}
=
nq\prob{\bar\eta_T^{-i}=k-1}$
therefore implies
\[
\begin{aligned}
H(z)
&=\frac{1}{nq}
\sum_{k=1}^{m}\prob{\bar\eta_T=k}
=\frac{\prob{1\le\bar\eta_T\le m}}{nq},\\
Q(z)
&=\frac{1}{nq}
\sum_{k=1}^{m}
k\prob{\bar\eta_T=k}=\frac{
\expect{\bar\eta_T
\indicate{1\le\bar\eta_T\le m}}
}{nq}.
\end{aligned}
\]
Taking their ratio yields
\[
\frac{H(z)}{Q(z)}
=
\frac{\prob{1\le\bar\eta_T\le m}}
{\expect{\bar\eta_T
\indicate{1\le\bar\eta_T\le m}}}
=
\frac{1}{
\expect{\bar\eta_T\mid1\le\bar\eta_T\le m}
}.
\]

Next, evaluating $\Phi(x)$ at $x=z$ gives
\[
\begin{aligned}
\Phi(z)
&=\frac{\rate\Payment(m)}{\cost}zH(z)
-(z+a)Q(z)\\
&=(z+a)Q(z)
\left[
\frac{\rate\Payment(m)}{\cost}
\frac{z}{z+a}
\frac{H(z)}{Q(z)}-1
\right].
\end{aligned}
\]
Since $p_T=z/(z+a)$, the
expenditure-equivalent prize
\eqref{eq:expenditure_equivalent_pool}
satisfies
\[
\frac{\rate\Payment(m)}{\cost}p_T
=
\expect{\min\{\bar\eta_T,m\}
\mid\bar\eta_T>0}.
\]
Substituting this identity and
\eqref{eq:contest_terminal_share}
establishes \eqref{eq:contest_terminal_ic}.

To establish the inequality, observe that
conditional on $\bar\eta_T>0$,
$\min\{\bar\eta_T,m\}$ equals $\bar\eta_T$
when $\bar\eta_T\le m$ and equals $m$
otherwise. Since
$
\expect{\bar\eta_T
\mid1\le\bar\eta_T\le m}\le m$,
it follows that
$
\expect{\min\{\bar\eta_T,m\}
\mid\bar\eta_T>0}
\ge
\expect{\bar\eta_T
\mid1\le\bar\eta_T\le m}$.
The inequality is strict when $1<m<n$,
because $\prob{\bar\eta_T>m}>0$ and
$\expect{\bar\eta_T
\mid1\le\bar\eta_T\le m}<m$.
When $m=1$ or $m=n$, the two expectations
coincide.
\end{proof}

\begin{proof}[Proof of Claim~\ref{lem:contest_integral_identities}]
Let $\bar\eta_t$ denote the total number of potential
breakthroughs by date $t$ under full effort, without
censoring at the contest's stopping time.

We first establish \eqref{eq:contest_Q_integral}.
At any date $t$, firm $i$ generates a breakthrough
with unconditional density
$\delta\rate e^{-\rate t}$.
Such a breakthrough increases
$\min\{\bar\eta_t,m\}$ by one if fewer than
$m$ rivals have already succeeded, an event
with probability $Q(e^{-\rate t})$.
By symmetry and independence across firms,
\[
\frac{d}{dt}
\expect{\min\{\bar\eta_t,m\}}
=
n\delta\rate e^{-\rate t}
Q(e^{-\rate t}).
\]
Integrating over $[0,T]$ and using
$\bar\eta_0=0$ gives
\[
\expect{\min\{\bar\eta_T,m\}}
=
n\delta\rate
\int_0^T e^{-\rate t}
Q(e^{-\rate t})\,dt.
\]
The change of variables $u=e^{-\rate t}$
yields \eqref{eq:contest_Q_integral}.

Next, consider \eqref{eq:contest_H_integral}.
Conditional on a breakthrough by firm $i$ at
date $t$, $H(e^{-\rate t})$ is its
probability-weighted expected prize share.
Thus, the firm's unconditional expected
prize share is
\[
\delta\rate\int_0^T e^{-\rate t}
H(e^{-\rate t})\,dt
=
\delta\int_z^1 H(u)\,du.
\]

Under the equal-sharing rule, the total prize
shares sum to one whenever at least one firm
succeeds, and to zero otherwise. Hence the
expected sum of prize shares is
$\prob{\bar\eta_T>0}$.
Since firms are symmetric, each firm's
expected share equals
$\prob{\bar\eta_T>0}/n$. Therefore,
\[
\delta\int_z^1 H(u)\,du
=
\frac{\prob{\bar\eta_T>0}}{n},
\]
which establishes \eqref{eq:contest_H_integral}.

Lastly,
\[
\begin{aligned}
H(1)
&=\expect{\frac{1}{\min\{m,1+\bar\eta_T^{-i}\}}}\\
&=\frac{1}{nq}
\expect{\frac{\bar\eta_T}
{\min\{m,\bar\eta_T\}}
\indicate{\bar\eta_T>0}}\\
&=\frac{\prob{\bar\eta_T>0}}{nq}
+\frac{\expect{(\bar\eta_T-m)^+}}{mnq}.
\end{aligned}
\]
The last equality uses
$k/\min\{m,k\}=1+(k-m)^+/m$ for $k\ge1$,
and $\expect{\bar\eta_T}=nq$.
\end{proof}

\begin{proof}[Proof of \cref{cor:fixed_prize_expenditure}]
For $1<m<n$, the expenditure-equivalent prize
implements full effort if and only if $T\ge T_m^*$,
by \cref{prop:expenditure_equivalent_contest_ic}.
If $T<T_m^*$, the expenditure-equivalent prize
violates the initial incentive constraint. Since
incentives are strictly increasing in the prize,
any fixed prize implementing full effort must
be strictly larger.
If $T=T_m^*$, the expenditure-equivalent prize
implements full effort and the initial incentive
constraint binds. Hence, it is the minimum
implementing prize.
If $T>T_m^*$, all dynamic incentive constraints
hold strictly at the expenditure-equivalent prize.
By continuity, a slightly smaller prize also
implements full effort, yielding strictly lower
expected expenditure.
Finally, for $m=1$ and $m=n$,
\cref{prop:expenditure_equivalent_contest_ic}
establishes that the expenditure-equivalent prize
is the minimum prize implementing full effort.
\end{proof}
\section{Appendix: Omitted Proof in \cref{sec:rank-contingent contests}}\label{apx:rank-contingent contests}

\bigskip

\begin{proof}[Proof of \cref{lem:rank-specific-incentive-targeting}]
We first establish the lower bound on success prizes under interdependent contracts. Fix an attainable success rank $r$ and a firm $i$. Since firm $i$ can occupy rank $r$ with positive probability, there are reachable histories at which $r-1$ other firms have succeeded, firm $i$ remains unresolved, and $t<\gamma_r(\taus)$. At every such history, firm $i$ must be willing to exert effort. By \cref{prop:zeroC}, we can restrict attention to contracts with zero termination compensation. Working costs $\cost$ per unit of time and produces a success at instantaneous rate $\rate p_t$. Therefore, the local incentive constraint is $\rate p_t\Payment_i\geq\cost$, or equivalently, $\Payment_i \geq \frac{\cost}{\rate p_t}$. Taking $t\uparrow\gamma_r(\taus)$ yields $\Payment_i \geq \frac{\cost}{\rate p_{\gamma_r(\taus)}} = \Payment_r(\taus)$. Since firm $i$ can occupy every attainable rank, $\Payment_i \geq \max_r\Payment_r(\taus) $, which establishes \eqref{eq:contract-common-lower-bound}.

We next construct the rank-contingent contest. Assign the firm producing the $r$-th success the rank-specific payment $\Payment_r(\taus) = \frac{\cost}{\rate p_{\gamma_r(\taus)}}$. Choose a symmetric public disclosure rule $\drule\in\dset$ that reveals the realized success count $\eta_t$ and calendar time relevant for firms' continuation decisions.




\begin{lemma}
\label{lem:rank-payment-global-IC}
Fix an admissible symmetric stopping rule $\taus\in\stopset$ whose
continuation region $\continuation$ depends only on the aggregate
success count $\eta_t$ and calendar time $t$, and under which
termination is absorbing.
For every attainable success rank $r$, define
$    \gamma_r(\taus) := \sup\{t\in[0,T]:(r-1,t)\in\continuation\}$.
Suppose $\gamma_r(\taus)$ is decreasing in $r$.
Then under the proposed rank-contingent contest, full effort until success or termination is globally incentive
compatible for every firm.
\end{lemma}

\cref{lem:rank-payment-global-IC} establishes that the contest induces exactly the stopping profile $\taus$ and implements the same full-effort profile before stopping. 
\end{proof}

\begin{proof}[Proof of \cref{lem:rank-payment-global-IC}]
Fix firm $i$ and suppose all other firms follow the prescribed
full-effort strategy. We show that full effort maximizes firm $i$'s
expected payoff against any admissible unilateral effort deviation.
Fix a realization $h_{-i}$ of the other firms' success histories.
Conditional on firm $i$ remaining unresolved, the stopping rule
depends only on the aggregate success count and calendar time.
Since the aggregate success count then coincides with the number
of successes among the other firms, the stopping time induced by
$h_{-i}$, denoted by $\tau_i(h_{-i})$, is independent of firm $i$'s
effort strategy.

Let $ R_i(t) :=\Payment_{\eta_{-i,t}+1}$
denote the success payment available to firm $i$ at time $t$.
Because $\eta_{-i,t}$ is nondecreasing and the rank-specific
payments are weakly decreasing, $R_i(t)$ is weakly decreasing
along every realization $h_{-i}$.

Recall that $S(x)$ is the probability that firm $i$ remains unresolved
after cumulative effort $x$, and $w(x)$ is the density of success
with respect to cumulative effort. And we have $   w(x)=\rate p_x S(x).$
Conditional on $h_{-i}$, firm $i$'s expected payoff is
\[
    U_i(\strategy_i\mid h_{-i})
    =
    \int_0^{\tau_i(h_{-i})}
    \strategy_i(t)
    \left[
        w(\Strategy_i(t))R_i(t)
        -\cost S(\Strategy_i(t))
    \right]dt.
\]

Let $\Strategy=\Strategy_i(\tau_i(h_{-i}))$ denote the total cumulative
effort exerted before termination, and let $t(x)$ denote the
calendar time at which the $x$-th unit of effort is exerted.
Since $\strategy_i(t)\leq1$, we have $t(x)\geq x$.
Changing variables from calendar time to cumulative effort gives
\begin{equation}\label{eq:rank-IC-frontloading}
    \begin{aligned}
        U_i(\strategy_i\mid h_{-i})& =
    \int_0^A
    \left[
        w(x)R_i(t(x))
        -\cost S(x)
    \right]dx \\
    & \leq
    \int_0^A
    \left[
        w(x)R_i(x)-\cost S(x)
    \right]dx,
    \end{aligned}
\end{equation}
where the second line comes from $R_i(t)$ weakly decreasing and $t(x)\geq x$.

We next show that the integrand on the right-hand side is
nonnegative whenever the contract remains active.

\begin{lemma}\label{claim:non-negative integrand}
    $ w(x)R_i(x)-\cost S(x)\geq 0$ for $0\leq x\leq \tau_i(h_{-i})$.
\end{lemma}

\begin{proof}[Proof of \cref{claim:non-negative integrand}]
    Fix $x<\tau_i(h_{-i})$, and let
$r=\eta_{-i,x^-}+1$ denote the attainable success rank.
Since the contract remains active at time $x$,
$    (r-1,x)\in\continuation$,
which implies $    x\leq\gamma_r(\taus)$.

Because $p_x$ is weakly decreasing in cumulative effort,
$    p_x\geq p_{\gamma_r(\taus)}$.
By the definition of the rank-specific payment, $ R_i(x)=\Payment_r = \frac{\cost}{\rate p_{\gamma_r(\taus)}}.$
Therefore,
\begin{equation*}
    w(x)R_i(x)-\cost S(x)=
    S(x)\left[\rate p_xR_i(x)-\cost\right]=
    \cost S(x)
    \left[
        \frac{p_x}{p_{\gamma_r(\taus)}}-1
    \right]\geq0.
\end{equation*}
\end{proof}

Since $A\leq\tau_i(h_{-i})$, it follows that
\[
    U_i(\strategy_i\mid h_{-i})
    \leq
    \int_0^{\tau_i(h_{-i})}
    [w(x)R_i(x)-\cost S(x)]dx:= U_i(\strategy_i^{FE}\mid h_{-i}).
\]
The right-hand side is precisely firm $i$'s conditional
expected payoff from exerting full effort until success or
termination. Since $    U_i(\strategy_i\mid h_{-i})
    \leq
    U_i(\strategy_i^{FE}\mid h_{-i})$ holds
for every admissible deviation $\strategy_i$, taking expectations over the other firms' success histories,
\[
    U_i(\strategy_i,\strategy_{-i}^{FE})
    \leq
    U_i(\strategy_i^{FE},\strategy_{-i}^{FE}).
\]
\end{proof}

\section{Appendix: Omitted Proofs for
\cref{sec:good-time-varying}}
\bigskip

\begin{proof}[Proof of \cref{lem:independent_termination_payments}]
\textbf{Step 1. Expected payoff in cumulative-effort time.}
We represent the firm's effort decisions in cumulative-effort
time. Suppose the firm plans to exert a total of
$0\le\Strategy\le T$ units of effort in the absence
of a breakthrough or termination. Let $t(x)$ denote the
calendar date at which its $x$-th unit of effort is performed.
Since effort cannot exceed one, every feasible effort plan
satisfies
\[
t(x)\ge x,
\qquad
t(x_2)-t(x_1)\ge x_2-x_1
\quad\text{for }x_2>x_1.
\]
Given the exogenous deadline $T$,
 $t(x)\le T$ for $x\in[0,\Strategy]$.
In particular, full effort corresponds to
$\Strategy=T$ and $t(x)=x$.

Under pure good-news learning, the probability that an approach
remains unresolved after $x$ units of effort is
$S(x)=1-\delta+\delta e^{-\rate x}$, and its posterior probability
of feasibility conditional on remaining unresolved is
$p_x=\delta e^{-\rate x}/S(x)$.
Denote the unconditional density of a breakthrough at cumulative
effort $x$ by
$w(x)\equiv -S'(x)=\delta\rate e^{-\rate x}
=\rate p_x S(x)$.
Let $Q(t)\equiv\prob{\sigma>t}$ denote
the probability that external termination has not
occurred by calendar time $t$.
For a given breakthrough-payment schedule $\Payment(t)$,
let $R(t)\equiv Q(t)\Payment(t)$ be the breakthrough payment
weighted by the probability that the experiment has not been
terminated at date $t$.
Notice that we allow $Q(t)=0$, for any $t\leq T$.
Because $\sigma$ is independent of the firm's experimentation
and the firm receives no signal about $\sigma$
other than whether termination has occurred, the expected
payoff from any feasible effort plan is
\begin{equation}
\label{eq:independent_horizon_utility}
U(\Strategy,t(\cdot);\Payment)
=
\int_0^\Strategy
\left[
    w(x)R(t(x))
    -\cost S(x)Q(t(x))
\right]dx.
\end{equation}
This is because at cumulative effort $x$, the unconditional
breakthrough density is $w(x)$, while the probability that
the firm remains unresolved is $S(x)$. Both payments and
effort costs are incurred only if
external termination has not occurred at calendar
time $t(x)$.
In particular, under full effort the probability
that the firm is both unresolved and active at $t$ is
$S(t)Q(t)$.

Under full effort, $t(x)=x$, the principal's expected
expenditure is
\begin{equation}
\label{eq:independent_horizon_expenditure}
\totalcost(\Payment)
=
\int_0^T w(t)R(t)\,dt.
\end{equation}

\textbf{Step 2. A necessary incentive condition.}
Consider a deviation in which the firm exerts full effort
until date $t<T$, pauses for
$0<\varepsilon<T-t$ units of time, and then resumes
full effort until $T$.
Relative to full effort, cumulative effort is reduced by
$\varepsilon$, and each unit of effort following the pause
is delayed by $\varepsilon$.

Using \eqref{eq:independent_horizon_utility}, the deviation's
payoff gain is
\begin{align*}
D(t,\varepsilon)
={}&
\int_{t+\varepsilon}^T
\left[
    w(s-\varepsilon)R(s)
    -\cost S(s-\varepsilon)Q(s)
\right]ds\\
&-
\int_t^T
\left[
    w(s)R(s)
    -\cost S(s)Q(s)
\right]ds.
\end{align*}
If full effort is incentive-compatible, then
$D(t,\varepsilon)\le0$.

Since $w'(s)=-\rate w(s)$ and $S'(s)=-w(s)$, for almost
every $t<T$,
\begin{equation}
\label{eq:independent_horizon_pause_slack}
\lim_{\varepsilon\downarrow0}
\frac{D(t,\varepsilon)}{\varepsilon}
=-M(t),
\end{equation}
where
\begin{equation}
\label{eq:independent_horizon_M}
\begin{aligned}
M(t)\equiv{}&
w(t)R(t)-\cost S(t)Q(t)\\
&-\int_t^T
\left[
    \rate w(s)R(s)-\cost w(s)Q(s)
\right]ds.
\end{aligned}
\end{equation}
Finite expected expenditure makes $wR$ integrable.
Together with the boundedness of $S$, $w$, and $Q$ on
$[0,T]$, this justifies the limit by Lebesgue differentiation
and dominated convergence.
Consequently, every payment schedule implementing full
effort must satisfy
\begin{equation}
\label{eq:independent_horizon_IC}
M(t)\ge0
\qquad\text{for almost every }t\in[0,T).
\end{equation}

\textbf{Step 3. A lower bound on expected expenditure.}
Use the deterministic-horizon cost function from the main text:
\[
\mathcal C(t)\equiv
\frac{\cost S(t)}{\rate}
\left(e^{\rate t}-1\right).
\]
We will show that
\begin{equation}
\label{eq:independent_horizon_lower_bound}
\totalcost(\Payment)
\ge \expect{\mathcal C(\sigma)}.
\end{equation}

Multiply \eqref{eq:independent_horizon_M} by
$e^{\rate t}$ and integrate over $[0,T]$.
We obtain
\begin{align*}
\int_0^T e^{\rate t}M(t)\,dt
={}&
\int_0^T w(t)R(t)\,dt
-\cost\int_0^T Q(t)
\left[
    e^{\rate t}S(t)
    -\frac{e^{\rate t}-1}{\rate}w(t)
\right]dt\\
={}&
\totalcost(\Payment)
-\int_0^T Q(t)\proofedit{\mathcal C'(t)}\,dt\\
={}&
\totalcost(\Payment)
-\expect{\proofedit{\mathcal C(\sigma)}}.
\end{align*}
The first equality follows from Fubini's theorem and
$\int_0^s e^{\rate t}\,dt=(e^{\rate s}-1)/\rate$.
The second uses
\[
\mathcal C'(t)
=
\cost\left[
e^{\rate t}S(t)
-\frac{e^{\rate t}-1}{\rate}w(t)
\right].
\]
For the third, $Q(t)=\prob{\sigma>t}$ and $\mathcal C(0)=0$
imply
\[
\int_0^T Q(t)\mathcal C'(t)\,dt
=
\expect{\int_0^\sigma\mathcal C'(t)\,dt}
=
\expect{\mathcal C(\sigma)}.
\]
Therefore, the principal's expected expenditure can be
expressed as
\begin{equation}
\label{eq:independent_horizon_cost_slack}
\totalcost(\Payment)
=
\expect{\proofedit{\mathcal C(\sigma)}}
+\int_0^T e^{\rate t}M(t)\,dt.
\end{equation}
By \eqref{eq:independent_horizon_IC}, the second term is
nonnegative for every schedule implementing full effort.
Hence, we establish \eqref{eq:independent_horizon_lower_bound}.

\textbf{Step 4. Attaining the lower bound.}
We now construct a payment schedule attaining
\eqref{eq:independent_horizon_lower_bound} and implementing
full effort. Consider the payment schedule
\[
\Payment^\sigma(t)
=
\expect{\left.
\frac{\cost}{\rate p_\sigma}
\,\right|\,\sigma>t}
\]
at dates $t<T$ for which $Q(t)>0$.
Set $\Payment^\sigma(t)=0$ at all other dates; these values
do not affect expected expenditure or incentives.
Let $F(s)\equiv\cost/(\rate p_s)$.
This is an auxiliary notation for the
deterministic-horizon incentive requirement.
By construction,
\begin{equation}
\label{eq:independent_horizon_weighted_payment}
Q(t)\Payment^\sigma(t)
=
\expect{
    F(\sigma)\indicate{\sigma>t}
}.
\end{equation}

The expected expenditure under full effort is
\begin{align*}
\totalcost(\Payment^\sigma)
&=
\int_0^T
w(t)\expect{
F(\sigma)\indicate{\sigma>t}
}\,dt\\
&=
\expect{
F(\sigma)\int_0^\sigma w(t)\,dt
}\\
&=
\expect{
\frac{\cost}{\rate p_\sigma}
\delta(1-e^{-\rate\sigma})
}\\
&=
\expect{\proofedit{\mathcal C(\sigma)}},
\end{align*}
where the last equality follows from the definitions
of $p_\sigma$ and $\proofedit{\mathcal C(\sigma)}$.
Thus the proposed schedule attains the expenditure
lower bound.
Its expenditure is finite because $\mathcal C$
is bounded on $[0,T]$.

It remains to verify global incentive compatibility.
For any cumulative effort $x$ and calendar time $t\ge x$,
define
\[
\begin{aligned}
g(x,t)
&\equiv
w(x)Q(t)\Payment^\sigma(t)
-\cost S(x)Q(t)\\
&=
\expect{
\indicate{\sigma>t}
\left[
w(x)F(\sigma)-\cost S(x)
\right]
}.
\end{aligned}
\]

Since $p_s$ is decreasing under pure good-news learning,
$F(s)$ is increasing. Moreover,
$w(s)F(s)=\cost S(s)$.
On the event $\{\sigma>t\ge x\}$, the monotonicity of
$F$ and this identity together imply that
\[
w(x)F(\sigma)
\ge w(x)F(x)
=\cost S(x).
\]
Therefore, $g(x,t)\ge0$ whenever $t\ge x$.
Furthermore, for $t_2\ge t_1\ge x$,
\begin{align*}
g(x,t_1)-g(x,t_2)
&=
\expect{
\indicate{t_1<\sigma\le t_2}
\left[
w(x)F(\sigma)-\cost S(x)
\right]
}\ge0.
\end{align*}
Hence $g(x,t)$ is nonincreasing in $t$ for $t\ge x$.
Combining these observations gives
\begin{align*}
U(\Strategy,t(\cdot);\Payment^\sigma)
&=\int_0^\Strategy g(x,t(x))\,dx\\
&\le\int_0^\Strategy g(x,x)\,dx\\
&\le\int_0^T g(x,x)\,dx\\
&=U(T,t^{*}(x);\Payment^\sigma),
\end{align*}
where $t^*(x)=x$ represents the full-effort plan.
Recall that any feasible effort plan satisfies
$t(x)\ge x$ and $\Strategy\le T$.
The first inequality follows because $g(x,t)$ is
non-increasing in $t\ge x$, and the second because
$g(x,x)\ge0$.
Thus, the proposed payment schedule implements full effort,
and its expected expenditure achieves the minimum in
\eqref{eq:independent_horizon_minimum_cost}.

\textbf{Step 5. Uniqueness.}
Suppose another full-effort-implementing schedule
$\widetilde{\Payment}$ attains the same minimum cost.
By \eqref{eq:independent_horizon_cost_slack},
its incentive slack must satisfy
$\widetilde M(t)=0$ almost everywhere.
The constructed schedule $\Payment^\sigma$ also has
zero slack almost everywhere.

Define
\[
\proofedit{
\Delta R(t)\equiv
Q(t)\left[
\widetilde{\Payment}(t)-\Payment^\sigma(t)
\right].
}
\]
Subtracting the two instances of
\eqref{eq:independent_horizon_M} gives
\[
\proofedit{
w(t)\Delta R(t)
=
\int_t^T \rate w(s)\Delta R(s)\,ds
}
\qquad\text{for almost every }t.
\]
\proofedit{
Let
\[
J(t)\equiv\int_t^T \rate w(s)\Delta R(s)\,ds.
\]
Finite expected expenditure under both schedules implies
that $J$ is absolutely continuous. Hence $J(T)=0$ and
\[
J'(t)
=
-\rate w(t)\Delta R(t)
=
-\rate J(t)
\qquad\text{almost everywhere}.
\]
The unique solution is $J(t)\equiv0$.
Since $w(t)>0$, this implies $\Delta R(t)=0$
almost everywhere, or equivalently,
}
$\widetilde{\Payment}(t)=\Payment^\sigma(t)$
for almost every $t$ with $Q(t)>0$.
Thus the minimizing payment schedule is unique
up to Lebesgue-null sets on dates with positive
continuation probability.
\end{proof}

\begin{proof}[Proof of \cref{prop:constant_payment_optimal}]
Apply \cref{lem:independent_termination_payments} with
$\sigma\equiv T$. Since the termination horizon is deterministic,
it is independent of the firm's feasibility, effort, and
breakthrough process. Moreover, for every $t<T$,
$\proofedit{\prob{\sigma>t}=1}$.
The optimal payment schedule characterized by the lemma
therefore reduces to
\[
    \Payment^*(t)
    =
    \proofedit{
    \expect{\left.
        \frac{\cost}{\rate p_\sigma}
        \,\right|\,\sigma>t}
    }
    =
    \frac{\cost}{\rate p_T},
    \qquad t<T.
\]
\proofedit{Taking the constant extension to $t=T$ gives an
admissible absolutely continuous schedule.}
The lemma establishes that this schedule implements full effort
and uniquely minimizes expected expenditure up to
Lebesgue-null sets. Since the admissible payment schedules
are absolutely continuous, equality almost everywhere
implies equality everywhere on $[0,T]$.
Thus the constant schedule is the unique minimizer.
\end{proof}

\begin{proof}[Proof of \cref{prop:stochastic_termination_optimal_payment}]
Apply \cref{lem:independent_termination_payments} to
$\sigma=\tau\wedge T$. For each $t<T$,
$\{\sigma>t\}=\{\tau>t\}$,
\proofedit{
so the lemma yields
\[
\Payment^*(t)
=
\expect{\left.
\frac{\cost}{\rate p_{\tau\wedge T}}
\,\right|\,\tau>t},
\qquad
t<T\ \text{with}\ \prob{\tau>t}>0.
\]
The minimum expected expenditure is
\[
\totalcost^{\min}(\tau\wedge T)
=
\expect{\mathcal C(\tau\wedge T)}.
\]
Implementation of full effort and uniqueness up to
Lebesgue-null sets on active dates follow directly from
the lemma.
}
\end{proof}

\begin{proof}[Proof of \cref{thm:optimal_interdependent_timevarying}]
\proofedit{
Maintain the information and compensation rules in the theorem:
intermediate rival successes are concealed, termination is
announced immediately, and no compensation is paid upon
termination. A success that triggers termination receives
its breakthrough payment.

For the comparison across stopping rules, let $\eta_t$ denote
the potential success-count process when all firms exert full
effort until their own success or $T$, before applying any
termination rule. Under a stopping rule $\tau\le T$,
the realized success count at date $t$ is $\eta_{t\wedge\tau}$.
All stopping rules below are evaluated on this common
potential process.
}

\textbf{Step 1. Minimum expenditure conditional on a stopping rule.}
Fix an arbitrary admissible symmetric stopping rule $\tau$.
Consider firm $i$ before its own breakthrough.
Because $\tau$ is based on observed breakthroughs,
applying the stopping rule with firm $i$ kept unresolved
gives a termination date $\sigma_i^\tau\in[0,T]$
determined solely by the other firms' breakthrough histories.
\proofedit{
It is therefore independent of firm $i$'s feasibility and
potential breakthrough process, and it is unaffected by
unilateral changes in the firm's effort.
}
Moreover, before firm $i$ succeeds, its contract is active
exactly when $\sigma_i^\tau$ has not occurred.
The other firms' success histories are not disclosed to
firm $i$, so the firm observes no additional signal about
this termination date beyond its survival.
Consequently, \cref{lem:independent_termination_payments}
applies with $\sigma=\sigma_i^\tau$.
\proofedit{Using the same cost function $\mathcal C$ as in
the main text,} we obtain the minimum expected payment
to firm $i$:
\begin{equation}
\label{eq:interdependent_individual_min_cost}
\totalcost_i^{\min}(\tau)
=
\proofedit{\expect{\mathcal C(\sigma_i^\tau)}}.
\end{equation}
\proofedit{This minimum is attained by}
\[
\Payment_i^\tau(t)
=
\proofedit{
\expect{\left.
\frac{\cost}{\rate p_{\sigma_i^\tau}}
\,\right|\,\sigma_i^\tau>t},
\qquad
t<T\ \text{with}\ \prob{\sigma_i^\tau>t}>0,
}
\]
\proofedit{
which implements full effort until success or termination.
Payments at all other dates may be set to zero.
}
Symmetry of $\tau$ and of the approaches implies that
this payment schedule is common to all firms.
Thus, after minimizing transfers for a fixed $\tau$,
the principal needs only to choose the stopping rule.

\textbf{Step 2. The principal's objective.}
Let $Q_i^\tau(t)\equiv\proofedit{\prob{\sigma_i^\tau>t}}$.
Since $\proofedit{\mathcal C(0)}=0$,
\begin{align*}
\sum_{i=1}^n \totalcost_i^{\min}(\tau)
&=
\proofedit{
\sum_{i=1}^n\expect{\mathcal C(\sigma_i^\tau)}
}\\
&=
\int_0^T \proofedit{\mathcal C'(t)}
\sum_{i=1}^n Q_i^\tau(t)\,dt\\
&=
\int_0^T \ell(t)S(t)
\sum_{i=1}^n Q_i^\tau(t)\,dt,
\end{align*}
where
\[
\ell(t)
\equiv
\frac{\proofedit{\mathcal C'(t)}}{S(t)}
=
\cost
\frac{\delta e^{-\rate t}+(1-\delta)e^{\rate t}}
{1-\delta+\delta e^{-\rate t}}
\]
\proofedit{
is the effective expenditure flow per unresolved active firm.
It is an auxiliary flow cost; dividing it by the breakthrough
intensity $\rate p_t$ gives the main-text threshold $\kappa(t)$.
}
Under full effort, a given firm has not succeeded by $t$
with probability $S(t)$.
Independence of its own potential breakthrough from
$\sigma_i^\tau$ and the definition of the counterfactual
horizon imply
\[
S(t)Q_i^\tau(t)
=
\proofedit{
\prob{i\text{ unresolved at }t,\ \tau>t}
}.
\]
Summing over firms gives
\begin{equation}
\label{eq:interdependent_survival_count_identity}
S(t)\sum_{i=1}^n Q_i^\tau(t)
=
\proofedit{
\expect{(n-\eta_t)\indicate{\tau>t}}
}.
\end{equation}
This implies that the minimum total expected payment under
$\tau$ is
\begin{equation}
\label{eq:interdependent_min_total_spending}
\sum_{i=1}^n \totalcost_i^{\min}(\tau)
=
\proofedit{
\expect{\int_0^\tau(n-\eta_t)\ell(t)\,dt}
}.
\end{equation}

Thus, the principal's objective for any fixed $\tau$ can
be reduced to
\begin{equation}
\label{eq:interdependent_reduced_objective}
\obj(\tau)
=
\proofedit{
\expect{
\val(\eta_\tau)
-\int_0^\tau(n-\eta_t)\ell(t)\,dt
}
}.
\end{equation}
Under full effort, conditional on $\eta_t=k$, each of the
$n-k$ unresolved firms has breakthrough intensity $\rate p_t$.
Hence $\eta_t$ is a pure-birth process with intensity
$(n-k)\rate p_t$ in state $k$.
Applying its compensator to the bounded stopping time
$\tau$ gives
\[
\proofedit{\expect{\val(\eta_\tau)}}
=
\val(0)
+
\proofedit{
\expect{\int_0^\tau
(n-\eta_t)\rate p_t\Delta\val(\eta_t)\,dt}
},
\]
where the integrand is taken to be zero at $\eta_t=n$,
and $\eta_{t-}$ may equivalently replace $\eta_t$.
\proofedit{The same zero-integrand convention applies below
whenever $\eta_t=n$.}
It follows that
\begin{equation}
\label{eq:interdependent_surplus_flow}
\obj(\tau)
=
\val(0)
+
\proofedit{
\expect{\int_0^\tau
(n-\eta_t)\rate p_t
\big[\Delta\val(\eta_t)-\kappa(t)\big]\,dt}
},
\end{equation}
where
\[
\kappa(t)
\equiv
\frac{\ell(t)}{\rate p_t}
=
\proofedit{
\frac{\mathcal C'(t)}
{\delta\rate e^{-\rate t}}
}
=
\frac{\cost}{\rate}
\left(1+\frac{1-\delta}{\delta}e^{2\rate t}\right).
\]
\proofedit{This is exactly the marginal expenditure per
additional expected success used in the main text.}

\textbf{Step 3. Optimal stopping time.}
Since $\val$ is concave, $\Delta\val(k)$ is nonincreasing
in $k$. The process $\eta_t$ is nondecreasing in calendar
time, and $\kappa(t)$ is nondecreasing.
Therefore, along every realized breakthrough path,
$\Delta\val(\eta_t)-\kappa(t)$ is nonincreasing while
$\eta_t<n$.
Once it is nonpositive, it cannot become positive at any
later date; at $\eta_t=n$, the entire surplus flow in
\eqref{eq:interdependent_surplus_flow} is zero.

Let
\[
G(t,k)\equiv
(n-k)\rate p_t[\Delta\val(k)-\kappa(t)],
\qquad k<n,
\quad G(t,n)\equiv0.
\]
The stopping time $\tau^*$ defined in
\eqref{eq:optimal_interdependent_stopping} is the first date
at which this surplus flow becomes nonpositive, or the
terminal deadline if no such date occurs.
Hence $G(t,\eta_t)>0$ for almost every $t<\tau^*$
and $G(t,\eta_t)\le0$ for almost every $t>\tau^*$,
\proofedit{with both statements restricted to $t\in[0,T]$}.
For every other admissible stopping time $\tau$,
on the same potential breakthrough path,
\[
\int_0^\tau G(t,\eta_t)\,dt
\le
\int_0^{\tau^*}G(t,\eta_t)\,dt.
\]
Taking expectations and using
\eqref{eq:interdependent_surplus_flow} establishes
$\obj(\tau)\le\obj(\tau^*)$.

Finally, take any admissible payment schedules implementing
full effort under any admissible symmetric stopping rule
$\tau$. By Step 1, their total expected payments are at least
$\proofedit{\sum_{i=1}^n\expect{\mathcal C(\sigma_i^\tau)}}$,
so their principal objective cannot exceed
$\obj(\tau)\le\obj(\tau^*)$.
\proofedit{
Taking $\Payment_i^*(t)=\Payment_i^{\tau^*}(t)$ gives
the schedules in
\eqref{eq:optimal_interdependent_timevarying_payment}.
}
These schedules attain the expenditure bound under
$\tau^*$ and globally implement full effort by
\cref{lem:independent_termination_payments}.
Thus the proposed symmetric stopping rule and common
time-varying payment schedule are jointly optimal.
\proofedit{
This also covers $\tau^*=0$: no experimentation occurs,
expected payments are zero, and there are no active dates
on which the conditional payment formula must be evaluated.
}
\end{proof}

\clearpage
\bibliographystyle{apalike}
\bibliography{references_corrected}

@techreport{gao2021,
  author       = {{U.S. Government Accountability Office}},
  title        = {{Operation Warp Speed}: Accelerated {COVID-19} Vaccine Development Status and Efforts to Address Manufacturing Challenges},
  institution  = {U.S. Government Accountability Office},
  year         = {2021},
  month        = feb,
  number       = {GAO-21--319},
  address      = {Washington, D.C.},
  url          = {https://www.gao.gov/products/gao-21-319},
}

@article{tabarrok2023,
  author       = {Tabarrok, Alex},
  title        = {{Operation Warp Speed}: Negative and Positive Lessons for New Industrial Policy},
  journal      = {Innovations: Technology, Governance, Globalization},
  year         = {2025},
  volume       = {14},
  number       = {1--2},
  pages        = {2--22},
  doi          = {10.1162/inov.a.268},
}

@article{aghion2005competition,
  author       = {Aghion, Philippe and Bloom, Nick and Blundell, Richard and Griffith, Rachel and Howitt, Peter},
  title        = {Competition and Innovation: An Inverted-{U} Relationship},
  journal      = {The Quarterly Journal of Economics},
  year         = {2005},
  volume       = {120},
  number       = {2},
  pages        = {701--728},
  doi          = {10.1162/0033553053970214},
}

@article{gilbert2006schumpeter,
  author       = {Gilbert, Richard},
  title        = {Looking for {Mr. Schumpeter}: Where Are We in the Competition--Innovation Debate?},
  journal      = {Innovation Policy and the Economy},
  year         = {2006},
  volume       = {6},
  pages        = {159--215},
  doi          = {10.1086/ipe.6.25056183},
}

@article{shan2017optimal,
  author       = {Shan, Yaping},
  title        = {Optimal Contracts for Research Agents},
  journal      = {The RAND Journal of Economics},
  year         = {2017},
  volume       = {48},
  number       = {1},
  pages        = {94--124},
  doi          = {10.1111/1756-2171.12169},
}

@article{poblete2017managing,
  author       = {Poblete, Joaqu{\'i}n and Spulber, Daniel F.},
  title        = {Managing Innovation: Optimal Incentive Contracts for Delegated {R\&D} with Double Moral Hazard},
  journal      = {European Economic Review},
  year         = {2017},
  volume       = {95},
  pages        = {38--61},
  doi          = {10.1016/j.euroecorev.2017.03.004},
}

@article{taylor1995digging,
  author       = {Taylor, Curtis R.},
  title        = {Digging for Golden Carrots: An Analysis of Research Tournaments},
  journal      = {American Economic Review},
  year         = {1995},
  volume       = {85},
  number       = {4},
  pages        = {872--890},
}

@article{che2003optimal,
  author       = {Che, Yeon-Koo and Gale, Ian},
  title        = {Optimal Design of Research Contests},
  journal      = {American Economic Review},
  year         = {2003},
  volume       = {93},
  number       = {3},
  pages        = {646--671},
  doi          = {10.1257/000282803322157025},
}

@article{bimpikis2019designing,
  author       = {Bimpikis, Kostas and Ehsani, Shayan and Mostagir, Mohamed},
  title        = {Designing dynamic contests},
  journal      = {Operations Research},
  year         = {2019},
  volume       = {67},
  number       = {2},
  pages        = {339--356},
  doi          = {10.1287/opre.2018.1823},
}

@article{zhang2024optimal,
  author       = {Zhang, Mengxi},
  title        = {Optimal contests with incomplete information and convex effort costs},
  journal      = {Theoretical Economics},
  year         = {2024},
  volume       = {19},
  number       = {1},
  pages        = {95--129},
  doi          = {10.3982/TE4259},
}

@article{halac2017contests,
  author       = {Halac, Marina and Kartik, Navin and Liu, Qingmin},
  title        = {Contests for experimentation},
  journal      = {Journal of Political Economy},
  year         = {2017},
  volume       = {125},
  number       = {5},
  pages        = {1523--1569},
  doi          = {10.1086/693040},
}

@article{halac2016optimal,
  author       = {Halac, Marina and Kartik, Navin and Liu, Qingmin},
  title        = {Optimal Contracts for Experimentation},
  journal      = {The Review of Economic Studies},
  year         = {2016},
  volume       = {83},
  number       = {3},
  pages        = {1040--1091},
  doi          = {10.1093/restud/rdw013},
}

@article{fan2024contest,
  author       = {Fan, Qintao and Johnson, Nicole},
  title        = {Contest Schemes and Dynamic Incentive Provision},
  journal      = {Management Science},
  year         = {2024},
  volume       = {70},
  number       = {10},
  pages        = {7265--7282},
  doi          = {10.1287/mnsc.2022.00035},
}

@article{Ely2023contest,
  author       = {Ely, Jeffrey C. and Georgiadis, George and Khorasani, Sina and Rayo, Luis},
  title        = {Optimal Feedback in Contests},
  journal      = {The Review of Economic Studies},
  year         = {2023},
  volume       = {90},
  number       = {5},
  pages        = {2370--2394},
  doi          = {10.1093/restud/rdac074},
  url          = {https://doi.org/10.1093/restud/rdac074},
}

@article{bergemann1998venture,
  author       = {Bergemann, Dirk and Hege, Ulrich},
  title        = {Venture capital financing, moral hazard, and learning},
  journal      = {Journal of Banking \& Finance},
  year         = {1998},
  volume       = {22},
  number       = {6--8},
  pages        = {703--735},
  doi          = {10.1016/S0378-4266(98)00017-X},
}

@article{bergemann2005financing,
  author       = {Bergemann, Dirk and Hege, Ulrich},
  title        = {The financing of innovation: Learning and stopping},
  journal      = {The RAND Journal of Economics},
  year         = {2005},
  volume       = {36},
  number       = {4},
  pages        = {719--752},
}

@article{GreenStokey1983,
  author       = {Green, Jerry R. and Stokey, Nancy L.},
  title        = {A Comparison of Tournaments and Contracts},
  journal      = {Journal of Political Economy},
  year         = {1983},
  volume       = {91},
  number       = {3},
  pages        = {349--364},
  doi          = {10.1086/261153},
}

@article{LazearRosen1981,
  author       = {Lazear, Edward P. and Rosen, Sherwin},
  title        = {Rank-Order Tournaments as Optimum Labor Contracts},
  journal      = {Journal of Political Economy},
  year         = {1981},
  volume       = {89},
  number       = {5},
  pages        = {841--864},
  doi          = {10.1086/261010},
}

@misc{frontier280earth2024,
  author       = {{Frontier Climate}},
  title        = {Frontier buyers sign \$40M in offtake agreements with {280 Earth}},
  year         = {2024},
  month        = jul,
  howpublished = {Frontier Climate},
  url          = {https://frontierclimate.com/writing/280-earth},
  urldate      = {2026-11-23},
  note         = {Accessed November 23, 2026},
}

@misc{frontierlithos2023,
  author       = {{Frontier Climate}},
  title        = {Frontier buyers sign world's first enhanced weathering offtake agreements with {Lithos Carbon}},
  year         = {2023},
  month        = dec,
  howpublished = {Frontier Climate},
  url          = {https://frontierclimate.com/writing/lithos},
  urldate      = {2026-11-23},
  note         = {Accessed November 23, 2026},
}

@misc{frontierexergi2024,
  author       = {{Frontier Climate}},
  title        = {Frontier buyers sign \$48.6M in offtake agreements with {Stockholm Exergi}, in first large {European} deal},
  year         = {2024},
  month        = jun,
  howpublished = {Frontier Climate},
  url          = {https://frontierclimate.com/writing/exergi},
  urldate      = {2026-11-23},
  note         = {Accessed November 23, 2026},
}

@misc{xprize2004,
  author       = {{XPRIZE Foundation}},
  title        = {{Mojave Aerospace Ventures} Wins the Competition That Started It All},
  year         = {2020},
  month        = jan,
  howpublished = {XPRIZE Foundation},
  url          = {https://www.xprize.org/news/mojave-aerospace-ventures-wins-the-competition},
  urldate      = {2026-11-23},
  note         = {Accessed November 23, 2026},
}

@misc{darpa2005,
  author       = {{Defense Advanced Research Projects Agency}},
  title        = {The {DARPA Grand Challenge}: Ten Years Later},
  year         = {2014},
  month        = mar,
  howpublished = {Defense Advanced Research Projects Agency},
  url          = {https://www.darpa.mil/news/2014/grand-challenge-ten-years-later},
  urldate      = {2026-11-23},
  note         = {Accessed November 23, 2026},
}

@article{benkert2020designing,
  author       = {Benkert, Jean-Michel and Letina, Igor},
  title        = {Designing dynamic research contests},
  journal      = {American Economic Journal: Microeconomics},
  year         = {2020},
  volume       = {12},
  number       = {4},
  pages        = {270--289},
  doi          = {10.1257/mic.20180263},
}

@article{moroni2022experimentation,
  author       = {Moroni, Sofia},
  title        = {Experimentation in organizations},
  journal      = {Theoretical Economics},
  year         = {2022},
  volume       = {17},
  number       = {3},
  pages        = {1403--1450},
  doi          = {10.3982/TE3069},
}

@article{strulovici2010learning,
  author       = {Strulovici, Bruno},
  title        = {Learning while voting: Determinants of collective experimentation},
  journal      = {Econometrica},
  year         = {2010},
  volume       = {78},
  number       = {3},
  pages        = {933--971},
  doi          = {10.3982/ECTA8011},
}

@article{fullerton1999auctioning,
  author       = {Fullerton, Richard L. and McAfee, R. Preston},
  title        = {Auctioning entry into tournaments},
  journal      = {Journal of Political Economy},
  year         = {1999},
  volume       = {107},
  number       = {3},
  pages        = {573--605},
  doi          = {10.1086/250072},
}

@article{letina2023optimal,
  author       = {Letina, Igor and Liu, Shuo and Netzer, Nick},
  title        = {Optimal contest design: Tuning the heat},
  journal      = {Journal of Economic Theory},
  year         = {2023},
  volume       = {213},
  pages        = {105616},
  doi          = {10.1016/j.jet.2023.105616},
}

@article{akcigit2016role,
  author       = {Akcigit, Ufuk and Liu, Qingmin},
  title        = {The role of information in innovation and competition},
  journal      = {Journal of the European Economic Association},
  year         = {2016},
  volume       = {14},
  number       = {4},
  pages        = {828--870},
  doi          = {10.1111/jeea.12153},
}

@article{kremer2022designing,
  author       = {Kremer, Michael and Levin, Jonathan and Snyder, Christopher M.},
  title        = {Designing advance market commitments for new vaccines},
  journal      = {Management Science},
  year         = {2022},
  volume       = {68},
  number       = {7},
  pages        = {4786--4814},
  doi          = {10.1287/mnsc.2021.4163},
}

@article{ely2025feedback,
  author       = {Ely, Jeffrey C. and Georgiadis, George and Rayo, Luis},
  title        = {Feedback design in dynamic moral hazard},
  journal      = {Econometrica},
  year         = {2025},
  volume       = {93},
  number       = {2},
  pages        = {597--621},
  doi          = {10.3982/ECTA21871},
}

@article{che2021prizes,
  author       = {Che, Yeon-Koo and Iossa, Elisabetta and Rey, Patrick},
  title        = {Prizes versus contracts as incentives for innovation},
  journal      = {The Review of Economic Studies},
  year         = {2021},
  volume       = {88},
  number       = {5},
  pages        = {2149--2178},
  doi          = {10.1093/restud/rdaa092},
}

@article{keller2015breakdowns,
  author       = {Keller, Godfrey and Rady, Sven},
  title        = {Breakdowns},
  journal      = {Theoretical Economics},
  year         = {2015},
  volume       = {10},
  number       = {1},
  pages        = {175--202},
  doi          = {10.3982/TE1670},
}

@article{board2013reputation,
  author       = {Board, Simon and Meyer-ter-Vehn, Moritz},
  title        = {Reputation for Quality},
  journal      = {Econometrica},
  year         = {2013},
  volume       = {81},
  number       = {6},
  pages        = {2381--2462},
  doi          = {10.3982/ECTA9039},
}

@misc{far522492,
  author       = {{Federal Acquisition Regulatory Council}},
  title        = {{Federal Acquisition Regulation}, 52.249-2: Termination for Convenience of the Government (Fixed-Price)},
  year         = {{n.d.}},
  howpublished = {Acquisition.gov},
  url          = {https://www.acquisition.gov/far/52.249-2},
  urldate      = {2026-11-23},
  note         = {Accessed November 23, 2026},
}

\end{document}